\documentclass[submission, Phys]{SciPost}

\pdfoutput=1
\usepackage[utf8]{inputenc} % input encodings (allow UTF-8 input)
\usepackage[T1]{fontenc} 	% selecting font encodings (use 8-bit T1 fonts)
\usepackage[english]{babel} % multilingual support (English language/hyphenation)

\usepackage[bitstream-charter]{mathdesign}
\usepackage{geometry} 		% flexible and complete interface to document dimensions
\usepackage{amsmath} 		% math package (American Mathematical Society)
\usepackage{mathtools} 		% math package (fixes various deficiencies of amsmath)
\usepackage{float} 			% floating objects such as figures and tables
\usepackage{graphicx} 		% enhanced support for graphics
\usepackage{tabularx} 		% tabulars with adjustable-width columns
\usepackage{booktabs} 		% professional-quality tables
\usepackage{color, xcolor} 	% foreground and background colour management
\usepackage{pdfpages} 		% inclusion of external multi-page PDF documents
\usepackage{extarrows} 		% extra arrows beyond those provided in amsmath
\usepackage{multirow} 		% create tabular cells spanning multiple rows
\usepackage{multicol} 		% intermix single and multiple columns
\usepackage{caption} 		% customising captions in floating environments
\usepackage{subcaption} 	% support for sub-captions (captions for subfigures)
\usepackage{enumitem} 		% control layout of itemize, enumerate, description
\usepackage{xspace} 		% define commands that appear not to eat spaces
\usepackage{stackrel} 		% enhancement to the \stackrel command
\usepackage{tikz} 			% create PostScript and PDF graphics
\usetikzlibrary{calc}
\usepackage{braket} 		% Dirac bra-ket notation
\usepackage{bm} 			% access bold symbols in maths mode
\usepackage{tensor} 		% typeset tensors (tensor-style super- and subscripts)
\usepackage{slashed} 		% slash through characters (Feynman slash notation)
\usepackage{siunitx} % SI units package (typesetting values with units)
\usepackage{lastpage} 		% reference last page
\usepackage{cite} 			% improved citation handling
\usepackage[normalem]{ulem} % package for underlining
\usepackage{fontawesome} 	% access to web-related icons
\usepackage{tocloft} 		% control over the typography of the TOC, LOF, and LOT
\usepackage{titlesec} 		% interface to sectioning commands (various title styles)
\usepackage{doi} 			% create correct hyperlinks for DOI numbers
\usepackage[most]{tcolorbox} 					% coloured and framed text boxes
\usepackage[capitalize]{cleveref} 	% intelligent cross-referencing
\usepackage[nottoc, notlot, notlof]{tocbibind} 	% add references/index/contents to TOC
\usepackage[ruled, vlined]{algorithm2e} 		% floating algorithm environment
\usepackage{makecell}
\usepackage{pifont}
\usepackage{stackrel}
\usepackage{rotating}

\hypersetup{
	pdftitle={A Global View of the EDM Landscape},
    linktoc=section,
	breaklinks=true,			% splits links across lines
	colorlinks=true, 			% false: boxed links, true: colored links
	linkcolor={red!50!black}, 	% color of internal links (sections, pages, etc.)
	citecolor={blue!50!black}, 	% color of citation links (links to bibliography)
	urlcolor={blue!80!black} 	% color of URL links (external links)
} 
\DeclareSymbolFont{usualmathcal}{OMS}{cmsy}{m}{n}
\DeclareSymbolFontAlphabet{\mathcal}{usualmathcal}

\SetArgSty{textnormal}
\SetKwComment{Comment}{{\small\#}~}{}
\SetCommentSty{mycommfont}

\setitemize{itemsep=2pt, parsep=0pt} 				% adjust itemize environment
\setenumerate{itemsep=2pt, parsep=0pt} 				% adjust enumerate environment
\crefname{section}{Sec.}{Secs.}
\crefname{equation}{Eq.}{Eqs.}
\crefname{figure}{Fig.}{Figs.}
\crefname{table}{Tab.}{Tabs.}
\Crefname{section}{Section}{Sections}
\Crefname{equation}{Equation}{Equations}
\Crefname{figure}{Figure}{Figures}
\Crefname{table}{Table}{Tables}
\newcommand{\Langle}{\bigl\langle}
\newcommand{\Rangle}{\bigr\rangle}

\newcommand{\qqquad}{\qquad\quad}
\newcommand{\qqqquad}{\qquad\qquad}

\DeclareMathOperator{\sgn}{sgn}
\newcommand{\sfitter}{\textsc{SFitter}\xspace}

\newcommand{\lag}{\mathscr{L}}
\newcommand{\ope}{\mathcal{O}}

\newcommand{\arXiv}[2][]{%
	\ifthenelse{\equal{#1}{}}%
	{\href{http://arxiv.org/abs/#2}{arXiv:#2}}%
	{\href{http://arxiv.org/abs/#2}{arXiv:#2~[#1]}}}

\def\slashchar#1{\setbox0=\hbox{$#1$}           % set a box for #1
   \dimen0=\wd0                                 % and get its size
   \setbox1=\hbox{/} \dimen1=\wd1               % get size of /
   \ifdim\dimen0>\dimen1                        % #1 is bigger
      \rlap{\hbox to \dimen0{\hfil/\hfil}}      % so center / in box
      #1                                        % and print #1
   \else                                        % / is bigger
      \rlap{\hbox to \dimen1{\hfil$#1$\hfil}}   % so center #1
      /                                         % and print /
   \fi}

\newcommand{\tikznode}[2]{%
\ifmmode%
\tikz[remember picture,baseline=(#1.base),inner sep=0pt] \node (#1) {$#2$};%
\else
\tikz[remember picture,baseline=(#1.base),inner sep=0pt] \node (#1) {#2};%
\fi}

\def\mathswitchr#1{\relax\ifmmode{\text{#1}}\else$\text{#1}$\xspace\fi}
\def\mathswitch#1{\relax\ifmmode#1\else$#1$\xspace\fi}

\graphicspath{{./figs/}}

\begin{document}
%\begin{fmffile}{feynman}

\begin{center}
{\Large \textbf{A Global View of the EDM Landscape}}
\end{center}

\begin{center}
Skyler Degenkolb\textsuperscript{1},
Nina Elmer\textsuperscript{2}, \\
Tanmoy Modak\textsuperscript{2},
Margarete M\"uhlleitner\textsuperscript{3}, and
Tilman Plehn\textsuperscript{2,4}
\end{center}

\begin{center}
{\bf 1} Physikalisches Institut, Universit\"at Heidelberg, Germany\\
{\bf 2} Institut f\"ur Theoretische Physik, Universit\"at Heidelberg, Germany\\
{\bf 3} Institute for Theoretical Physics, Karlsruhe Institute of Technology, Karlsruhe, Germany\\
{\bf 4} Interdisciplinary Center for Scientific Computing (IWR), Universit\"at Heidelberg, Germany
\end{center}

%\begin{center}
%\today
%\end{center}

\section*{Abstract}
{\bf Permanent electric dipole moments (EDMs) are sensitive probes of the symmetry
  structure of elementary particles, which in turn is closely tied to
  the baryon asymmetry in the universe.  A meaningful
  interpretation framework for EDM measurements has to be based on
  effective quantum field theory. We interpret the measurements performed to date
  in terms of a hadronic-scale Lagrangian, using the SFitter global
  analysis framework. We find that part
  of this Lagrangian is constrained very well, while some of the 
  parameters suffer from too few high-precision measurements.
  Theory uncertainties lead to weaker model constraints, but
  can be controlled within the global analysis.}

% For convenience during refereeing: line numbers
%\linenumbers

%\vspace{10pt}
\vspace{1.8pt}
\noindent\rule{\textwidth}{1pt}
\tableofcontents\thispagestyle{fancy}
%\noindent\rule{\textwidth}{1pt}
%\vspace{10pt}
\vspace{2pt}

\newpage
%%%%%%%%%%%%%%%%%%%%%%%%%%%%%%%%%%%%%%%%%%%%%%%%%%%%%%%%%%%%%%%%%%%%%%%%
\section{Introduction}
\label{sec:intro}

While the Standard Model (SM) is structurally complete, it fails to
explain key observations and therefore fails to qualify as a complete
theory of elementary particles. The two leading shortcomings are a
missing dark matter agent and a missing explanation of the baryon
asymmetry in the Universe. For the latter, the Sakharov
conditions~\cite{Sakharov:1967dj} tell us precisely which structures
would be required: (i) C- and CP-violation, (ii) baryon number
violation, and (iii) a deviation from thermal equilibrium. The first
condition is especially interesting, because one can read it off the
fundamental Lagrangian and build, for instance, models for electroweak
baryogenesis around
it~\cite{Kuzmin:1985mm,Shaposhnikov:1987tw,Nelson:1991ab,Morrissey:2012db,Bodeker:2020ghk}. In
the SM, CP symmetry is violated through the fermion mixing among three
generations and through the adjoint gluon field strength. Measurements
of the neutron EDM, consistent with zero, clearly show that CP
violation in QCD is too small to explain the observed baryon
asymmetry~\cite{Cohen:1993nk,Gavela:1993ts,Huet:1994jb,Gavela:1994ds,Gavela:1994dt,Riotto:1999yt}. Physics
beyond the Standard Model (BSM), explaining the baryon asymmetry,
should then violate CP to a greater extent and with observable
consequences.

Over recent years, EDM measurements have been performed on a wide
range of particles, atoms, and
molecules~\cite{Pospelov:2005pr,Engel:2013lsa,Yamanaka:2017mef,Chupp:2017rkp}. None
has been able to confirm a signal for CP violation.  To judge
their combined impact on BSM physics, we need to combine them in a
consistent framework. Such a global analysis must start from a 
Lagrangian and express the experimental limits on all EDMs in terms of
its fundamental parameters. One choice is a hadronic-scale
Lagrangian, describing the interactions of nucleons, pions, and
electrons at the GeV
scale~\cite{Pospelov:2005pr,Mereghetti:2010kp,Chupp:2014gka,Chupp:2017rkp,Dekens:2018bci}. Alternatively,
we can combine EDM measurements at the weak scale, using an effective 
extension of the renormalizable SM
Lagrangian~~\cite{McKeen:2012av,Engel:2013lsa,Dwivedi:2015nta,Chien:2015xha,Cirigliano:2016njn,Cirigliano:2016nyn,Dekens:2018bci}. Going
beyond effective field theories (EFTs), an ultimate link to the
cosmological motivation requires a UV-complete extension of the SM at
the weak scale, for instance leptoquark models~\cite{Dekens:2018bci},
extended Higgs
sectors~\cite{Inoue:2014nva,Fontes:2017zfn,Altmannshofer:2020shb}, left-right symmetry~\cite{RAMSEYMUSOLF2021136136}, or
supersymmetry~\cite{Barger:2001nu,Ellis:2008zy,Li:2010ax,Cheung:2011wn,King:2015oxa,Dao:2022rui}. Technically,
all global analyses~\cite{Gaul:2023hdd} face similar challenges which we will tackle in
this paper for the hadronic-scale Lagrangian.

We perform a global EDM analysis with the \sfitter analysis
tool~\cite{Lafaye:2004cn,Lafaye:2007vs,Lafaye:2009vr,Brivio:2022hrb,Elmer:2023wtr},
with its focus on the comprehensive treatment of uncertainties.  Our
analysis provides state-of-the-art limits on   the multi-dimensional
model parameter space (with no assumptions made about the underlying sources of CP violation). It also allows us to judge the impact of new
or proposed measurements and to identify shortcomings in relating
measurements to fundamental parameters\footnote{Only a proper
  fundamental physics interpretation can make full use of experimental
  limits.}, while remaining easy to repeat or adapt in response to new
theoretical or experimental inputs. For this purpose it is crucial
to interpret all measurements in the same fundamental physics
framework and to include all uncertainties and correlations, including theory
uncertainties, even though these typically lack a statistical
interpretation~\cite{Ghosh:2022lrf}.

We start by introducing a consistent hadronic-scale Lagrangian, with
properly chosen Wilson coefficients for our global analysis, in
Sec.~\ref{sec:ops}. We then use this Lagrangian to provide predictions
for the measured EDMs, as detailed in Sec.~\ref{sec:global}. We start
our global analysis without theory uncertainties in
Sec.~\ref{sec:results}, to understand the relations among different EDM
measurements in terms of the hadronic-scale Lagrangian.  To extract
correlations and limits on single model parameters we employ a profile
likelihood. We find that the current EDM measurements
define a subspace of well-constrained model parameters and an
orthogonal subspace of poorly constrained parameters with narrow
correlation patterns. Adding theory uncertainties on the relations
between Lagrangian parameters and observables in
Sec.~\ref{sec:results_th} degrades the interpretation in terms of
fundamental physics.  We emphasize that this degradation does not cut
into the discovery potential of EDM measurements, as probes of
fundamental symmetries of elementary particles, but it hampers their
interpretation as limits in fundamental physics.

%%%%%%%%%%%%%%%%%%%%%%%%%%%%%%%%%%%%%%%%%%%%%%%%%%%%%%%%%%%%%%%%%%%%%%%%
\section{EDM Lagrangian}
\label{sec:ops}

Because new sources of CP violation are motivated by cosmology and can be related to
physics beyond the SM at and above the weak scale, we start by
introducing CP violation into the weak-scale Lagrangian in
Sec.~\ref{eq:ops_weak}, relate this to the GeV-scale Lagrangian
in Sec.~\ref{eq:ops_had}, and use simple matching arguments to
simplify this hadronic-scale Lagrangian which we use as the interpretation
framework for our global analysis in Sec.~\ref{eq:ops_match}.  A
detailed analysis of CP-violating new physics at the electroweak scale
includes a renormalization group evolution from the GeV-scale to the 
electroweak scale and is not part of this first study.  Instead,
we focus here on a global analysis at the hadronic scale and the
role of correlations and theory uncertainties.

%%%%%%%%%%%%%%%%%%%%%%%%%%%%%%%%%%%%%%%%%%%%%%%%%%%%%%%%%%%%%%%%%%%%%%%%
\subsection{Weak-scale Lagrangian}
\label{eq:ops_weak}

The operators generating CP violation within and beyond the
SM-Lagrangian, neglecting CP violation in the neutrino
sector~\cite{Engel:2013lsa}, appear in the Lagrangian
\begin{align}
  \lag_\text{CPV} 
    = \lag_\text{CKM} 
    + \lag_{\bar{\theta}} 
    + \lag_\text{dipole} 
    + \lag_\text{Weinberg} 
    + \lag_\text{EFT} \;.
  \label{eq:lag_ew}
\end{align}
The first term represents CP violation at mass dimension four, from
the complex phases in the CKM matrix.  The second arises from the
gluon field strength, also at dimension four,
\begin{align}
  \lag_{\bar{\theta}} = \frac{g_3^2}{32 \pi^2} \bar{\theta} \; \text{Tr} (G^{\mu\nu} \tilde{G}_{\mu\nu} ) \; ,
  \label{eq:lag_theta}
\end{align}
where $g_3$ is the strong coupling, $G^{\mu\nu}$ is the gluon field
strength, $\widetilde{G}^{\mu\nu}=
\epsilon^{\mu\nu\lambda\sigma}G_{\lambda\sigma}/2$ is its dual, and
$\bar{\theta}$ is the re-scaled CP-violating parameter in QCD. The bar
notation indicates that corrections from the quark mass matrix are
included. In principle $\bar\theta$ can also be included as a model parameter, 
for instance in testing a specific BSM model~\cite{RAMSEYMUSOLF2021136136}. 
However, we take the view that the neutron EDM experimentally constrains $\bar\theta$ to 
have such a small value that this fine-tuning problem requires a proper explanation.
This means that at present there is little to be learned from including $\bar\theta$ as a model 
parameter in our global analysis. 

The other three contributions in Eq.\eqref{eq:lag_ew} are
higher-dimensional and not part of a renormalizable extension of the
SM-Lagrangian.  Electric dipole moments of fermions $d_f^E$, and
chromoelectric dipole moments of quarks $d_q^C$, appear at mass
dimension five:
\begin{align}
  \lag_\text{dipole}
  = -\frac{i}{2} F^{\mu\nu} \sum_{f=q,\ell} d_f^E  \; \left( \bar f \sigma_{\mu\nu}\gamma_5 f \right) 
  -\frac{i}{2} g_3  G^a_{\mu\nu} \sum_{f=q} d_q^C  \; \left( \bar q \sigma^{\mu\nu}\gamma_5 T^a q \right) \; ,
  \label{eq:lag_dipole}
\end{align}
where the indices $q$ and $\ell$ denote quarks and leptons of all
three generations. The electromagnetic field strength is
$F^{\mu\nu}$. We chose the metric convention $(1,-\mathbb{1})$ with
$\gamma_5 = - i \gamma_0 \gamma_1 \gamma_2 \gamma_3$. The fermion
spins are $\sigma_{\mu\nu}=i\left[ \gamma_\mu,\gamma_\nu \right]/2$,
and $T^a$ are the $SU(3)$ generators.  The Weinberg operator is again
built out of the gluon field strength and introduces the gluonic
chromo-electric dipole moment $d ^G$,
\begin{align}
  \lag_\text{Weinberg}
  = \frac{1}{3} d^G \; f_{abc} G^a_{\mu\nu} \widetilde{G}^{b \nu\rho} G^{c\;\mu}_\rho \; .
  \label{eq:lag_weinberg}
\end{align}
Additional CP-violation occurs at mass dimension six and higher,
generated at a new physics scale larger than the Higgs vacuum expectation
value, $\Lambda > v$,
\begin{align}
  \lag_\text{EFT}
  = \sum_i \frac{C_i^{(6)}}{\Lambda^2} \ope_i^{(6)} + \mathcal{O}(\Lambda^{-3}) \; .
  \label{eq:lag_eft}
\end{align}
Relevant dimension-6 operators that generate EDMs include the following
semileptonic and quark 4-fermion operators,
\begin{align}
  \lag_\text{EFT} 
  \supset& C_{\ell e qd} \; \left( \bar L^j e_R \right) \; \left( \bar d_R Q_j \right) 
  + C^{(1)}_{\ell e q u} \; \left( \bar L^j e_R \right) \epsilon_{jk} \left( \bar Q^k u_R \right)
  + C^{(3)}_{\ell e q u} \; \left( \bar L^j \sigma_{\mu\nu} e_R \right) \epsilon_{jk} \left( \bar Q^k  \sigma_{\mu\nu} u_R \right)
  \notag\\
  &+ C^{(1)}_{quqd} \; \left( \bar Q^j u_R \right) \epsilon_{jk} \left( \bar Q^k d_R \right)
  + C^{(8)}_{quqd} \; \left( \bar Q^j T^a u_R \right) \epsilon_{jk} \left( \bar Q^k T^a d_R \right) + \text{h.c.}
  \label{eq:fourf}
\end{align}
The dipole moments, the Weinberg operator, and the additional
4-fermion interactions can serve as the basis for an EDM analyses in
the SMEFT
framework~\cite{Pospelov:2005pr,Ellis:2008zy,Li:2010ax,Cheung:2011wn,McKeen:2012av,Engel:2013lsa,Inoue:2014nva,Dwivedi:2015nta,Chien:2015xha,Cirigliano:2016njn,Cirigliano:2016nyn,Dekens:2018bci}.
As always, the set of higher-dimensional SMEFT operators that turn out to be 
most relevant depends on the high-scale BSM model that the SMEFT
represents.  For instance, in supersymmetric models there are no
contributions from $\ope^{(8)}_{quqd}$ and $\ope^{(3)}_{\ell equ}$ at
tree level, and the relative size of down-type and up-type quark
couplings is affected by a potentially large $\tan \beta$ enhancement.

%%%%%%%%%%%%%%%%%%%%%%%%%%%%%%%%%%%%%%%%%%%%%%%%%%%%%%%%%%%%%%%%%%%%%%%%
\subsection{Hadronic-scale Lagrangian}
\label{eq:ops_had}

The challenge with EDMs in view of the Lagrangian of
Eq.\eqref{eq:lag_ew} is that they are measured far below the
electroweak scale, where the propagating degrees of freedom are
leptons, non-relativistic nucleons $N = (p,n)^T$ with average mass
$m_N$, and pions $\vec
\pi=(\pi^+,\pi^0,\pi^-)^T$~\cite{Pospelov:2005pr,Ellis:2008zy,Mereghetti:2010kp,Cheung:2011wn,Chupp:2017rkp,Dekens:2018bci}. We take
the particle physics convention for strong isospin, $\tau_3 \left |n \right> = - \left |n
\right>$

When we evolve our EFT to the experimentally relevant GeV scale, only the lepton
EDMs $d^E_\ell \equiv d_\ell$ in Eq.\eqref{eq:lag_dipole} of the
electroweak Lagrangian in Eq.\eqref{eq:lag_ew} remain unchanged. While
for the weak-scale Lagrangian the relation between the three charged
leptons, for instance the scaling with the lepton mass, raises
interesting questions, we factorize the muon and tau EDMs from the
hadronic-scale Lagrangian. They can be included in the same framework, but given
the systems for which experimental limits are available today, the
indirect constraints on EDMs of heavy leptons are many orders of
magnitude weaker than for the electron and there is little interplay among
the relevant model parameters.

We split the hadronic-scale Lagrangian describing EDMs at the
experimentally relevant GeV scale into
\begin{align}
\lag_\text{had} \supset 
\lag_{N, \text{sr}} + \lag_{\pi N} + \lag_{eN} 
- \frac{i}{2} F^{\mu \nu} d_e \; \bar{e} \sigma_{\mu \nu} \gamma_5 e \; .
\label{eq:lag_had}
\end{align}
While the observable nucleon EDMs $d_N$ can be included directly (as for the electron), it is also possible to separate out the "short-range" parameters $d_N^\text{sr}$ by explicitly calculating pion loop contributions within chiral perturbation theory. The dipole moments of the nucleons in that case now read
\begin{align}
  \lag_{N,\text{sr}}
%  =& -2 \bar N \left[ d_0 +d_1 \tau_3 \right] S_\mu N v_\nu F^{\mu\nu}  \notag \\
%  =& -2 \bar N \left[ (d_0 + d_1) \frac{1 + \tau_3}{2} + (d_0 - d_1) \frac{1 - \tau_3}{2} \right] S_\mu N v_\nu F^{\mu\nu}  \notag \\
  =& -2 \bar N \left[ d_p^\text{sr} \frac{1 + \tau_3}{2} + d_n^\text{sr} \frac{1 - \tau_3}{2} \right] S_\mu N v_\nu F^{\mu\nu}  \; ,
  \label{eq:nucl-sr}
\end{align}
where $S_\mu$ and $v_\mu$ are the spin and velocity of the (non-relativistic) nucleon.
The isoscalar and isovector contributions then determine either
$d_N$ or $d_N^\text{sr}$, as the Lagrangian parameters for the neutron and proton EDMs: we will come back to this choice of Lagrangian parameters for the nucleon EDMs in Sec.~\ref{sec:global_nucl} and App.~\ref{sec:alternative}.

Next come the interactions of pions and nucleons,
\begin{align}
  \lag_{\pi N}
%  =& \textcolor{orange}{-2\bar{N}(\bar{d}_0+\bar{d}_1 \tau_3) S_\mu N v_\nu F^{\mu\nu}}
  =& \bar N \Big[
      g^{(0)}_\pi \vec{\tau}\cdot\vec{\pi}
      + g^{(1)}_\pi \pi^0
      + g^{(2)}_\pi \left( 3\tau_3 \pi^0- \vec{\tau} \cdot \vec{\pi} \right)
      \Big] N \notag \\
  &+ C_1 \; \left( \bar N N \right) \; \partial_\mu\left( \bar N S^\mu \bar N \right)
   + C_2 \; \left( \bar N \vec{\tau} N \right) \cdot \partial_\mu \left( \bar N S^\mu \bar N \vec{\tau} \right) + \cdots 
 \label{eq:piN}
\end{align}
where $\tau$ are the internal-space Pauli matrices and we neglect, for example,
interactions with more than one pion.  The contribution involving
$g^{(2)}_\pi$ is suppressed relative to $g^{(0,1)}_\pi$ by one order
in the chiral expansion~\cite{deVries:2012ab,Maekawa:2011vs}, but can
be taken into account in similar fashion.  In our parameterization these interactions
always contribute to nuclear EDMs, and when $d_N^\text{sr}$ are chosen as model parameters they additionally contribute to nucleon EDMs through calculable pion
loops~\cite{Mereghetti:2010kp} that appear as additional terms in $\lag_{\pi N}$ as discussed in Sec.~\ref{sec:global_nucl} below.

Naive dimensional
analysis~\cite{Manohar:1983md} suggests that short-range nucleon
interactions enter only at NNLO~\cite{deVries:2020loy} in the chiral
expansion~\cite{Weinberg:1990rz} and can be
neglected~\cite{Maekawa:2011vs}.  In Eq.\eqref{eq:piN} this applies to
all interactions in the second line.  One caveat is that a consistent
treatment of long-range effects may require additional short-distance
counter terms~\cite{deVries:2020loy} that appear as effective
short-range nucleon-nucleon forces. These are not presently taken into account, and theoretical coefficients for the sensitivity of specific systems to these forces are largely unavailable at present.

Finally, the higher-dimensional
operators in Eq.\eqref{eq:fourf} induce effective interactions that
can be organized according to their tensor structure, isospin
character, and their dependence on the electron and nucleon fields and
spins~\cite{Chupp:2017rkp,Dekens:2018bci}:
\begin{align}
  \lag_{eN}
  =& -\frac{G_F}{\sqrt{2}} \; \left( \bar e i \gamma_5 e \right) \;
  \bar N \left( C_S^{(0)} + C_S^{(1)} \tau_3 \right) N
  \notag \\
  &+ \frac{8G_F}{\sqrt{2}} \; v_\nu \; \left( \bar e \sigma^{\mu\nu} e \right) \;
  \bar N \left( C_T^{(0)} + C_T^{(1)} \tau_3 \right) S_\mu N
  \notag\\
  &-\frac{G_F}{\sqrt{2}} \; \left( \bar e  e \right) \;
  \frac{\partial^\mu}{m_N} 
  \left[ \bar N \left( C_P^{(0)}   + C_P^{(1)}   \tau_3 \right) S_\mu N \right] \; .
  \label{eq:e-nucl}
\end{align}
In a heavy baryon expansion, the last line can be dropped at
leading order~\cite{Engel:2013lsa}. However, we retain all three terms
since (1) a pion pole enhancement of the isovector contribution
somewhat offsets this hierarchy and (2) contributions of heavy quarks
can also render it relevant for some new physics models.  At this
stage, the independent Lagrangian parameters for our global EDM
analysis at the hadronic scale are
\begin{align}
  \Big\{ \; 
   d_e, C_S^{(0)}, C_S^{(1)}, C_T^{(0)},C_T^{(1)}, C_P^{(0)}, C_P^{(1)}, g_\pi^{(0)}, g_\pi^{(1)}, d_{n}, d_{p} \; 
  \Big\} \; ,
\label{eq:low_paras}
\end{align}
where the observable $d_{n,p}$ can be replaced by the short-range
nucleon EDMs $d_{n,p}^\text{sr}$ for an alternate parameterization, 
as described in the Appendix.

%%%%%%%%%%%%%%%%%%%%%%%%%%%%%%%%%%%%%%%%%%%%%%%%%%%%%%%%%%%%%%%%%%%%%%%%
\subsection{Matched Lagrangians for semileptonic interactions}
\label{eq:ops_match}

The set of model parameters defined in Eq.\eqref{eq:low_paras} can be
further simplified by matching the semileptonic part of the hadronic-scale
Lagrangian Eq.\eqref{eq:e-nucl} to the corresponding weak-scale
4-fermion interactions of Eq.\eqref{eq:fourf}, both evaluated for
external nucleons. The light quark content in the nucleons is related
to the nucleon Lagrangian as
\begin{align}
  g_S^{(0)} \; \bar \psi_N \psi_N 
  &= \frac{1}{2} \Langle N \left| \bar u u + \bar d d \right| N \Rangle \notag \\
  g_S^{(1)} \; \bar \psi_N \tau_3 \psi_N 
  &= \frac{1}{2} \Langle N \left| \bar u u - \bar d d \right| N \Rangle \notag \\
  g_T^{(0)} \; \bar \psi_N \sigma_{\mu\nu} \psi_N 
  &= \frac{1}{2} \Langle N \left| \bar u \sigma_{\mu\nu} u + \bar d \sigma_{\mu\nu} d \right| N \Rangle \notag \\
  g_T^{(1)} \; \bar \psi_N \sigma_{\mu\nu} \tau_3 \psi_N 
  &= \frac{1}{2} \Langle N \left| \bar u \sigma_{\mu\nu} u - \bar d \sigma_{\mu\nu} d \right| N \Rangle \notag \\
  g_P^{(0)} \; \bar \psi_N \gamma_5 \psi_N 
  &= \frac{1}{2} \Langle N \left| \bar u \gamma_5 u + \bar d \gamma_5 d \right| N \Rangle \notag  \\
  g_P^{(1)} \; \bar \psi_N \gamma_5 \tau_3 \psi_N 
  &= \frac{1}{2} \Langle N \left| \bar u \gamma_5 u - \bar d \gamma_5 d \right| N \Rangle \; ,
  \label{eq:gp1}
\end{align}
which relations define the scalar, tensor, and pseudoscalar nucleon
form factors $g_{S,T,P}^{(0,1)}$.  Using these, the hadronic-scale Wilson
coefficients in Eq.\eqref{eq:e-nucl} can be matched to the SMEFT
Wilson coefficients in Eq.\eqref{eq:fourf}
as~\cite{Engel:2013lsa,Chupp:2017rkp},
\begin{alignat}{9}
  C_S^{(0)}
  &= - g_S^{(0)}\frac{v^2}{\Lambda^2} \; \text{Im}\left( C_{\ell e d q} - C^{(1)}_{\ell e qu} \right)
  &\qqquad 
  C_S^{(1)}
  &= \phantom{-}  g_S^{(1)}\frac{v^2}{\Lambda^2} \; \text{Im}\left( C_{\ell e d q} + C^{(1)}_{\ell e qu} \right)
  \notag \\
  C_T^{(0)}
  &= - g_T^{(0)}\frac{v^2}{\Lambda^2} \; \text{Im}\left( C^{(3)}_{\ell e qu} \right)
  &\qqquad 
  C_T^{(1)}
  &= -g_T^{(1)} \frac{v^2}{\Lambda^2} \; \text{Im} \left( C^{(3)}_{\ell e qu} \right)
  \notag \\
  C_P^{(0)}
  &= \phantom{-}   g_P^{(0)}\frac{v^2}{\Lambda^2} \; \text{Im} \left( C_{\ell e d q} + C^{(1)}_{\ell e qu} \right)
  &\qqquad 
  C_P^{(1)}
  &= - g_P^{(1)}\frac{v^2}{\Lambda^2} \; \text{Im} \left( C_{\ell e d q} - C^{(1)}_{\ell e qu} \right) \; .
  \label{eq:CSPT}
\end{alignat}
Here the six couplings $C_{S,T,P}^{(0,1)}$ are expressed in terms of
only three SMEFT Wilson coefficients, implying
\begin{align}
\frac{C_P^{(0)}}{g_P^{(0)}} = \frac{C_S^{(1)}}{g_S^{(1)}} 
\qqqquad 
\frac{C_T^{(0)}}{g_T^{(0)}} = \frac{C_T^{(1)}}{g_T^{(1)}}
\qqqquad 
\frac{C_S^{(0)}}{g_S^{(0)}} = \frac{C_P^{(1)}}{g_P^{(1)}} \; .
  \label{eq:relations}
\end{align}
Only three independent semileptonic parameters actually enter the
hadronic-scale global analysis.  We choose them as $C_{S,T,P}^{(0)}$ and
combine them using the known ratios of hadronic matrix elements to
construct the full Lagrangian.  In addition to the light
quarks described by Eq.\eqref{eq:CSPT}, the relations in
Eq.\eqref{eq:relations} also must include the contributions of
heavy quarks. These contributions are contained in the nucleon form
factors, renormalized at an appropriate mass scale and accounting for
the corresponding anomaly relations.

Note that in this way our global analysis preserves the full dependence
on $C_{S,T,P}^{(0,1)}$, although only $C_{S,T,P}^{(0)}$ appear as
model parameters. This remains the case regardless of
isospin-violating effects, and also when including several experimental systems for
which the coefficients of $C_S^{(1)}$ differ significantly, as discussed below.

The implementation of Eq.\eqref{eq:CSPT} for $C_S^{(0,1)}$ is
particularly straightforward, because for each experimental system the
effective parameter that combines the isoscalar and isovector terms is
independent of the nuclear spin,
\begin{alignat}{8}
  C_S &=  C_S^{(0)} + \frac{Z-N}{Z+N}  \; C_S^{(1)}
  &\qquad  \notag \\
  &= C_S^{(0)} + \frac{Z-N}{Z+N} \frac{g_S^{(1)}}{g_P^{(0)}}  \; C_P^{(0)}
  &\qquad &\text{with} \qquad 
  \frac{g_S^{(1)}}{g_P^{(0)}} &\approx 0.1\, .
  \label{eq:Cs}
\end{alignat}
In the second step we replace $C_S^{(1)}$ with $C_P^{(0)}$, reflecting
Eq.\eqref{eq:CSPT}.  Since $g_S^{(1)}$ is already suppressed relative
to $g_S^{(0)}$ by the small isospin violation of the nucleon matrix
element, one could argue that $C_S^{(1)} \ll C_S^{(0)}$. Moreover, in
the heavy nuclei of all atomic and molecular systems for which EDMs
have been measured so far, the isoscalar and isovector contributions
occur in approximately the same ratio, $(Z-N)/(Z+N) \approx -0.2$, so
the effective parameter $C_S$ is approximately system-independent. As
noted above, we do not rely on these assumptions.

Next, we relate the pseudoscalar and tensor semileptonic interactions
in a similar fashion.  We start with the linear combinations for
nucleons, $C^{(n,p)}_{P,T} = C_{P,T}^{(0)} \mp C_{P,T}^{(1)}$, where the upper sign refers to $n$ according to our isospin convention. From those, the coefficients for a given nucleus can be
constructed according to the sum over spins of the constituent
nucleons, where $\left<\sigma_{p,n}\right>$ represents the expectation value for
neutrons or protons, evaluated via Pauli operators for the measured nuclear
state~\cite{Stadnik:2014xja,Berengut:2011tz}:
\begin{align}
C_{P,T} &= \frac{C^{(n)}_{P,T}\left< \sigma_n \right> + C^{(p)}_{P,T}\left< \sigma_p \right>}{\left< \sigma_n \right> + \left< \sigma_p \right>}  \;.
\label{eq:cpt_np1}
\end{align}
For $C_T$ we can see from Eq.\eqref{eq:CSPT} that the isoscalar and
isovector couplings differ only through the corresponding nucleon form
factors. These are calculated with small theoretical uncertainties in
lattice QCD~\cite{FlavourLatticeAveragingGroupFLAG:2021npn}, allowing
us to write
\begin{align}
C_T = \left( 1 -  \frac{g_T^{(1)}}{g_T^{(0)}} \;  \frac{\left< \sigma_n \right> - \left< \sigma_p \right>}{\left< \sigma_n \right> + \left< \sigma_p \right>}  \right) \; C_T^{(0)}
\qquad \text{with} \qquad 
%\epsilon_T = 
\frac{g_T^{(1)}}{g_T^{(0)}} \approx 1.7 \; .
\label{eq:cpt_np2}
\end{align}
Similarly, for $C_P$ it can be shown that
\begin{align}
C_P = C_P^{(0)} - \frac{g_P^{(1)}}{g_S^{(0)}} \; \frac{\left< \sigma_n \right> - \left< \sigma_p \right>}{\left< \sigma_n \right> + \left< \sigma_p \right>}   
\; C_S^{(0)}
\qquad \text{with} \qquad 
%\epsilon_{SP} = 
\frac{g_P^{(1)}}{g_S^{(0)}} \approx 20.2\; .
\label{eq:cpt_np}
\end{align}

For the derivation of $g_P^{(1)}$ in this ratio we follow
Ref.~\cite{Engel:2013lsa}. We consider only the first generation as
relevant light quarks, such that $g_P^{(1)}$ is dominated by the
$\pi$-pole contribution
\begin{align}
  g_P^{(1)} = \frac{g_A \bar{m}_N}{\bar{m}} \; \frac{m_\pi^2}{m_\pi^2 - q^2}
  + \text{heavy quarks} %\sum_{Q\in\{ s,c,b,t \}}g_P^{Q(1)}
  \qqqquad \
  \bar{m} = \frac{m_u + m_d}{2} \; .
  \label{gp1_pionPole}
\end{align}
Here $\bar{m}_N\approx940$~MeV is the average nucleon mass, $\bar{m}$
the average light quark mass, and $g_A$ is the (isovector) axial
vector coupling~\cite{HERMES:2006jyl}. The coupling $g_P^{(0)}$,
appearing in Eq.\eqref{eq:Cs}, involves the isoscalar axial coupling
$g_A^{(0)}$, obtained from the sum rather than the difference of the
light quark axial charges \cite{ANSELM1985116}.  We extend this by including a light
$s$-quark, where the $\pi$-pole dominance is replaced by an octet
$\eta$-pole with an appropriately modified average light quark mass
$m^*$,
\begin{align}
  g_P^{(0)} = \frac{g_A^{(0)} \bar{m}_N}{m^*} \; \frac{m_\eta^2}{m_\eta^2 - q^2}
  + \text{heavy quarks} %\sum_{Q\in\{ c,b,t \}}g_P^{Q(0)}
  \qqqquad
  m^* = \frac{\bar{m} + 2 m_s}{3} \; .
  %m^* = \frac{m_u + m_d + 2 m_s}{3} \; .
  \label{gp1_etaPole}
\end{align}
Note that heavy quark contributions enter differently in $g_P^{(0)}$
and $g_P^{(1)}$, and can be derived using the $U(1)_A$ axial anomaly
together with the divergence of the anomaly-free axial current
$J^q_{\mu5} = \bar q \gamma_\mu \gamma_5 q$ for all quarks $q$
\cite{SHIFMAN1978443,ANSELM1985116,Barr:1992cm,Bauer:2017qwy,Dienes:2013xya}. These represent a
relatively minor contribution to $g_P^{(1)}$, but due to suppression
of the $\eta$-pole relative to the $\pi$-pole by the factor
$m^*/\bar{m}$, contribute to $g_P^{(0)}$ at approximately the same
level as the light quarks.

Our final, simplifying assumption is not strictly needed, but is
effective in reducing the number of model parameters by one and 
removing a poorly constrained direction in model space,
\begin{align}
 d_p \approx -d_n \; .
 \label{eq:same_pn}
\end{align}
Alternatively, we could set $d_{n,p}^\text{sr}$, as described further in App.~\ref{sec:alternative}. The nuclei of the measured closed-shell systems, which apart from the neutron itself provide
the strongest constraints on nucleon EDMs, are typically dominated
either by a valence proton (in the case of TlF) or a valence neutron (all others).  In analogy with the anomalous
magnetic moment, the short-range nucleon EDMs are assumed to be
dominated by the isovector contribution. This assumption can be
relaxed for the purposes of a rigorous global analysis, provided
improved theory uncertainties concerning the contributions of all
nucleons to the overall EDM of each measured system. Of the present
experimental limits, TlF has the leading sensitivity to $d_p$. (Note that Cs also has a valence proton, whose contribution can be somewhat enhanced through a magnetic quadrupole moment \cite{Khriplovich:1997ga,Engel:2025uci}).

With the simplifications described in this section, the set of independent low-energy parameters given in
Eq.\eqref{eq:low_paras} reduces to
\begin{align}
  c_j \in \Big\{ \; 
  d_e, C_S^{(0)}, C_T^{(0)}, C_P^{(0)}, g_\pi^{(0)}, g_\pi^{(1)}, d_n  \;  \; 
  \Big\} \; ,
\label{eq:low_paras1}
\end{align}
which are the model parameters for our global EDM analysis.

%%%%%%%%%%%%%%%%%%%%%%%%%%%%%%%%%%%%%%%%%%%%%%%%%%%%%%%%%%%%%%%%%%%%%%%%
\section{EDM Measurements}
\label{sec:global}

In terms of the Lagrangian parameters in Eq.\eqref{eq:low_paras1} we
can predict the measured EDMs $d_i$ as linear combinations with
system-specific coefficients $\alpha_{i,c_j}$,
\begin{align}
    d_i = \sum_{c_j} \alpha_{i,c_j} c_j \; .
    \label{eq:linear0}
\end{align}
The measurements we analyze are listed in Tab.~\ref{tab:meas} and
discussed below. Unless otherwise indicated, the isotopes and charge
states that we discuss are those given in Tab.~\ref{tab:meas}. The
isotopes that are relevant for these molecular systems are $^{180}$Hf,
$^{232}$Th, $^{174}$Yb, $^{205}$Tl, $^{16}$O, and $^{19}$F. We neglect
constraints from the comparatively weaker experimental bounds from
$^{85}$Rb~\cite{PhysRev.164.270.2,PhysRev.153.36},
Xe$^m$~\cite{MAPlayer_1970}, PbO~\cite{Eckel:2013lsa},
Eu$_{0.5}$Ba$_{0.5}$TiO$_3$~\cite{Eckel:2012aw}, and the $\Lambda$
hyperon~\cite{Pondrom:1981gu}. The experimental bounds for the $\mu$
and $\tau$ leptons constrain the corresponding Lagrangian parameters
and factorize from the hadronic-scale Lagrangian. We do not include them in
our first global analysis.

As discussed in the last section, the $\alpha$-values for \textit{all}
$C_{S,P,T}^{(0,1)}$ are fixed by the corresponding values for the basis of Eq.\eqref{eq:low_paras1}. For this we follow
Eqs.~\eqref{eq:Cs}-\eqref{eq:cpt_np} and find for example
\begin{align}
    \alpha_{C_S^{(0)}} &= \alpha_{C_S} - \alpha_{C_P} \frac{g_P^{(1)}}{g_S^{(0)}} \frac{\left< \sigma_n \right> - \left< \sigma_p \right>}{\left< \sigma_n \right> + \left< \sigma_p \right>}  \notag \\
    \alpha_{C_P^{(0)}} &= \alpha_{C_P} + \alpha_{C_S} \frac{g_S^{(1)}}{g_P^{(0)}} \frac{Z-N}{Z+N} \notag \\
    \alpha_{C_T^{(0)}} &= \left( 1 -  \frac{g_T^{(1)}}{g_T^{(0)}} \;  \frac{\left< \sigma_n \right> - \left< \sigma_p  \right>}{\left< \sigma_n \right> + \left< \sigma_p \right>}  \right) \; \alpha_{C_T} \; ,
    \label{eq:alpha_CS}
\end{align}
with the $\alpha_{C_{S,P,T}}$ given in Tab.~\ref{tab:LEC}. 
In this table we also show
\begin{align}
 \left< \sigma_z \right>^{(0)}
 =\left< \sigma_n \right> + \left< \sigma_p \right> \; ,
\end{align}
which is proportional to the isoscalar sum of neutron
and proton spin projections in a shell model of the
nucleus, a factor of two being given by the usual relation of the spin and Pauli operators.
The shell model is not
expected to be reliable for the deformed nuclei $^{171}$Yb and
$^{225}$Ra. The values given in Tab.~\ref{tab:LEC} are chosen to optimize 
agreement of the calculated and measured nuclear magnetic moments within this 
framework. Literature values are available for these nuclear 
species~\cite{Berengut:2011tz}. The spin fractions contributing to semileptonic
coefficients in TlF only take into account the $^{205}$Tl nucleus,
though see Ref.~\cite{Coveney:1983ht} for some consideration of
contributions from the $^{19}$F nucleus. For $\alpha_{^{129}\text{Xe},C_S}$ 
we use a scaling relation to derive a value from 
$\alpha_{^{129}\text{Xe},C_T}$ of the cited reference.

With the exception of $\alpha_{^{129}Xe,C_{S}}$, we use only explicitly calculated values and employ the established phenomenological scaling relations only as consistency checks. For this one case, however, although a precise explicit calculation is available it deviates from both the scaling relations \cite{Fleig:2020obr} and other explicit calculations \cite{Gaul:2023hdd} for presently unknown reasons. We therefore employ a scaled value of the explicitly calculated $\alpha_{^{129}Xe,C_{T}}$ \cite{Dzuba:2009kn}, using the semianalytical relations of \cite{Fleig:2020obr,Khriplovich:1997ga,Kozlov:1988qn,Flambaum:1985gx}.

%--------------------------------------------
\begin{table}[t]
\centering
\begin{small}
\begin{tabular}{c|rcr}
\toprule
System $i$&  Measured $d_i$ $\left[\text{$e$\,cm}\right]$& Upper limit on $|d_i|$ $\left[\text{$e$\,cm}\right]$& Reference \\
\midrule
$n$
 & $ (0.0\pm 1.1_\text{stat}\pm 0.2_\text{syst})\cdot10^{-26}$
 & $2.2\cdot 10^{-26}$
 &  \cite{Abel:2020pzs}\\ [0mm]
% $\mu$
% & $(0.0\pm 0.2_\text{stat}\pm 0.9_\text{syst})\cdot10^{-19}$
% & $1.8\cdot 10^{-19}$
% &  \cite{Muong-2:2008ebm}\\[0mm]
% $\tau$
% & $\Re(d_\tau) = (-6.2\pm 6.3)\cdot10^{-18}$
% & $1.7\cdot 10^{-17}$
% & \cite{Belle:2021ybo} \\[0mm]
% &$\Im(d_\tau) = (-4.0\pm 3.2)\cdot10^{-18}$
% & $9.3\cdot 10^{-18}$
% & \cite{Belle:2021ybo} \\[0mm]
\midrule
$^{205}$Tl
& $ (-4.0\pm4.3)\cdot 10^{-25} $
& $1.1\cdot 10^{-24}$
& \cite{Regan:2002ta}\\[0mm] 
$^{133}$Cs
& $ (-1.8\pm 6.7_\text{stat}\pm 1.8_\text{syst})\cdot10^{-24}$
& $1.4\cdot 10^{-23}$
&  \cite{Murthy:1989zz}\\
\midrule
HfF$^+$
& $ (-1.3\pm 2.0_\text{stat}\pm 0.6_\text{syst})\cdot10^{-30}$
%$2\pi (0.10 \pm 0.87 \pm 0.20) \dfrac{\text{mrad}}%{\text{s}}$~\cite{Cairncross:2017fip}
& $4.8\cdot 10^{-30}$
&  \cite{Roussy:2022cmp}\\[0mm]
ThO
& $ (4.3\pm 3.1_\text{stat}\pm 2.6_\text{syst})\cdot10^{-30}$ %$-510 \pm 373_\text{stat} \pm 310_\text{syst} \dfrac{\mu \text{rad}}{\text{s}}$~\cite{ACME:2018yjb} 
& $1.1\cdot 10^{-29}$
& \cite{ACME:2018yjb} \\[0mm]
YbF
& $ (-2.4\pm 5.7_\text{stat}\pm 1.5_\text{syst})\cdot10^{-28}$ %$5.30 \pm 12.6 \pm 3.30 \dfrac{\text{mrad}}{\text{s}}$~\cite{Hudson:2011zz}
& $1.2\cdot 10^{-27}$
& \cite{Hudson:2011zz}\\[0mm]
\midrule
$^{199}$Hg
&  $(2.20\pm 2.75_\text{stat}\pm 1.48_\text{syst})\cdot10^{-30}$
& $7.4\cdot 10^{-30}$
&  \cite{Graner:2016ses,PhysRevLett.119.119901}\\[0mm]
$^{129}$Xe
& $ (-1.76\pm 1.82)\cdot10^{-28}$ %$ (1.4\pm 6.6_\text{stat}\pm 2.0_\text{syst})\cdot10^{-28}$
& $4.8\cdot 10^{-28}$ %$1.4\cdot 10^{-27}$
&  \cite{Sachdeva:2019rkt,PhysRevA.100.022505}\\[0mm]
$^{171}$Yb
& $ (-6.8\pm 5.1_\text{stat}\pm 1.2_\text{syst})\cdot10^{-27}$
& $1.5\cdot 10^{-26}$
& \cite{Zheng:2022jgr} \\[0mm]
$^{225}$Ra
& $ (4\pm 6_\text{stat}\pm 0.2_\text{syst})\cdot10^{-24}$
& $1.4\cdot 10^{-23}$
& \cite{Bishof:2016uqx} \\[0mm]
TlF
& $ (-1.7\pm 2.9)\cdot10^{-23}$
& $6.5\cdot 10^{-23}$
&  \cite{Cho:1991ig}
\\ \midrule
&  Measured $\omega_i$ $\left[\text{$\text{mrad}/s$}\right]$& Rescaling factor $x_i$ for $d_i$ & Reference
\\ \midrule
HfF$^+$
& $\phantom{-}(-0.0459 \pm 0.0716_\text{stat} \pm 0.0217_\text{syst})^\dagger$%$\phantom{-}2\pi\times(-0.0146 \pm 0.0228_\text{stat} \pm 0.0069_\text{syst})$
& 0.999
&   \cite{Roussy:2022cmp}
 \\[0mm]
ThO
& $\phantom{-}(-0.510 \pm 0.373_\text{stat} \pm 0.310_\text{syst}$) 
& 0.982
&   \cite{ACME:2018yjb}\\[0mm]
YbF
&  $\phantom{-}(5.30 \pm 12.60_\text{stat} \pm 3.30_\text{syst}$) 
% this value is only given in e*cm in the cited paper
& 1.12
&   \cite{Hudson:2011zz}\\
\bottomrule
\end{tabular}
\end{small}
\caption{Measured EDM values and $95\%$~CL for upper limits on their absolute values.  For $^{129}$Xe we
  combine two independent results with similar precision, using
  inverse-variance weighting. For the open-shell molecules, we also
  provide the measured angular frequencies and the rescaling factor
  which allows us to use $x_i d_i$ for each experimentally reported
  $d_i$. For the definition of $x_i$, see text.  $^\dagger$The
  frequency for HfF$^+$ is scaled by a factor of 2 relative to
  Ref.\cite{Roussy:2022cmp}, to consistently use
  Eq.\eqref{eq:paramole2} for all systems.}
\label{tab:meas}
\end{table}
%-----------------------------------------------

%-------------------------------------------
\begin{table}[t]
\centering
\begin{small}
\begin{tabular}{c|rrrrrr}
\toprule
System $i$
& $\left< \sigma_n \right>$
& $\left< \sigma_p \right>$
& $\left< \sigma_z \right>^{(0)}$
& $\alpha_{i,C_S} \left[ \text{$e$ cm} \right]$ 
& $\alpha_{i,C_P} \left[ \text{$e$ cm} \right]$
& $\alpha_{i,C_T} \left[ \text{$e$ cm} \right]$\\
\midrule
Tl
%& $d_\text{Tl} = (-4.00\pm4.30) \cdot 10^{-25} \text{$e$cm}$
& $0.274$
& $0.726$
& $1$
& $-6.77\cdot10^{-18}$ \cite{Fleig:2019alx}
& $1.5\cdot 10^{-23}$ \cite{GINGES200463}% n/a
& $5\cdot 10^{-21}$ \cite{GINGES200463}% n/a
\\
Cs
%& $d_\text{Cs} = (-1.80\pm 6.70_\text{stat}\pm 1.80_\text{syst}) \cdot10^{-24}  \text{$e$cm}$
& $-0.206$
& $-0.572$
& $-0.778$
& $7.8\cdot10^{-19}$ \cite{Engel:2013lsa}
& $2.2\cdot 10^{-23}$ \cite{GINGES200463}% - 
& $9.2\cdot 10^{-21}$ \cite{GINGES200463}% -
\\ \midrule
$^{199}$Hg
%& $d_\text{$^{199}$Hg} = (2.20\pm 2.75_\text{stat}\pm 1.48_\text{syst}) \cdot10^{-30}  \text{$e$cm}$
& $-0.302$
& $-0.032$
& $-0.334$
& $-2.8\cdot 10^{-22}$ \cite{Fleig:2018bsf}
& $6\cdot10^{-23}$ \cite{Dzuba:2009kn}
& $1.7\cdot 10^{-20}$ \cite{Dzuba:2009kn}
%& $-1.56\cdot10^{-4}$ & $-1.56\cdot10^{-5}$
\\
$^{129}$Xe
%& $d_\text{$^{129}$Xe} = (1.4\pm 6.6_\text{stat}\pm 2.0_\text{syst}) \cdot10^{-28}  \text{$e$cm}$
& $0.73$
& $0.27$
& $1$
& $-6.28\cdot 10^{-23}$ \cite{Dzuba:2009kn}
& $1.6\cdot 10^{-23}$ \cite{Dzuba:2009kn}
& $5.7\cdot 10^{-21}$ \cite{Dzuba:2009kn}
%& $1.70\cdot 10^{-5}$ & $3.51\cdot 10^{-6}$
\\
$^{171}$Yb
%& $d_\text{$^{171}$Yb} = (-6.8\pm 5.1_\text{stat}\pm 1.2_\text{syst}) \cdot10^{-27}  \text{$e$cm}$
& $-0.3$
& $-0.034$
& $-0.334$
& $-7.34\cdot 10^{-22}$ \cite{Gaul:2023hdd}
& $3.60\cdot 10^{-23}$ \cite{Gaul:2023hdd}
& $1.04\cdot 10^{-20}$ \cite{Gaul:2023hdd}
%& $-1.128\cdot10^{-4}$ & $-1.128\cdot10^{-5}$
\\
$^{225}$Ra
%& $d_\text{$^{225}$Ra} = (4\pm 6_\text{stat}\pm 0.2_\text{syst}) \cdot10^{-24}  \text{$e$cm}$
& $0.72$
& $0.28$
& $1$
& $5.63\cdot 10^{-21}$ \cite{Gaul:2023hdd}
& $-6.4\cdot 10^{-22}$ \cite{Dzuba:2009kn}
& $-1.8\cdot 10^{-19}$ \cite{Dzuba:2009kn}
%& $-5.36\cdot 10^{-4}$ & $-1.11\cdot 10^{-4}$
\\
TlF
%& $d_\text{TlF} = (-1.7\pm 2.9) \cdot10^{-23}  \text{$e$cm}$
& $0.274$
& $0.726$
& $1$
& $1.09\cdot 10^{-16}$ \cite{Gaul:2023hdd}
& $3.8\cdot 10^{-18}$ \cite{Gaul:2023hdd}
& $1.06\cdot 10^{-15}$ \cite{Fleig:2023tnh}
%& $-9.47\cdot 10^{-2}$ & $-4.59\cdot 10^{-1}$
\\ \bottomrule
\end{tabular}
\end{small}
\caption{Effective parameters used as input to our global analysis, as
  summarized in Tab.~\ref{tab:LEC2}.}
%  Within a shell model of the
%  nucleus, the quantity $\left< \sigma_z \right>^{(0)}=\left< \sigma_n
%  \right> + \left< \sigma_p \right>$ is proportional to the isoscalar sum of neutron
%  and proton spin projections, a factor of two being given by the usual %relation of the spin and Pauli operators. Note that the shell model is not
%  expected to be reliable for the deformed nuclei $^{171}$Yb and
%  $^{225}$Ra; the given values are chosen to optimize agreement of the %calculated and measured nuclear magnetic moments within this framework. Literature values are also available for these nuclear species \cite{Berengut:2011tz}. The spin fractions contributing to semileptonic
%  coefficients in TlF only take into account the $^{205}$Tl nucleus,
%  though see Ref.~\cite{Coveney:1983ht} for some consideration of
%  contributions from the $^{19}$F nucleus. For $\alpha_{^{129}\text{Xe},C_S}$ we use a phenomenological scaling relation to derive a value from $\alpha_{^{129}\text{Xe},C_T}$ of the cited reference.}
\label{tab:LEC}
\end{table}
%-------------------------------------------

%%%%%%%%%%%%%%%%%%%%%%%%%%%%%%%%%%%%%%%%%%%%%%%%%%%%%%%%%%%%%%%%%%%%%%%%
\subsection{Nucleons}
\label{sec:global_nucl}

We start with the simplest hadronic systems for which we can measure
EDMs, the nucleons. Their EDMs can be written directly as a Lagrangian term
\begin{align}
  -\frac{i}{2} F^{\mu\nu} d_N  \; \left( \bar N \sigma_{\mu\nu}\gamma_5 N \right) \; ,
  \label{eq:lag_nucl_dipole}
\end{align}
but can also be described by heavy baryon chiral perturbation theory, based on the
hadronic-scale Lagrangian. In that case we can consider the nucleon EDMs as
observables that receive contributions from: short-range contributions
$d_N^\text{sr}$, NNLO pion-loop contributions, and potential direct
contributions to Eq.\eqref{eq:lag_nucl_dipole} within and beyond the
SM~\cite{deVries:2021sxz},
\begin{align}
\label{eq:d_np}
d_n & = d_n^\text{sr}
  - \frac{e g_A}{8\pi^2 F_\pi} \left[
    g_\pi^{(0)}
   \left( \ln \frac{m_\pi^2}{m_N^2} - \frac{\pi m_\pi}{2 m_N}\right)
    -  \frac{g_\pi^{(1)}}{4} \left( \kappa_0 - \kappa_1 \right)
    \frac{m_\pi^2}{m_N^2} \ln \frac{m_\pi^2}{m_N^2}
    \right] \\
d_p &= d_p^\text{sr}
  + \frac{e g_A}{8\pi^2 F_\pi} \left[
    g_\pi^{(0)}
    \left(\ln \frac{m_\pi^2}{m_N^2} - \frac{2\pi m_\pi}{m_N}\right)
    -  \frac{g_\pi^{(1)}}{4}
    \left( \frac{2 \pi m_\pi}{m_N}+ \left(\frac{5}{2}+\kappa_0 + \kappa_1 \right)\frac{m_\pi^2}{m_N^2} \ln \frac{m_\pi^2}{m_N^2}\right)
    \right] \; ,
  \notag 
\end{align}
where $F_\pi = 92$~MeV is the pion decay
constant~\cite{Workman:2022ynf}, $m_\pi = 139$~MeV, $m_N = 940$~MeV,
$g_A \approx 1.27$ is the nucleon isovector axial charge, and the
isoscalar and isovector nucleon anomalous magnetic moments are
$\kappa_0 = -0.12$ and $\kappa_1 = 3.7$.  We set the renormalization
scale to the nucleon mass, and the splitting between proton and neutron
masses leads to a higher-order effect that is presently negligible in
relation to other uncertainties. In terms of weak-scale parameters,
finite values of $g_\pi^{(0,1)}$ can be related to a CKM phase or
$\bar\theta$, but also to 4-fermion quark operators. The input from
the presently most sensitive neutron EDM measurement to our global
analysis is given in Tab.~\ref{tab:meas}, alongside the other measurements we use.

From Eq.\eqref{eq:d_np} is it clear that we have a choice of defining
either $d_{n,p}^\text{sr}$ or $d_{n,p}$ as our model parameters. Constraints on these
Lagrangian parameters, including a renormalization condition, are the
output of our global analysis; but their real purpose is to compute
yet other observables, and compare them to measurements within
a consistent theory.

The two choices are not equivalent any
longer, when we remove $d_p$ or $d_p^\text{sr}$.  Choosing $d_{n,p}$ as
model parameters implies that we use measured values as Lagrangian
parameters. The corresponding renormalization condition is referred to
as on-shell renormalization in collider physics.  In that case, we can
extract $d_n$ and $d_p$ from data and use Eq.\eqref{eq:d_np} to
translate them into $d_{n,p}^\text{sr}$. We use this scheme in the
main body of the paper. The alternative of using $d_n^\text{sr}$ as
model parameters may be a better motivation to use the approximate
relation $d_p^\text{sr} = - d_n^\text{sr}$ rather than Eq.\eqref{eq:same_pn}. We present a complete set of results using this approach in the 
Appendix.

Neither
of the two renormalization conditions allows for a simple running of
the Lagrangian parameters to higher scales. For this third option we
must renormalize $d_N$ in the $\overline{\text{MS}}$ scheme and
compute the difference, in analogy to Eq.\eqref{eq:d_np}, in
perturbation theory.

%To be able to use the approximate iso-vector symmetry between the proton and
%neutron short-range contributions and provide easily interpretable 
%results we choose a hybrid model and instead of $d_n^\text{sr}$ show 
%$d_n$, such that $d_p$ can be computed as\mpar{SMD: I don't think this makes sense...}
%%
%\begin{align}
%d_p 
%&= - d_n  
%  - \frac{e g_A}{8\pi^2 F_\pi} \left[
%    g_\pi^{(0)}
%    \frac{\pi m_\pi}{2 m_N}
%      +  \frac{g_\pi^{(1)}}{4}
%    \left( \frac{2 \pi m_\pi}{m_N}+ \left(\frac{5}{2} + 2\kappa_1 \right)\frac{m_%\pi^2}{m_N^2} \ln \frac{m_\pi^2}{m_N^2}\right)
%    \right] \; .
%\end{align}
%%

%-------------------------------------------
\begin{sidewaystable}[ph!]
%\centering
\begin{footnotesize}
\resizebox{\textwidth}{!}{
\begin{tabular}{@{}c|rrrrrrrr}
\toprule
System $i$
& $\alpha_{i,d_e}$
& $\alpha_{i,C_S^{(0)}} \left[ \text{$e$ cm} \right]$ 
& $\alpha_{i,C_P^{(0)}} \left[ \text{$e$ cm} \right]$
& $\alpha_{i,C_T^{(0)}} \left[ \text{$e$ cm} \right]$
& $\alpha_{i,g_\pi^{(0)}} \left[ \text{$e$ cm} \right]$
& $\alpha_{i,g_\pi^{(1)}} \left[ \text{$e$ cm} \right]$
& $\alpha_{i,d_n}$
& $\alpha_{i,d_p}$\\
\midrule
$n$
& $-$ & $-$ & $-$ & $-$
& \textcolor{black}{$0$} 
& \textcolor{black}{$0$} 
& \textcolor{black}{$1$} &\textcolor{black}{ ($-1$)}
\\ \midrule
$^{205}$Tl
& $-558^{\pm28}$ \cite{sym12040498}
& $-6.77^{\pm0.34}\cdot10^{-18}$
& $1.4^{+2.5}_{-0.8}\cdot10^{-19}$
& $8.8^{+4.0}_{-1.2}\cdot10^{-21}$
& \textcolor{black}{$-6.41^{+3.57}_{-4.51}\cdot10^{-18}$}
& \textcolor{black}{$2.27^{+6.13}_{-1.19}\cdot10^{-19}$}
& \textcolor{black}{$-5.75^{+3.44}_{-4.44}\cdot10^{-6}$}
& \textcolor{black}{$1.61^{+3.35}_{-7.56}\cdot10^{-5}$}
\\
$^{133}$Cs
& $123^{\pm4}$ \cite{Fleig:2018bsf,Chupp:2017rkp}
& $7.80^{+0.2}_{-0.8}\cdot10^{-19}$
& $-1.4^{+0.8}_{-2.2}\cdot10^{-20}$ 
& $1.7^{\pm0.4}\cdot10^{-20}$ 
& \textcolor{black}{$8.09^{+151.1}_{-2.63}\cdot10^{-19}$} 
& \textcolor{black}{$2.70^{+3.27}_{-1.79}\cdot10^{-18}$} 
& \textcolor{black}{$1.0^{\pm 1}\cdot10^{-5}$}  
& \textcolor{black}{$6.7^{\pm 2}\cdot10^{-4}$} 
\\ \midrule
$^{199}$Hg
%& $-0.012^{+0.0094}_{-0.002}$ \cite{Martensson-Pendrill_1987,Gaul:2023hdd}
& $11.6^{+10}_{-18}\cdot10^{-3}$ \cite{Martensson-Pendrill_1987,Gaul:2023hdd}
& $-1.26_{-1.6}^{+0.7}\cdot10^{-21}$
& $6.6_{-1.6}^{+2.4}\cdot10^{-23}$ 
& $-6.4_{-4}^{+3}\cdot10^{-21}$ 
& \textcolor{black}{$-3.05_{-14.5}^{+1.75}\cdot10^{-18}$}
& \textcolor{black}{$-6.10_{-25.5}^{+16.6}\cdot 10^{-18}$}
& \textcolor{black}{$-1.36^{+0.38}_{-3.45}\cdot10^{-4}$}
& \textcolor{black}{$-1.36^{+0.38}_{-5.12}\cdot10^{-5}$}
\\
$^{129}$Xe
& $-8^{+0}_{-8}\cdot 10^{-4}$ \cite{Martensson-Pendrill_1987,Gaul:2023hdd} 
& $-2.1_{-2.5}^{+1.2}\cdot10^{-22}$
& $1.7_{-0.4}^{+0.5}\cdot10^{-23}$ 
& $1.24_{-0.61}^{+0.78}\cdot10^{-21}$
& \textcolor{black}{$-3.91_{-23.4}^{+1.75}\cdot 10^{-19}$}
& \textcolor{black}{$2.93_{-1.64}^{+24.4}\cdot 10^{-19}$}
& \textcolor{black}{$2.29_{-0.58}^{+0.77}\cdot10^{-5}$} 
& \textcolor{black}{$4.89^{+1.64}_{-1.25}\cdot10^{-6}$}
\\
$^{171}$Yb
& $1.44^{+165}_{-4.5}\cdot10^{-3}$ \cite{Gaul:2023hdd}
& $-1.31_{-1.06}^{+0.55}\cdot10^{-21}$
& $4.93_{-1.55}^{+3.54}\cdot10^{-23}$
& $-3.68_{-2.43}^{+1.86}\cdot10^{-21}$
& \textcolor{black}{$-2.84^{+0.43}_{-6.05}\cdot 10^{-18}$}
& \textcolor{black}{$-5.67^{+7.74}_{-10.3}\cdot 10^{-18}$}
& \textcolor{black}{$-1.13^{+0.32}_{-0.28}\cdot10^{-4}$}
& \textcolor{black}{$-1.13^{+0.32}_{-0.28}\cdot10^{-5}$}
\\
$^{225}$Ra
%& $-0.054_{-0.02}^{+1.34}$ \cite{Gaul:2023hdd}
& $-5.4_{-2.0}^{+134}\cdot10^{-2}$ \cite{Gaul:2023hdd}
& $1.13_{-0.51}^{+2.94}\cdot10^{-20}$ 
& $-7.63_{-3.88}^{+2.05}\cdot10^{-22}$ 
& $-4.5_{-2.5}^{+2.0}\cdot10^{-20}$
& \textcolor{black}{$1.72_{-0.67}^{+5.73}\cdot10^{-15}$}
& \textcolor{black}{$-6.89_{-22.9}^{+2.66}\cdot10^{-15}$}
& \textcolor{black}{$-5.38^{+1.20}_{-1.58}\cdot10^{-4}$} 
& \textcolor{black}{$-1.19^{+0.27}_{-0.35}\cdot10^{-4}$}
\\
TlF
& $1.36_{-0.32}^{+0.36}\cdot 10^{3}$ \cite{Gaul:2023hdd} 
& $1.44_{-0.5}^{+0.8}\cdot10^{-16}$ 
& $1.51_{-5.6}^{+2.2}\cdot10^{-18}$ 
& $1.87_{-0.17}^{+0.19}\cdot10^{-15}$ 
& \textcolor{black}{$-7.00_{-2.17}^{+1.39}\cdot 10^{-13}$}
& \textcolor{black}{$2.48^{+4.58}_{-0.34}\cdot 10^{-14}$}
& \textcolor{black}{$-6.28^{1.71}_{-2.28}\cdot10^{-1}$}
& \textcolor{black}{$1.76^{+2.41}_{-6.76}$}
\\ \midrule
HfF$^+$
& 1
& $9.17^{\pm0.06} \cdot 10^{-21}$
& $-$ & $-$ & $-$ & $-$ & $-$ & $-$
\\
ThO
& 1
& $1.51^{+0}_{-0.2} \cdot 10^{-20}$
& $-$ & $-$ & $-$ & $-$ & $-$ & $-$
\\
YbF
& 1
& $8.99^{\pm0.70} \cdot 10^{-21}$ 
& $-$ & $-$ & $-$ & $-$ & $-$ & $-$
\\ \midrule
& $\eta^{(m)}_{i,d_e} \left[  \dfrac{\text{mrad}}{\text{s $e$ cm}} \right]$
& $k^{(m)}_{i,C_S} \left[  \dfrac{\text{mrad}}{\text{s}} \right]$
& $\alpha_{i,C_P}$
& $\alpha_{i,C_T}$
& $\alpha_{i,g_\pi^{(0)}}$
& $\alpha_{i,g_\pi^{(1)}}$
& $\alpha_{i,d_n}$ 
& $\alpha_{i,d_p}$ 
\\ \midrule
%\multirow{3}{*}{molecules}
HfF$^+$
& $3.49^{\pm0.14} \cdot 10^{28}$ \cite{Fleig:2018bsf,10.1063/1.4993622,Fleig:2017mls,Petrov:2007zz,PhysRevA.73.062108}
& $3.2^{+0.1}_{-0.2} \cdot 10^{8}$ \cite{Fleig:2018bsf,10.1063/1.4993622,Fleig:2017mls}
& $-$ & $-$ & $-$ & $-$ & $-$ & $-$
\\
ThO
& $-1.21^{+0.05}_{-0.39} \cdot 10^{29}$ \cite{Fleig:2018bsf,10.1063/1.4968229,Meyer:2008gc,10.1063/1.4968597}$^\dagger$
& $-1.82^{+0.42}_{-0.27} \cdot 10^{9}$ \cite{10.1063/1.4968229,Fleig:2018bsf,PhysRevA.84.052108,PhysRevA.85.029901,10.1063/1.4968597}$^\dagger$
& $-$ & $-$ & $-$ & $-$ & $-$ & $-$
\\
YbF
& $-1.96^{\pm0.15} \cdot 10^{28}$ \cite{Fleig:2018bsf,PhysRevA.84.052108,PhysRevA.85.029901,PhysRevA.93.042507,PhysRevA.90.022501}
& $-1.76^{\pm0.2} \cdot 10^{8}$ \cite{Fleig:2018bsf,PhysRevA.84.052108,PhysRevA.93.042507,PhysRevA.85.029901}
& $-$ & $-$ & $-$ & $-$ & $-$ & $-$
\\ \bottomrule
\end{tabular}
}
\end{footnotesize}
\caption{Central values and theory uncertainties for the $\alpha$-parameters defined in
  Eq.\eqref{eq:linear0}, using $d_{n,p}$ as model parameters. The parentheses around $\alpha_{n,d_p}$ indicate our assumption Eq.\eqref{eq:same_pn}, i.e., this value should not be construed as entering Eq.\eqref{eq:linear0}. (For the implementation using $d_{n,p}^{\text{sr}}$, see Appendix~\ref{sec:alternative} and in particular Tab.\ref{tab:LEC3_app}.) Here entries given only as ``$-$'' indicate that we neglect the dependence in
  our global analysis, e.g., because that coefficient arises only for nuclei with spin. $^\dagger$There appears to be an overall sign
  error in the coefficients reported for ThO in Table~4 of
  Ref.~\cite{Fleig:2018bsf}.   The values for Tl and Cs of
  $\alpha_{i,C_T^{(0)}}$ are estimated by simple analytical
  calculations~\cite{GINGES200463}, and the uncertainties quoted here
  are estimated as approximately twice those arising from the relevant
  hadronic matrix elements.}
\label{tab:LEC2}
\end{sidewaystable} 
%-------------------------------------------

%%%%%%%%%%%%%%%%%%%%%%%%%%%%%%%%%%%%%%%%%%%%%%%%%%%%%%%%%%%%%%%%%%%%%%%%
\subsection{Open-shell (paramagnetic) systems}
\label{eq:global_para}

Atoms and molecules with open electronic shells, sometimes also referred to as paramagnetic\footnote{We prefer the nomenclature of open-shell and closed-shell systems, rather than paramagnetic or diamagnetic, since this more clearly indicates the properties that are relevant for determining the leading contributions to the total system EDM.} systems, are primarily sensitive to the
electron EDM and the scalar electron-nucleon couplings $C_S^{(0,1)}$
of Eq.\eqref{eq:e-nucl}.  It is common to distinguish the case of
open-shell atoms with the atomic EDMs
\begin{align}
  d_i = \alpha_{i,d_e} d_e  + \alpha_{i,C_S} C_S + \dots \qquad \text{for} \qquad i \in \{ \text{Tl},\text{Cs} \}\;%\sum_{c_j} \alpha_{i, c_j} c_j
  \label{eq:paraatom}
\end{align}
from that of open-shell molecules, although in both cases the terms
involving the Lagrangian parameters $d_e$ and $C_S$ dominate. The
experimentally relevant quantity is a phase difference accumulated
over some measurement time, interpreted as a frequency. Measurements
with open-shell molecules are often reported as a parity- and
time-reversal-violating frequency shift $\omega_i$,
\begin{align}
  \omega_i = \eta^{(m)}_{i,d_e}  d_e  + k^{(m)}_{i,C_S}  C_S + \dots
  \qquad \text{for} \qquad 
  i\in\{\text{HfF$^+$},\text{ThO},\text{YbF}\} \; .%\sum_{c_j} \alpha_{i,c_j}^{(m)} c_j
  \label{eq:paramole}
\end{align}
The superscript $(m)$ indicates the molecular systems.  We also give
these measured angular frequencies in Tab.~\ref{tab:meas}.  This
emphasizes that $\omega_i$ does not scale in a simple way with the
experimentally applied electric field; it rather depends on the
molecular structure via a so-called effective electric field
$E_\text{eff}$ that saturates when the molecule is polarized.  Note
that $\alpha_{i,d_e}$ is dimensionless, while $\eta^{(m)}_{i,d_e}$ is
not; similarly $\alpha_{i,C_S}$ has the units of an EDM, while
$k^{(m)}_{i,C_S}$ has units of angular frequency.

The remaining contributions to open-shell molecular EDMs have weak
dependences on other low-energy constants, and we note that in
particular the EDMs of the $\mu$ and $\tau$ leptons have been
indirectly constrained via the experimental limits from ThO and
Hg~\cite{Ema:2021jds}. However, we are not aware of any established
values for the coefficients of semileptonic or hadronic parameters. (Note that due to the vanishing nuclear spin of $^{232}$Th, $^{180}$Hf, and $^{174}$Yb there is no leading contribution to the system EDM from $C_P$ or $C_T$ for the heavy nucleus in any of the measured open-shell molecules.) We
thus take $\alpha_{i,c_j}^{(m)}=0$ in Eq.\eqref{eq:paramole}, while for
the open-shell atoms we obtain values for semileptonic coefficients
either from the literature or by scaling arguments, as reported in
Tabs.~\ref{tab:LEC} and~\ref{tab:LEC2}.

Now starting with the truncated version of Eq.\eqref{eq:paramole}, we
adopt the convention that
\begin{align}
  \omega_i 
  &= \eta^{(m)}_{i,d_e}  d_e  + k^{(m)}_{i,C_S}  C_S \notag \\
  &= -\frac{E_{\text{eff}, i}}{\hbar} d_e + \frac{W_{i}}{\hbar} C_S \; ,
    \label{eq:paramole2}
\end{align}
where all signs, $g$-factors, spin magnitudes, etc. are absorbed into
the coefficients on the right-hand side.  It is also common to refer
to $\mathcal{E}_{\text{eff}}$ as the effective electric field, where
\begin{align}
    E_{\text{eff}} 
    &= \mathcal{E}_\text{eff} \sgn(\vec J \cdot \hat n) \left< \hat n \cdot \hat z \right>,\;
\end{align}
and $\vec J$ is the total electronic angular momentum, $\hat n$ is the
direction of the internuclear axis, and $\hat z$ is the direction of
the externally applied electric field. Here $\Omega = \vec J \cdot
\hat n$ is a quantum number used in molecular term symbols, and
$\left< \hat n \cdot \hat z \right>$ indicates the degree to which the
molecule is electrically polarized by the applied electric field. Some
of the conventions used for the orientation of the internuclear axis,
the internal electric field, etc. are summarized in Appendix~A.3 of
Ref.~\cite{ACME:2016sci}.

The truncated expression for the frequency shift thus provides us with
physical expressions for the two constants,
\begin{align}
    \eta^{(m)}_{i,d_e} = - \frac{E_{\text{eff}, i}}{\hbar}
    \qquad \text{and} \qquad 
    k^{(m)}_{i,C_S} = \frac{W_{i}}{\hbar}  \; .
\end{align}
The point is that this frequency measurement cannot be directly
translated into a permanent molecular dipole moment, since the
molecules are substantially or entirely electrically polarized during
the measurement.  The linear energy shift due to the lab-frame
electric field saturates as $\left< \hat n \cdot \hat z
\right>\rightarrow 1$, and the limiting dependence on the external
fields is that of an induced dipole moment.  To express the frequency
difference $\omega_i$ in terms of a physically meaningful permanent
electric dipole moment, we define the molecule's $d_i$ in relation to
the effective electric field,
\begin{align}
   - E_{\text{eff},i} d_i 
   \equiv -E_{\text{eff},i} d_e + W_{i}C_S \; .
\label{eq:assumption}
\end{align}
We can use this form to define, in analogy to Eq.\eqref{eq:paraatom},
\begin{align}
   d_i = d_e  + \alpha_{i,C_S} C_S
   \qquad \Leftrightarrow \qquad 
   \alpha_{i,d_e} = 1 
   \quad \text{and} \quad  
   \alpha_{i,C_S} = - \frac{W_{i}}{E_{\text{eff}, i}} =  \frac{k^{(m)}_{i,C_S}}{\eta^{(m)}_{i,d_e}}\; .
    \label{eq:paraatom2}
\end{align}
For our analysis we re-cast the limits from open-shell molecules in
units of $e\text{cm}$, such that the coefficients for all systems can
be expressed in the same units, with the side effect that
$\alpha_{i,d_e}=1$ for all open-shell molecules. This approach also
serves two additional purposes: (i) Many different conventions are in
use for these coefficients, sometimes using the same symbols for
different quantities. Since measured quantities are related in publications to the
electron EDM, our choice makes comparisons across different works
relatively straightforward. Of course, all quantities must be
ultimately connected to an experimentally measured phase. (ii) As
first pointed out in
Refs.~\cite{PhysRevA.84.052108,PhysRevA.85.029901}, while there is
considerable variation in the literature values for
$\eta^{(m)}_{i,d_e}$ and $k^{(m)}_{i,C_S}$ in a given system, there is
much less variation in their ratio. Dividing Eq.\eqref{eq:paramole}
by $\eta^{(m)}_{i,d_e}$ should pass this uncertainty on to all
semileptonic and hadronic coefficients, which are of much less relevance in open-shell systems.

Note that $\eta^{(m)}_{i,d_e}$ is used both to obtain single-source
limits on $d_e$ from the experimentally measured $\omega_i$, and to
convert the experimentally measured frequencies into EDM units for
Tab.~\ref{tab:meas}. The experimental EDM limits for open-shell
molecules in Tab.~\ref{tab:meas} are rescaled by the indicated factor
$x_i$, which differs from unity when updated values for
$\eta^{(m)}_{i,d_e}$ differ from the cited publication. This is
typically the case when improved molecular structure calculations have
become available since the experimental limit was published.

Recommended values for many of the $\alpha_{i,c_j}$ are given in Tables
III-V of Ref.~\cite{Chupp:2017rkp} and in Table 4 of
Ref.~\cite{Fleig:2018bsf}, including in many cases the ranges
corresponding to theory uncertainties (or at least the ranges of reported values, which is often what we are forced to use in lieu of true theory uncertainties). We give our choices for
$\alpha_{i,c_j}$ in Tab.~\ref{tab:LEC2}.

%%%%%%%%%%%%%%%%%%%%%%%%%%%%%%%%%%%%%%%%%%%%%%%%%%%%%%%%%%%%%%%%%%%%%%%%
\subsection{Closed-shell (diamagnetic) systems}
\label{eq:global_dia}

In contrast to the open-shell systems, where sub-leading
contributions are largely neglected due to lack of theory inputs, the
small contributions of $d_e$ and $C_S^{(0)}$ are taken into account for all
closed-shell systems. These are typically smaller contributions to the
observable EDM in cases where all electron spins are paired, with the
main contributions coming rather from nucleon EDMs, nuclear forces
mediated by pion exchange with strengths $g_\pi^{(0,1,2)}$, and the
nuclear-spin-dependent semileptonic interactions $C_T^{(0)}$ (and possibly
$C_P^{(0)}$).  The experimental precision, especially of Hg, is nevertheless
high enough to contribute meaningful constraining power for $d_e$ and
$C_S^{(0)}$.

Unfortunately it appears that arguments~\cite{Fleig:2018bsf} to the
effect that closed-shell systems add significant complementary
constraining power in the $d_e-C_S$ subspace due to a different sign
of the ratio $\alpha_{i,d_e}/\alpha_{i,C_S}$ in comparison to
open-shell systems, do not in fact hold. While recent
calculations~\cite{Gaul:2023hdd} may indicate that $\alpha_{i,d_e}$ and
$\alpha_{i,C_S}$ in some highly-correlated closed-shell systems could indeed have different sign, the theory uncertainties for these cases are large enough to cross zero. Numerical calculations for specific systems so far only support ratios of the same sign \cite{PhysRevA.103.012807}. Two subtleties are important to consider in this context: (1) the value for $\alpha_{i,d_e}$ includes contributions from multiple different interactions \cite{Gaul:2023hdd,Martensson-Pendrill_1987,Dzuba:2009kn,flambaum1985new} whose contributions must be summed, and (2) our implementation of $\alpha_{C_S^{(0)}}$ could generate sign changes via the difference of terms in equation \ref{eq:alpha_CS}.

At present, as can
be seen from Tab.s~\ref{tab:LEC} and ~\ref{tab:LEC2}, the only measured system where
$\alpha_{i,d_e}$ and $\alpha_{i,C_S}$ may have opposite sign are Hg, Yb, and Ra. On the other hand, the allowed ranges for $\alpha_{i,d_e}$ cross zero, and thus admit also a positive ratio as in all other systems discussed here. These may thus represent special cases
among closed-shell systems, in that reducing the error of experimental
and theory inputs could have significant impact on the $d_e-C_S^{(0)}$
subspace that is complementary to the dominating open-shell
constraints. The supporting arguments for any such claims should, however, be examined in detail.

The contributions of nuclear forces and nucleon EDMs are frequently
interpreted via the Schiff moment of a given
nucleus~\cite{Schiff:1963zz,Khriplovich:1997ga,Chupp:2017rkp}, which is more easily related to nuclear
structure parameters~\cite{GINGES200463,Engel:2025uci}.  In terms of our model
coefficients, the Schiff moment $S_i$ of system $i$ and a
system-specific coefficient $k_{i,S}$ can be expressed related to the
corresponding EDM via
\begin{align}
   k_{i,S} S_i & = \sum_{c_j \in \{ d_{n,p},g_\pi^{(0,1,2)} \}} \alpha_{i,c_j}c_j  \notag \\
   &\approx k_{i,S}\left[ s_{i,n} d_n + s_{i,p} d_p + \frac{m_N g_A}{F_\pi}\left( a_{i,0}g_\pi^{(0)} + a_{i,1}g_\pi^{(1)}  + a_{i,2}g_\pi^{(2)}\right)\right] \notag \\%\; . 
   &= k_{i,S}\left[ s_{i,n} d_n^\text{sr} + s_{i,p} d_p^\text{sr} + \frac{m_N g_A}{F_\pi}\left( \tilde a_{i,0}g_\pi^{(0)} + \tilde a_{i,1}g_\pi^{(1)}  + \tilde a_{i,2}g_\pi^{(2)}\right)\right]\; .
\label{eq:Schiff}
\end{align}
The coefficients $s_{i,N}$ ($N=n,p$) indicate the contributions
from EDMs of unpaired nucleons and the coefficients $a_{i,m}$
($m=0,1,2$) parameterize the strength of CP-violating pion exchange,
organized by isospin, for the nucleus of system $i$.
%Between the
%second and third lines
We drop the $g_\pi^{(2)}$ term as discussed
above and note that the coefficients in front of $g_\pi^{(0)}$ and
$g_\pi^{(1)}$ change when we replace $d_{n,p}$ with $d_{n,p}^\text{sr}$ using Eq.\eqref{eq:d_np} (i.e., $a_{i,j}\neq\tilde a_{i,j}$). While we continue to neglect $C_1$ and $C_2$ from Eq.\eqref{eq:piN}, these also contribute to the Schiff moment (the sensitivity coefficients for most measured species have not, however, been calculated). The coefficients $k_{i,S}$ are calculated for many systems of interest, and give the degree of electronic screening that suppresses an observable system EDM relative to its underlying nuclear Schiff moment. The Schiff moment itself can, in principle, be large in heavy and especially deformed nuclei such as $^{225}$Ra.  This can be viewed as a nuclear-structure
induced enhancement of the observable EDM, which affects not only the pion-exchange forces but also the nucleon EDMs and other contributions to $S_i$.
%and use Eq.(\ref{eq:d_np}) to rewrite the nucleon EDMs
%themselves and absorb their contributions from $g_\pi^{(0)}$ and
%$g_\pi^{(1)}$ into the implicitly re-defined coefficients $\tilde
%a_{i,m}$.

%Nuclear magnetic quadrupole moments, which also induce a higher-order correction to $S_i$, are themselves \textit{not} screened, and represent an experimentally useful signature of CP violation that we do not consider here in detail.

For nuclei with spin $I>1/2$, a CP-violating nuclear magnetic
quadrupole moment (MQM) can in principle exist in analogy to a Schiff-moment-induced EDM and be analyzed within a common global analysis (the MQM actually induces a higher-order correction to $S_i$). The MQM is \textit{not} screened, and itself represents an experimentally useful signature of CP violation that we do not consider here in detail (but see \cite{Khriplovich:1997ga,GINGES200463,Flambaum:2014jta}). At present, of the experimental systems already measured, this effect is relevant essentially only for Cs.

Among the
potentially leading contributions from our six hadronic-scale model
parameters $C_{P,T}^{(0)}$, $d_{n,p}$, and $g_\pi^{(0,1)}$,
very different weights can arise according to the electronic and
nuclear structure of the various closed-shell systems. Inspecting
Tab.~\ref{tab:LEC2} reveals that to a limited extent, this is indeed
already the case for the systems already measured. However, this complementarity is not yet fully
exploited and should be the target of further study.

Despite the pion pole enhancement, $C_P^{(0)}$ appears suppressed
relative to $C_T^{(0)}$ in all measured systems. The coefficients of
$g_\pi^{(0,1)}$ are typically of comparable size for a given system,
although possibly different in sign. Unfortunately the coefficients of
$d_{n,p}$ are not well known for most nuclei, although in principle
these can be calculated.  For the pion-exchange forces, even in nuclei
that have been the object of many studies, the theory uncertainties
are large. This is especially true for soft nuclei such as our Hg and Xe, in which non-static deformations can present special challenges.

As part of a global analysis, a large number of experimental
measurements from complementary closed-shell systems can disentangle
contributions from model parameters that are relevant at some level
for all of them. In this sense the role of $d_e$ and $C_S^{(0)}$ takes on a
new importance, not only to further constrain these parameters
themselves, but also as an additional contribution to the EDM that
brings along uncertainties which dilute the constraining power for
other model parameters. Closed-shell molecular systems such as TlF, or
molecular systems containing Schiff-enhanced nuclides, introduce
complementary constraining power to our global analysis.

We finally note that closed-shell systems unite theory inputs from
several quite different communities. Different conventions for assigning a negative
isospin projection in the nucleon doublet have to be carefully noted
when combining calculated coefficients from different sources,
especially when we include semileptonic interactions together with
nuclear forces. Reference~\cite{Yanase:2020agg} makes an effort to
disambiguate a part of this issue for $\alpha_{g_\pi^{(0,1,2)},^{129}\text{Xe}}$.

It is only possible to establish meaningful constraints in a context
where the notation and conventions are clear, and where the
uncertainties associated with theory inputs have been clearly
assessed. Table~\ref{tab:hadronic_details_app} in the appendix gives the literature values and references, adapted as needed for consistency within our framework here, that we have used to produce Tab.s \ref{tab:LEC2} and \ref{tab:LEC3_app} for actually performing a global analysis.

%%%%%%%%%%%%%%%%%%%%%%%%%%%%%%%%%%%%%%%%%%%%%%%%%%%%%%%%%%%%%%%%%%%%%%%%
\subsection{Single parameter ranges}
\label{sec:global_onepara}

In this section we provide single-source ranges for the individual model parameters of Eq.~\eqref{eq:low_paras1} plus the proton EDM,
\begin{align}
  c_j \in \Big\{ \; 
  d_e, C_S^{(0)}, C_T^{(0)}, C_P^{(0)}, g_\pi^{(0)}, g_\pi^{(1)}, d_n, d_p  \;  \; 
  \Big\} \; ,
\label{eq:low_paras1c}
\end{align}
They are derived from each measured system by neglecting all other parameters that could contribute to that system's EDM.
To extract these single-parameter ranges, we use isolated terms from Eq.\eqref{eq:linear0} for each
EDM measurement listed in Tab.~\ref{tab:meas} and the model
dependence from Tab.~\ref{tab:LEC2}.  For the likelihood we assume a
Gaussian form, which means that all limits are quoted as symmetric
one-sigma error bars around the central, best-fit value. While 
these single-parameter limits allow us to compare the reach of 
different measurements for CP violation as a whole, they do not 
give the allowed ranges of the
individual model parameters. The reason is that contributions from different
model parameters to the same measurement can cancel. Because an EFT builds on
the assumption that many higher-dimensional operators are induced by a
given physics model, the apparent constraining power of single-parameter limits is overly
optimistic.

%-------------------------------------------
\begin{table}[b!]
\centering
\begin{small}
\begin{tabular}{c|rrrr}
\toprule
System $i$
& $d_e\left[ \text{$e$ cm} \right]$
& $C_S^{(0)}$ 
& $C_P^{(0)}$
& $C_T^{(0)}$ \\
\midrule
Tl
& $\left( 7.2 \pm 7.7 \right)\cdot 10^{-28}$
& $\left( 5.9 \pm 6.4 \right)\cdot 10^{-8}$
& $\left(-2.9 \pm 3.1 \right)\cdot 10^{-6}$
& $\left(-4.5 \pm 4.9 \right)\cdot 10^{-5}$
\\
Cs
& $\left(-1.5 \pm 5.6 \right)\cdot 10^{-26}$
& $\left(-2.3 \pm 8.9 \right)\cdot 10^{-6}$
& $\left( 1.3 \pm 5.0   \right)\cdot 10^{-4}$
& $\left(-1.1 \pm 4.1 \right)\cdot 10^{-4}$
\\ \midrule
$^{199}$Hg
& $\left(1.9 \pm 2.7 \right)\cdot 10^{-28}$
& $\left(-1.7 \pm 2.5 \right)\cdot 10^{-9}$
& $\left( 3.3 \pm 4.7 \right)\cdot 10^{-8}$
& $\left(-3.4 \pm 4.9 \right)\cdot 10^{-10}$
\\
$^{129}$Xe
& $\left( 2.2 \pm 2.3 \right)\cdot 10^{-25}$ 
& $\left( 8.4 \pm 8.7 \right)\cdot 10^{-7}$ 
& $\left(-1.0 \pm 1.1 \right)\cdot 10^{-5}$ 
& $\left(-1.4 \pm 1.5 \right)\cdot 10^{-7}$ 
\\
$^{171}$Yb
& $\left( -4.7 \pm 3.6 \right)\cdot 10^{-24}$
& $\left( 5.2 \pm 4.0 \right)\cdot 10^{-6}$
& $\left(-1.4 \pm 1.1 \right)\cdot 10^{-4}$
& $\left( 1.8 \pm 1.4 \right)\cdot 10^{-6}$
\\
$^{225}$Ra
& $\left(-0.7 \pm 1.1 \right)\cdot 10^{-22}$
& $\left( 3.5 \pm 5.3   \right) \cdot 10^{-4}$
& $\left(-5.2 \pm 7.9 \right)\cdot 10^{-3}$
& $\left(-0.9 \pm 1.3  \right)\cdot 10^{-4}$
\\
TlF
& $\left(-1.3 \pm 2.1 \right)\cdot 10^{-26}$
& $\left(-1.2 \pm 2.0 \right)\cdot 10^{-7}$
& $\left(-1.1 \pm 1.9  \right)\cdot 10^{-5}$
& $\left(-0.9 \pm 1.6 \right)\cdot 10^{-8}$
\\ \midrule
HfF$^+$
& $\left(-1.3 \pm 2.1 \right)\cdot 10^{-30}$
& $\left(-1.4 \pm 2.3 \right)\cdot 10^{-10}$
\\
ThO
& $\left( 4.3 \pm 4.0 \right)\cdot 10^{-30}$
& $\left( 2.8 \pm 2.7 \right)\cdot 10^{-10}$
\\
YbF
& $\left(-2.4 \pm 5.9 \right)\cdot 10^{-28}$
& $\left(-2.7 \pm 6.6 \right)\cdot 10^{-8}$
\\ \bottomrule
& $g_\pi^{(0)}$
& $g_\pi^{(1)}$
& $d_n\left[ \text{$e$ cm} \right]$ 
& $d_p\left[ \text{$e$ cm} \right]$\\
\midrule
%$n$
%& $\left(0 \pm 8.1 \right)\cdot 10^{-13}$ 
%& $\left(0 \pm 4.1 \right)\cdot 10^{-11}$
%& $\left(0 \pm 1.1 \right)\cdot 10^{-26}$ 
%& $\left(0 \pm 1.1 \right)\cdot 10^{-26}$\\
%\midrule
Tl
& $\left(6.2 \pm 6.7 \right)\cdot 10^{-8}$ 
& $\left(-1.8 \pm 1.9 \right)\cdot 10^{-6}$
& $\left(7.0 \pm 7.5 \right)\cdot 10^{-20}$ 
& $\left(-2.5 \pm 2.7  \right)\cdot 10^{-20}$
\\
Cs
& $\left(-2.2 \pm 8.6 \right)\cdot 10^{-6}$
& $\left(-0.7 \pm 2.6 \right)\cdot 10^{-6}$
& $\left( -1.8 \pm 6.9   \right)\cdot 10^{-19}$
& $\left(-0.3 \pm 1.0 \right)\cdot 10^{-20}$
\\ 
\midrule
$^{199}$Hg
& $\left(-0.7 \pm 1.0 \right)\cdot 10^{-12}$
& $\left(-3.6 \pm 5.1 \right)\cdot 10^{-13}$
& $\left(-1.6 \pm 2.3 \right)\cdot 10^{-26}$
& $\left(-1.6 \pm 2.3 \right)\cdot 10^{-25}$
\\
$^{129}$Xe
& $\left( 4.5 \pm 4.7 \right)\cdot 10^{-10}$ 
& $\left( 6.0  \pm 6.2 \right)\cdot 10^{-10}$
& $\left(-7.7 \pm 7.9 \right)\cdot 10^{-24}$
& $\left(-3.6 \pm 3.7 \right)\cdot 10^{-23}$
\\
$^{171}$Yb
& $\left( 2.4 \pm 1.8 \right)\cdot 10^{-9}$
& $\left(1.2\pm 0.9 \right)\cdot 10^{-9}$
& $\left( 6.0 \pm 4.6 \right)\cdot 10^{-23}$
& $\left( 6.0 \pm 4.6 \right)\cdot 10^{-22}$
\\
$^{225}$Ra
& $\left( 2.3 \pm 3.5 \right)\cdot 10^{-9}$
& $\left(-5.8 \pm 8.7 \right)\cdot 10^{-10}$
& $\left(-0.7 \pm 1.1 \right)\cdot 10^{-20}$
& $\left(-3.4 \pm 5.0 \right)\cdot 10^{-20}$
\\
TlF
& $\left(2.4 \pm 4.1  \right)\cdot 10^{-11}$
& $\left(-0.7 \pm 1.2 \right)\cdot 10^{-9}$
& $\left( 2.7 \pm 4.6 \right)\cdot 10^{-23}$
& $\left( -1.0 \pm 1.6 \right)\cdot 10^{-23}$
\\ \bottomrule
\end{tabular}
\end{small}
\caption{Single-parameter ranges allowed by each of the EDM
  measurements given in Tab.~\ref{tab:meas} and coefficients from Tab.~\ref{tab:LEC2}. The neutron EDM itself is best constrained by the direct experimental measurement using neutrons, see Tab.~\ref{tab:meas}.}
\label{tab:onepara}
\end{table} 
%------------------------------------------

Using the single-parameter constraints in Tab.~\ref{tab:onepara} we
can compare the impact of different EDM measurements on a given model
parameter, with the caveat that multi-dimensional correlations might
change that picture. Starting with the electron EDM $d_e$, the
open-shell molecules HfF$^+$ and ThO provide the strongest
constraints, while the open-shell YbF molecule has a similar
constraining power as the closed-shell Hg. The same two open-shell
molecules are most constraining for the scalar electron-nucleon
coupling $C_S^{(0)}$, again followed by Hg, YbF, and Tl.

The pseudoscalar and tensor electron-nucleon couplings $C_{P,T}^{(0)}$ can
be probed by the open-shell atoms and the closed-shell systems. This
is done at present by far most efficiently with Hg and then, with much
reduced constraining power, by Tl, Xe, and TlF. It is difficult to
estimate the impact of these measurements on the combination of $C_P^{(0)}$
and $C_T^{(0)}$, and we will see in the next section how this more complex
dependence affects the full global analysis.

The four hadronic parameters, $g_\pi^{(0,1)}$ and $d_{n,p}$,
are most strongly constrained by the neutron EDM measurement and again
Hg, suggesting that there will be significant correlations between the
leptonic and hadronic model parameters in the global analysis. The
other closed-shell systems lead to much weaker limits, but are still needed to constrain the 3-dimensional hadronic model space. Note
that we fix the relation between the short-range neutron and
proton parameters in the Lagrangian, Eq.\eqref{eq:same_pn}, but from
Tab.~\ref{tab:LEC2} we know that different measurements probe
different admixtures of these two parameters.

%%%%%%%%%%%%%%%%%%%%%%%%%%%%%%%%%%%%%%%%%%%%%%%%%%%%%%%%%%%%%%%%%%%%%%
\section{Global analysis}
\label{sec:results}

To jointly analyze the available EDM measurements in terms of
the hadronic-scale Lagrangian and its parameters, Eq.\eqref{eq:low_paras1},
we use the established \sfitter analysis tool. It constructs a global
likelihood with a comprehensive uncertainty treatment and analyses it
in terms of high-dimensional correlations. Lower-dimensional and
one-dimensional likelihoods for the individual model parameters can be
derived by profiling or marginalization, depending on the preferred
statistical framework. If we assume that experimental uncertainties
are Gaussian, profiling and marginalization have to lead to identical
results. For the theory uncertainties, discussed in
Sec.~\ref{sec:results_th}, the difference between the two approaches
makes a formal, but not significant difference.

%%%%%%%%%%%%%%%%%%%%%%%%%%%%%%%%%%%%%%%%%%%%%%%%%%%%%%%%%%%%%%%%%%%%%%%
\subsection{SFitter framework}
\label{sec:results_sfitter}

\sfitter~\cite{Lafaye:2004cn,Lafaye:2007vs,Lafaye:2009vr} has been
developed for global analyses of LHC measurements in the context of
BSM physics~\cite{Butter:2016tjc} and Higgs and top
properties~\cite{Klute:2012pu,Biekotter:2018ohn,Elmer:2023wtr,Brivio:2019ius},
including comprehensive studies of the connection between EFTs and
UV-completions of the SM~\cite{Lopez-Val:2013yba,Brivio:2021alv}. It
has a focus on its uncertainty treatment, including theory
uncertainties~\cite{Ghosh:2022lrf}, the connection between Bayesian
and frequentist approaches~\cite{Lafaye:2007vs,Brivio:2022hrb}, and
published experimental likelihoods~\cite{Elmer:2023wtr}.

This first \sfitter analysis of EDMs relates the 11 measurements from
Tab.~\ref{tab:meas} to the seven model parameters in
Eq.\eqref{eq:low_paras1}. In general, \sfitter includes statistical,
systematic, and theory uncertainties. At the heart of \sfitter is the
fully exclusive likelihood as a function of model and nuisance
parameters.  All measurements are described as uncorrelated, with the
individual statistical uncertainties described by Poisson or Gaussian
likelihood. Statistical uncertainties are, usually, uncorrelated as
well and described by a Poisson distribution, turning into a Gaussian
for high statistics. Experimental systematics are assumed to
have a Gaussian shape, but can be described by any nuisance
parameter. This Gaussian shape is justified for parameters which are
measured elsewhere. We use it for all experimental uncertainties, assuming that the measurements are at least as much dominated by statistical uncertainty, with error distribution appropriate for count-rate limited frequency measurements.

For the EDM analysis the situation is then relatively simple. First, from
Eq.\eqref{eq:linear0} we know that all observables depend on the model
parameters linearly.  Second, we can combine the statistical and
systematic experimental uncertainties into the symmetric Gaussian
error bars given in Tab.~\ref{tab:meas}. Finally, we do not have to
consider nuisance parameters, if we assume that the likelihood has a
Gaussian form for each independent measurement. This Gaussian
assumption also implies that for uncorrelated uncertainties a profile
likelihood and a Bayesian marginalization will give the same result.

In our implementation theory uncertainties have no well-defined likelihood shape, and no
maximum: they can be thought of as a range of allowed values~\cite{Ghosh:2022lrf}. A
flat theory uncertainty is not parametrization-invariant, as one would expect from a fixed range, but without a preferred central value, we consider it conservative. With this implementation in \sfitter following a flat distribution, the central value of the parameter can be shifted within this range at no cost in the likelihood. Therefore, compared to analyses without considering theory uncertainties, the impact of the central value of certain parameters is negligible for larger ranges allowed by theory uncertainties. For the EDM global analysis, theory uncertainties significantly affect most
$\alpha$-values: the central values given in
Tab.~\ref{tab:LEC2} mainly impact implementations that neglect theory uncertainties, while the corresponding ranges are by far more relevant when theory uncertainties are taken into account.

To construct the exclusive likelihood, \sfitter evaluates EDM
predictions over the entire model parameter space. It uses a Markov
chain to encode the likelihood in the distribution of points covering
the model space. A helpful aspect, common to many BSM analyses, is
that we can safely assume the global minimum of the likelihood to be
at the SM parameter point. To remove nuisance parameters or to extract
limits on a reduced number of model parameters, \sfitter can employ a
profile likelihood or a Bayesian
marginalization~\cite{Lafaye:2007vs,Brivio:2022hrb}. These two methods
give different results, with the exception of uncorrelated Gaussians.
Profiling over flat theory uncertainties and Gaussian experimental
uncertainties leads to the RFit\cite{Hocker:2001xe} prescription,
profiling over two parameters with flat likelihood leads to linearly
added uncertainties even for uncorrelated parameters.

%While Eq.\eqref{eq:linear0} suggests a homogeneous set of model parameters, 
The typical sizes of the model parameters in
Eq.\eqref{eq:low_paras1} and the $\alpha$-values in
Tab.~\ref{tab:LEC2} can be extremely different.  For numerical
robustness, we internally re-scale each model parameter and each
$\alpha$-value such that all model parameters are evaluated with
similar size (while each term in Eq.~\eqref{eq:linear0} remains invariant).  Concretely, this means rescaling $d_e$ by a factor $10^{29}$, $C_S^{(0)}$
by $10^9$, $g_\pi^{(1)}$ and $g_\pi^{(0)}$ by $10^{10}$, $C_T^{(0)}$ by
$10^8$, $C_P^{(0)}$ by $10^6$, and $d_{n/p}$ by $10^{23}$. These
rescalings are also reflected in the way we present our results.

%%%%%%%%%%%%%%%%%%%%%%%%%%%%%%%%%%%%%%%%%%%%%%%%%%%%%%%%%%%%%%%%%%%%%%
\subsection{Well-constrained model sub-space}
\label{sec:results_well}

%-------------------------------------------
\begin{figure}[t]
\centering
\includegraphics[width=0.90\textwidth]{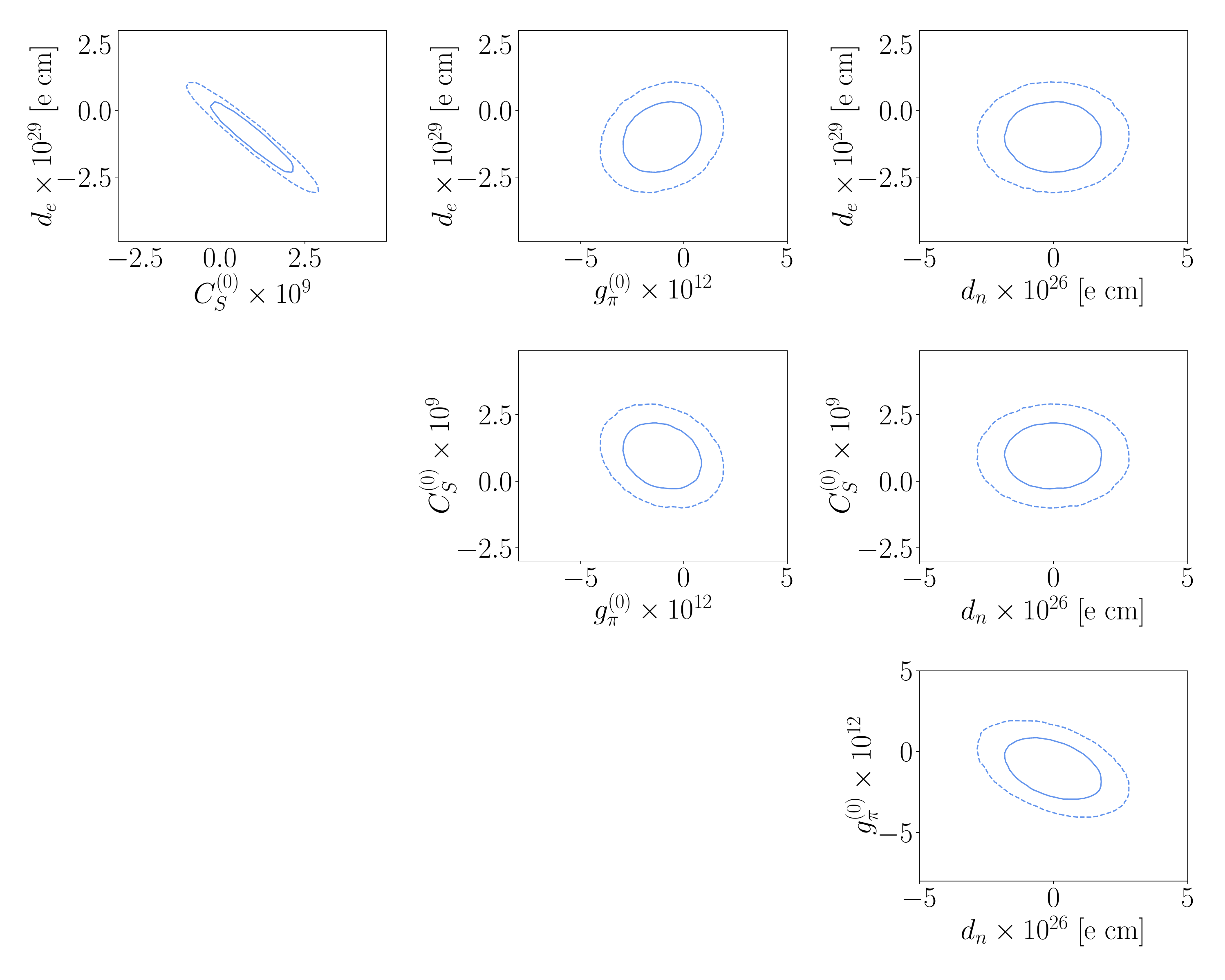}
\caption{Correlations from the 4-dimensional analysis of $\{ d_e, C_S^{(0)},
  g_\pi^{(0)}, d_n \}$, based on all EDM measurements but
  neglecting theory uncertainties. The ellipses indicate $68\%$ and 
  $95\%$~CL.}
\label{fig:4d}
\end{figure}
%-------------------------------------------

As a starting point of the global analysis and to understand the main
features, we consider a subspace of relatively well-constrained parameters.
Following our discussion in Sec.~\ref{sec:global_onepara} we expect
$d_e$ and $C_S^{(0)}$ to be constrained well by
the open-shell molecules HfF$^+$ and ThO. Similarly, the hadronic
parameters $g_\pi^{(0)}$ and $d_n$ are strongly constrained
by the neutron and Hg EDMs. This means the model subspace
\begin{align}
  \Big\{ \; 
  d_e, C_S, g_\pi^{(0)}, d_n \; 
  \Big\} 
\label{eq:4d_paras}
\end{align}
should be constrained well by the full set of measurements given in
Tab.~\ref{tab:meas}.

It is instructive to consider how the constraints on these four model
parameters are correlated.  In Fig.~\ref{fig:4d} we show these
correlations extracted as 2-dimensional profile likelihoods from the
fully exclusive, 4-dimensional likelihood. Three structural aspects
are evident: (i) a strong anti-correlation between $d_e$ and $C_S^{(0)}$; (ii)
a very slight anti-correlation between $g_\pi^{(0)}$ and $d_n$;
and (iii) essentially no correlations linking the $\{ d_e, C_S^{(0)}\}$ 
and $\{ g_\pi^{(0)}, d_n \}$ parameter subsets.

The strong correlation between $d_e$ and
$C_S^{(0)}$ and its independence from the remaining parameter space is
expected to remain for the full global analysis. It
is induced by the strongest measurements, HfF$^+$ and ThO, and according
to our parameterization as shown in Tabs.~\ref{tab:LEC2} and~\ref{tab:onepara} those two measurements are not affected by
any other model parameter. This means the upper-left panel of
Fig.~\ref{fig:4d} factorizes from our global EDM analysis, and we can
consider the remaining model parameters separately and without the
HfF$^+$ and ThO measurements.

For the hadronic parameters the situation is different. Again the neutron and Hg measurements are three orders-of-magnitude more constraining than the other measurements. However, as the two model
parameters we could as well have chosen $g_\pi^{(0)}$ vs
$g_\pi^{(1)}$, without any change in the conclusion. This means that
we have to expand the hadronic parameter space next, to see what
patterns emerge. The parameters $d_e$ and $C_S^{(0)}$ will be kept factorized for
the rest of our global analysis.

To understand the role of approximately flat directions in model space, we can
diagonalize and invert the $\alpha$-matrix given in Tab.~\ref{tab:LEC2} for the
well-constrained subsystem.
%. We use all seven independent model parameters
%%
%\begin{align}
%    c_j \in \Big\{ \; 
%  d_e, C_S^{(0)}, C_T^{(0)}, C_P^{(0)}, g_\pi^{(0)}, g_\pi^{(1)}, d_n^\text{sr} \;  \; 
%  \Big\} \;   
%\end{align}
%%
%Following the approximately factorized form of the $\alpha$-matrix, we start with the $d_e - C_S^{(0)}$ subsector. 
To be able to invert the $\alpha$-matrix, we have to truncate it to a square form.  
We know that the leading constraints in the $d_e - C_S^{(0)}$ plane come from the
ThO and HfF$^+$ measurements and can be described by the invertible relation
\begin{align}
\begin{pmatrix}
d_\text{HfF$^+$} \\ d_\text{ThO}
\end{pmatrix}
= 
\begin{pmatrix}
\alpha_{\text{HfF$^+$},d_e} & \alpha_{\text{HfF$^+$},C_S^{(0)}} \\
\alpha_{\text{ThO},d_e} & \alpha_{\text{ThO},C_S^{(0)}} 
\end{pmatrix} \; 
\begin{pmatrix}
d_e \\ C_S^{(0)} 
\end{pmatrix} \; .
\end{align}
We can diagonalize the $\alpha$-submatrix, find the eigenvalues $1.0$ and 
$5.93\cdot 10^{-21}$, and invert it to give the model parameters as a function of the 
measurements, 
\begin{align}
\begin{pmatrix}
d_e \\ C_S^{(0)} 
\end{pmatrix}
&= 
\begin{pmatrix}
    2.55 & -1.55\\
 -1.69\cdot 10^{20} & 1.69\cdot 10^{20}\\
\end{pmatrix} \;   
\begin{pmatrix}
d_\text{HfF$^+$} \\ d_\text{ThO}
\end{pmatrix}  \notag \\
&= 
\begin{pmatrix} 
2.55 d_\text{HfF$^+$} - 1.55 d_\text{ThO}\\
( -1.68 d_\text{HfF$^+$} +1.69 d_\text{ThO} ) \cdot 10^{20} 
\end{pmatrix} \; .
\end{align}
Approximately flat directions in model space appear because the measurements
are uncorrelated. We could determine $d_e$ much more
precisely and without any effect from $C_S^{(0)}$ if we could measure the fully 
correlated combination
$(2.55 d_\text{HfF$^+$} - 1.55 d_\text{ThO})$, which is not experimentally realistic. On the other hand, this provides a clear motivation to more carefully consider the role of deliberately-correlated experiments, such as comagnetometer comparisons in which neither EDM can be neglected, within a global context.

Deliberately correlated measurements can be envisioned, for instance
via comagnetometry, as already used for the neutron and Xe
measurements. In this interpretation it could be advantageous to
perform correlated measurements of two systems, with comparable
sensitivity to both system EDMs; unlike the usual comagnetometer
implementation, where it is usually assumed that one can be neglected
entirely.

%%%%%%%%%%%%%%%%%%%%%%%%%%%%%%%%%%%%%%%%%%%%%%%%%%%%%%%%%%%%%%%%%%%%%%
\subsection{Hadronic parameters from closed-shell systems}
\label{sec:results_dia}

We define a second restricted model parameter space, for the purely hadronic sector:
\begin{align}
  \Big\{ \; 
   g_\pi^{(0)}, g_\pi^{(1)}, d_n \; 
  \Big\} \; .
\label{eq:3d_paras}
\end{align}
All three parameters are constrained by the neutron and closed-shell
EDMs, while we know from the above discussion that the constraints
from closed-shell systems on $d_e$ and $C_S^{(0)}$ are weaker
than those from their open-shell counterparts.

From Tab.~\ref{tab:onepara} we see that the neutron and and Hg
measurements strongly constrain two of the three hadronic model
parameters in Eq.\eqref{eq:3d_paras}.  To understand the correlation
structure, we show the correlated constraints of the six possible remaining pairs of closed-shell measurements on the three different two-dimensional subspaces of
Eq.\eqref{eq:3d_paras}.

%-------------------------------------------
\begin{figure}[b!]
\includegraphics[width=0.99\textwidth]{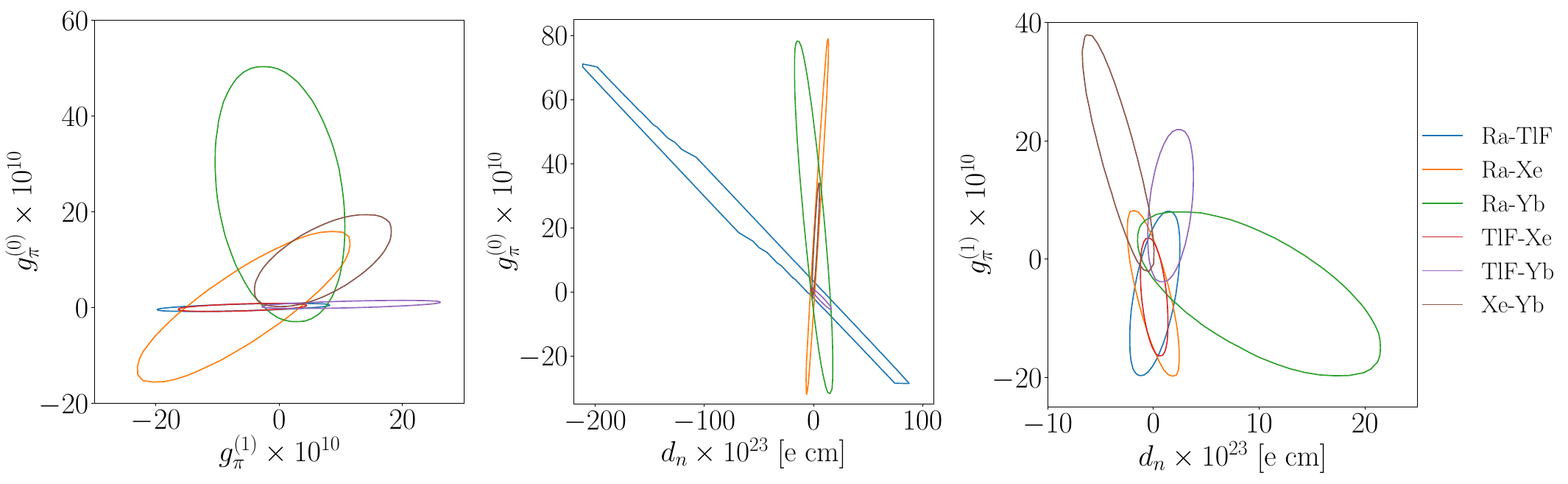}
\caption{Correlations from three 2-dimensional analyses in the $\{
  g_\pi^{(0)} ,g_\pi^{(1)}, d_n\}$ parameter space, each
  based on a different pair of closed-shell EDM measurements, as indicated by the color. The
  ellipses indicate $68\%$~CL, neglecting theory uncertainties.}
\label{fig:diamag}
\end{figure}
%-------------------------------------------

Starting with the left panel of Fig.~\ref{fig:diamag}, the different
pairs of measurements constrain the $g_\pi^{(0)}$ vs. $g_\pi^{(1)}$
plane with different correlation patterns, implying that a global
analysis will constrain the 3-dimensional hadronic subspace much
better than any single pair of measurements.  In the center panel, the
situation changes when we look at the correlations with the neutron
EDM $d_n$. Three combinations are 
aligned to similar negative correlations between $d_n$ and
$g_\pi^{(0)}$. In the right panel, the situation is similar for
$g_\pi^{(1)}$, with a positive correlation and less striking. In both cases the exception are the
combinations of Xe respectively with Ra, TlF, and Yb, which constrain $d_n$
extremely well and without any correlation with $g_\pi^{(0,1)}$. 
While this may sound like an advantage, we observe from
Tab.~\ref{tab:onepara} and in Fig.~\ref{fig:4d} that the corresponding limits
on $d_n$ are \textit{three orders of magnitude weaker} than would be expected from including
the neutron and and Hg measurements.

Altogether, Fig.~\ref{fig:diamag} confirms that the constraints of the
sub-leading four closed-shell measurements on the 3-dimensional
hadronic parameter space of Eq.\eqref{eq:3d_paras} are correlated in a
non-trivial manner.  Evaluating these correlations requires a global
analysis of the, formally, over-constraining set of closed-shell
measurements. Inclusion of $d_p,\,C_1,\,C_2,$ etc. as independent model parameters would further increase the need for complementary experimental constraints.

%%%%%%%%%%%%%%%%%%%%%%%%%%%%%%%%%%%%%%%%%%%%%%%%%%%%%%%%%%%%%%%%%%%%%%
\subsection{Poorly constrained model parameters}
\label{sec:results_poor}

%-------------------------------------------
\begin{figure}[b!]
\includegraphics[width=0.99\textwidth]{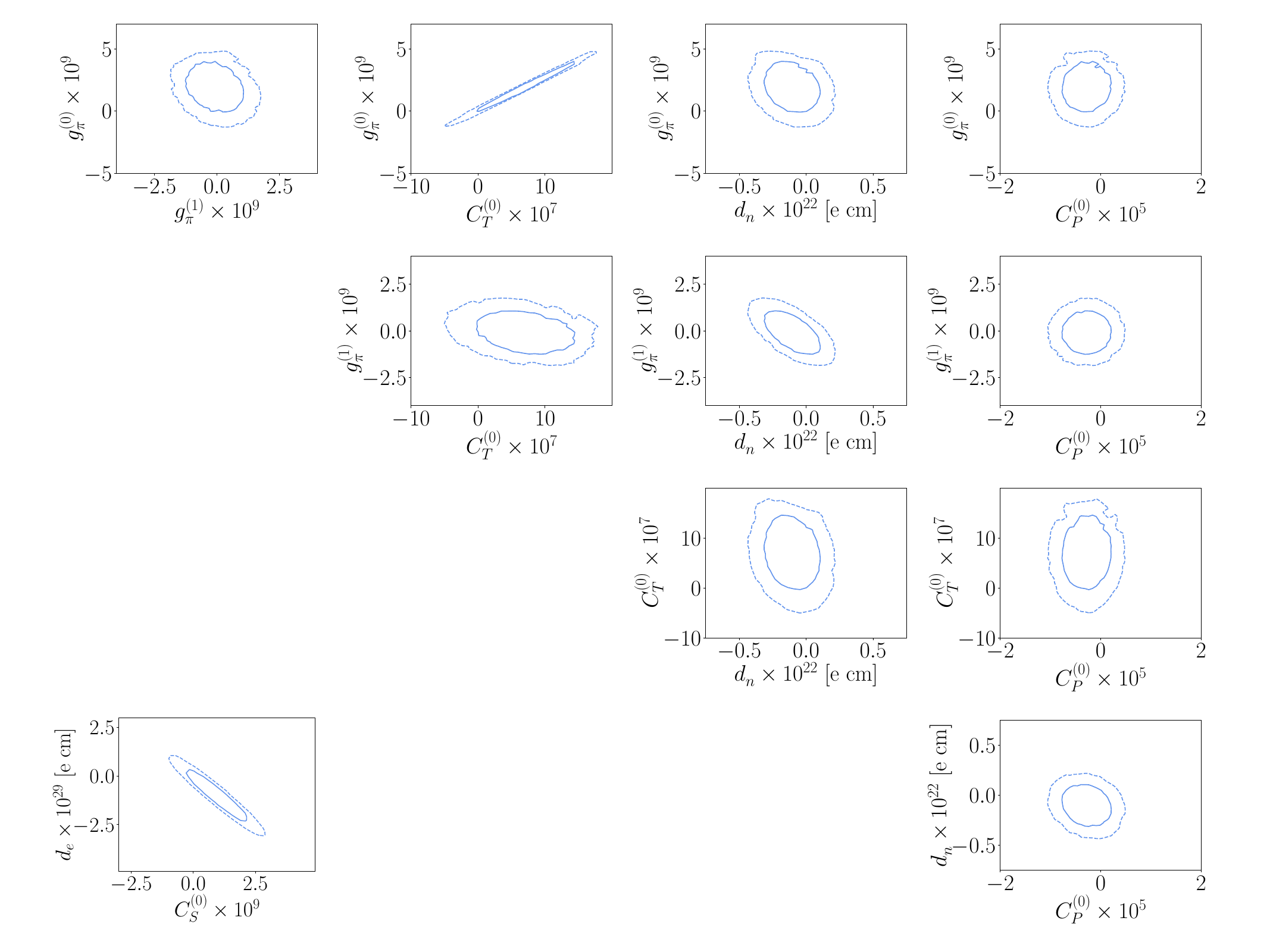}
\caption{Correlations from the 5-dimensional analysis of $\{ C_T^{(0)}, C_P^{(0)},
  g_\pi^{(0)}, g_\pi^{(1)}, d_n \}$ and the factorized $d_e
  - C_S^{(0)}$ plane from Fig.~\ref{fig:4d}.  We ignore the neutron and Hg
  measurements, which induce narrow correlation patterns in the
  5-dimensional parameters space and do not affect the profiled
  2-dimensional correlations. The ellipses indicate $68\%$ and
  $95\%$~CL, neglecting theory uncertainties. }
\label{fig:badmeas}
\end{figure}
%-------------------------------------------

Finally, we combine the effects of all remaining parameters,
\begin{align}
  \Big\{ \; 
 C_T^{(0)}, C_P^{(0)}, g_\pi^{(0)}, g_\pi^{(1)}, d_n \; 
  \Big\} \; ,
\label{eq:low_paras2}
\end{align}
ignoring the ThO and HfF$^+$ measurements, which constrain the
factorized $d_e - C_S^{(0)}$ subspace, and also ignoring the neutron and Hg
measurements. The latter do constrain the above parameters, but because
they are much stronger than all other measurements, they will induce
narrow correlations in the allowed 5-dimensional parameter space.

Narrow correlations in a higher-dimensional parameter space vanish
when we profile the likelihood onto two-dimensional correlations or even
single model parameters. As an example, consider a two-dimensional
parameter space constrained by one strong and one weak
measurement. The strong measurement leads to a narrow correlation
between the two parameters. When we extract the profile likelihood for
one model parameter we can adjust the other model parameter such that
the two parameters trace this narrow correlation.  This way, the
entire length of the correlation pattern gets projected onto the
one-dimensional profile likelihoods. The weaker measurement dominates the
individual profile likelihoods, while the stronger measurement allows us
to link the second model parameter precisely from a value for the first model parameter.

In our case this implies that as long as the correlations from the
neutron and Hg measurements cross the entire parameter
space, we can ignore these two measurements and their correlation
patterns in the following discussion of two-dimensional correlations and
single-parameter profile likelihoods. Losing the best few measurements
contributing to the global analysis is an unfortunate effect of the
conservative profile likelihood approach, but for standard error
ellipses it should also exist for Bayesian marginalization. The main
difference is that this kind of effect is numerically extremely challenging 
to compute using marginalization, while it is essentially trivial for the 
profile likelihood. 

Given these considerations, we are left with five model parameters,
constrained by seven measurements of comparable constraining power.
The only exception we see in Fig.~\ref{fig:badmeas} is a strong
leading correlation between $g_\pi^{(0)}$ and $C_T^{(0)}$, induced by
the fact that for both parameters the TlF measurement is now the
strongest by an order of magnitude.  All other model parameters are
nicely constrained. The allowed range, for instance for $g_\pi^{(0)}$
is of the order $10^{-9}$. This can be compared to the constraints
from Fig.~\ref{fig:4d}, of the order $10^{-12}$. The same hierarchy of
measurements can be observed for $g_\pi^{(1)}$ and for $d_n$, as
confirmed by Tab.~\ref{tab:onepara}. This means that the 5-dimensional
allowed parameter space illustrated by Fig.~\ref{fig:badmeas} is
crossed by two correlation patterns, roughly three orders of magnitude
more narrow than the full parameter space. We emphasize that
explaining this extremely narrow correlation poses a fine-tuning
problem in model parameter space, which the profile likelihood does
not address.

As before, we can truncate the number of available measurements for the 
poorly constrained 5-dimensional subspace given in Eq.\eqref{eq:low_paras2},
and invert the corresponding $\alpha$-submatrix to find
\begin{align}
\begin{pmatrix}
C_T^{(0)} \\ C_P^{(0)} \\ g_\pi^{(0)} \\ g_\pi^{(1)} \\ d_n 
\end{pmatrix}
=
\begin{pmatrix}
7.14\cdot 10^{18} & -3.21\cdot 10^{18} & 6.50\cdot 10^{19} & -8.77\cdot 10^{13} & -1.94\cdot 10^{15}\\
-1.03\cdot 10^{17} & -5.40\cdot 10^{19} & 1.09\cdot 10^{21} & -3.74\cdot 10^{14} & -3.26\cdot 10^{16}\\
-2.49\cdot 10^{14} & -1.45\cdot 10^{17} & 2.83\cdot 10^{18} & -2.37\cdot 10^{12} & -8.60\cdot 10^{13}\\
3.10\cdot 10^{14} & -3.46\cdot 10^{16} & -2.56\cdot 10^{18} & 1.58\cdot 10^{12} & 5.49\cdot 10^{13}\\
0. & 0. & 0. & 0. & 1.\\
\end{pmatrix} \; 
\begin{pmatrix}
d_\text{Tl} \\ d_\text{Hg}\\ d_\text{Xe}\\ d_\text{TlF}\\ d_\text{n}
\end{pmatrix} \; .
\end{align}
%

%%%%%%%%%%%%%%%%%%%%%%%%%%%%%%%%%%%%%%%%%%%%%%%%%%%%%%%%%%%%%%%%%%%%%%
\subsection{Theory uncertainties}
\label{sec:results_th}

%-------------------------------------------
\begin{figure}[t]
\centering
\includegraphics[width=0.90\textwidth]{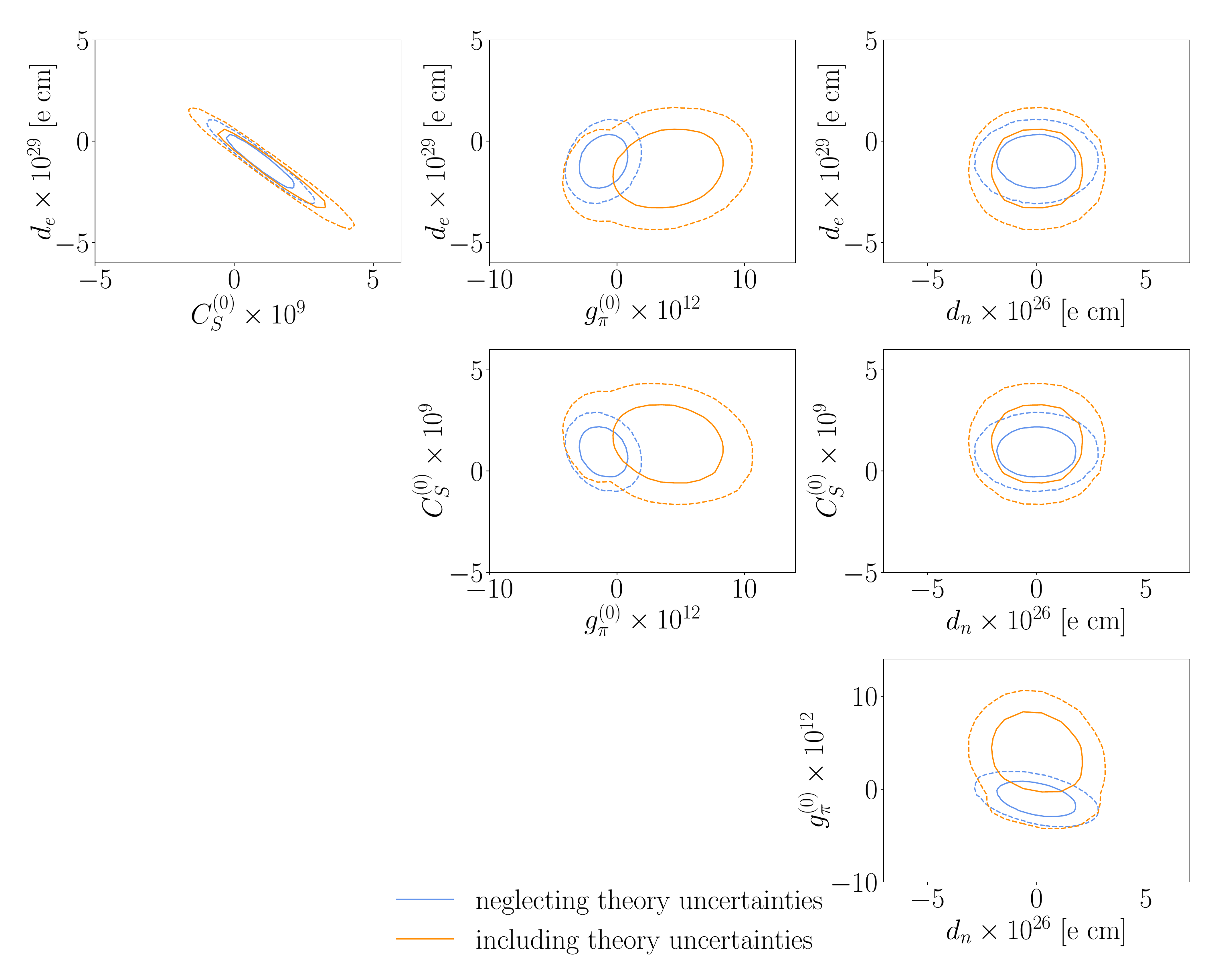}
\caption{Correlations from the 4-dimensional analysis of $\{ d_e, C_S^{(0)},
  g_\pi^{(0)}, d_n \}$. The orange curves show the effect
  of theory uncertainties on the results of Fig.~\ref{fig:4d}. The
  ellipses indicate $68\%$ and $95\%$~CL.}
%\textbf{tp: Where does this shift in $g_\pi$ come from? Without theory uncertainties the Hg measurement prefers small negative values at the $10^{-12}$ level, as we see in the blue curve. With theory uncertainties the limit moves to small positive values, why is that? In the plot I do not see any correlations which could lead to that shift...}}
\label{fig:4d_th}
\end{figure}
%-------------------------------------------

Theory uncertainties always appear when using quantum field theory to
predict observables, like EDMs, from Lagrangian parameters. No calculation method is arbitrarily precise, and a variety of systematic errors can affect accuracy. While there is some hope for estimating and controlling uncertainties for small
expansion parameters, the uncertainties associated with QCD observables at low energies (whether from lattice calculations or
sum-rule estimates) are challenging to quantify.  The precision
of nuclear physics calculations, and their links to effective quantum
field theory, also present difficulties. On the other hand we have
to estimate all of these uncertainties, and can only ignore them in cases where it can be shown
that they are significantly smaller than the experimental
uncertainties of the associated measurements.

%-------------------------------------------
\begin{figure}[t]
\includegraphics[width=0.99\textwidth, page=1]{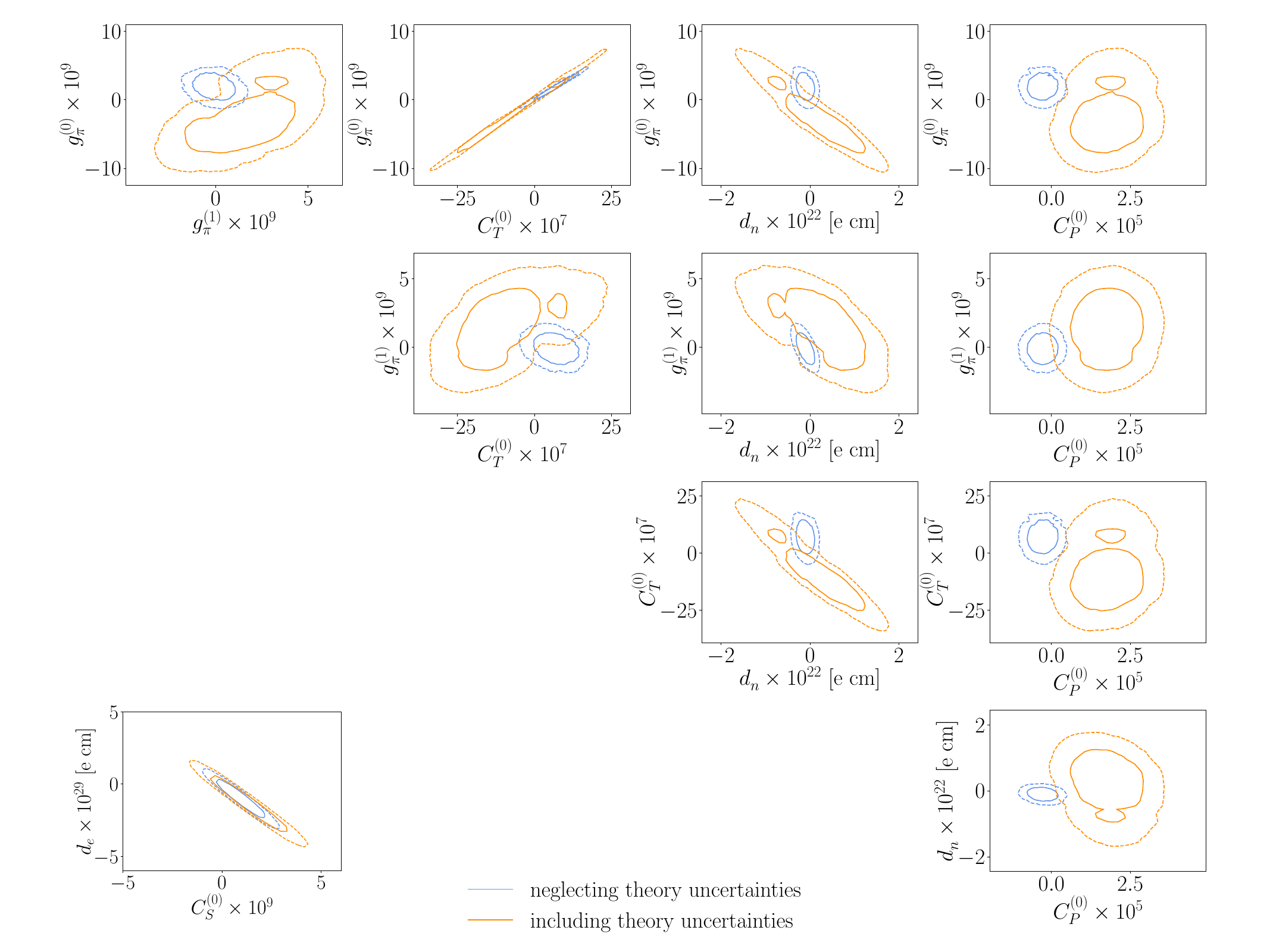}
\caption{Correlations from the 5-dimensional analysis of $\{ C_T^{(0)}, C_P^{(0)},
  g_\pi^{(0)}, g_\pi^{(1)}, d_n \}$, and the factorized $d_e
  - C_S^{(0)}$ plane from Fig.~\ref{fig:4d_th}.  The orange curves show
  the effect of theory uncertainties on the results of
  Fig~\ref{fig:badmeas}. The ellipses indicate $68\%$ and $95\%$~CL.}
\label{fig:5d_th}
\end{figure}
%-------------------------------------------

For our global EDM analysis, theory uncertainties affect the
$\alpha_{i,c_j}$ in Eq.\eqref{eq:linear0}. The estimated allowable ranges of each
coefficient are given in Tab.~\ref{tab:LEC2}. We include them in
\sfitter as uncorrelated theory uncertainties, an assumption that can
be modified if necessary. Because of the flat likelihood as a function
of the theory nuisance parameter, the profile likelihood approach
leads to the theory uncertainties adding linearly, weighted by the
respective model parameter. Profiling over independent $\alpha$-ranges
simplifies the numerical evaluation in two ways: first, any
parameter--observable pair for which $\alpha$ is compatible with zero
will effectively be removed from the global analysis, because the
optimal choice of $\alpha$ will remove all contributions from the
corresponding model parameter; second, even if we cannot choose
$\alpha$ such that measurement and prediction agree, we can choose it
to maximize the likelihood and to minimize the impact of the
measurement, which means we choose the smallest allowed absolute value
of $\alpha$.

As in Sec.~\ref{sec:results_well} we start with the well-constrained
4-dimensional subspace $\{ d_e, C_S^{(0)}, g_\pi^{(0)}, d_n
\}$. Again, $d_e$ and $C_S^{(0)}$ are constrained by
the open-shell molecules HfF$^+$ and ThO, just as without theory
uncertainties. From Tab.~\ref{tab:LEC2} we see that we can ignore the theory
uncertainty in relating the electron EDM parameter $d_e$ to these systems. In the hadronic sector, the theory uncertainties affecting the
relation of $g_\pi^{(0)}$ and $d_n$ to the neutron and Hg
EDMs is either trivial or reasonably small, although for the neutron this situation should not be over-interpreted. The quoted uncertainty arises from propagating ranges for the constants within Eq.\eqref{eq:d_np}, and not from careful evaluations of the chiral expansion itself.

In Fig.~\ref{fig:4d_th} we show the numerical impact of the theory
uncertainties on the 2-dimensional correlations. The slightly stronger
HfF$^+$ measurement, which determines the width of the correlation
pattern, is only minimally affected by the theory uncertainties, while
the larger theory uncertainties on the ThO measurement extend the
length of the ellipse visibly. The main effect of the theory
uncertainties is on $g_\pi^{(0)}$, where they shift the allowed ranges
from slightly negative to sizeable positive values. This analysis
includes all measurements, so according to Tab.~\ref{tab:onepara}
$g_\pi^{(0)}$ is most strongly constrained by the Hg measurement (with
a negative central value) and TlF (with a large and positive central
value). Adding the theory uncertainties weakens in particular the Hg
measurement, which means the two leading constraints get balanced
differently, and the entire range moves to the positive values
preferred by the TlF measurement.

Moving on, we can now look at the effect of the theory uncertainties
on the less-constrained hadronic sector for closed-shell systems,
discussed in Sec.~\ref{sec:results_dia}. Here, Tab.~\ref{tab:LEC2}
shows sizeable, order-one theory uncertainties. In addition, some of
the $\alpha$-values include an allowed zero value when we include
theory uncertainties. Specifically, $g_\pi^{(1)}$ is no longer
constrained by the Yb measurement and also loses the Hg
constraint.
In addition some theory uncertainties for closed-shell systems are not consistent with $\alpha=0$, but are large, which significantly impacts the global analysis.

Finally, we can look at all seven EDM parameters.  As for the case
without theory uncertainties, the neutron and Hg limits are much more
constraining for hadronic parameters than the other measurements. Following the argument given in Sec.~\ref{sec:results_poor},
this means that two-dimensional correlations and single-parameter limits
extracted by profiling the likelihoods will not be significantly impacted by these
strong measurements. The shift in the constrained two-dimensional
correlations are shown in Fig.~\ref{fig:5d_th}. In comparison their
effect on the factorized parameters $d_e$ and $C_S^{(0)}$ (copied from
Fig.~\ref{fig:4d_th}) is mild, and the constraints on the hadronic
model parameters are significantly weaker. The leading correlation in
this parameter extraction we found to come from the TlF measurement
contraining $g_\pi^{(0)}$ and $C_T^{(0)}$ in a correlated manner. This
correlation expands almost to a flat direction when we allow for the
additional theory uncertainties. In turn, this large effect extends to
the entire space of hadronic parameters, weakening and shifting essentially all
constraints.
The large shifts addressed in Fig.~\ref{fig:5d_th}, mostly seen for $C_P^{(0)}$, originate in the treatment and implementation of theory uncertainties. The dominant measurement for $C_P^{(0)}$ suffers from large theory uncertainties, allowing for a shift in estimating the central parameter value, as discussed in Sec.~\ref{sec:results_sfitter}. This clearly emphasizes the impact of theory uncertainties on the correlations and allowed ranges of the model parameters. To obtain a more coherent picture of these parameter correlations and ranges, theory uncertainties must be reduced (ideally including also a more nuanced treatment of the corresponding likelihoods). There are already some measurements and parameter combinations, like the $d_e$ and $C_S^{(0)}$ subspace, for which theory uncertainties are reasonably well under control (as compared to other elements of Tab.~\ref{tab:LEC2}). For this case, theory uncertainties produce no significant change of the correlation behavior. 

%%%%%%%%%%%%%%%%%%%%%%%%%%%%%%%%%%%%%%%%%%%%%%%%%%%%%%%%%%%%%%%%%%%%%%
\section{Outlook}

EDMs are extremely sensitive, targeted probes of one of the most
important symmetries of elementary particles, directly related to
the observed baryon asymmetry in the Universe. The number of EDM
measurements in very different systems has grown rapidly in recent
years, leading to the question of how the different measurements can
contribute to constraining and understanding CP violation in terms of
a fundamental Lagrangian.
  
\begin{figure}[t]
\centering
\includegraphics[width=0.60\textwidth]{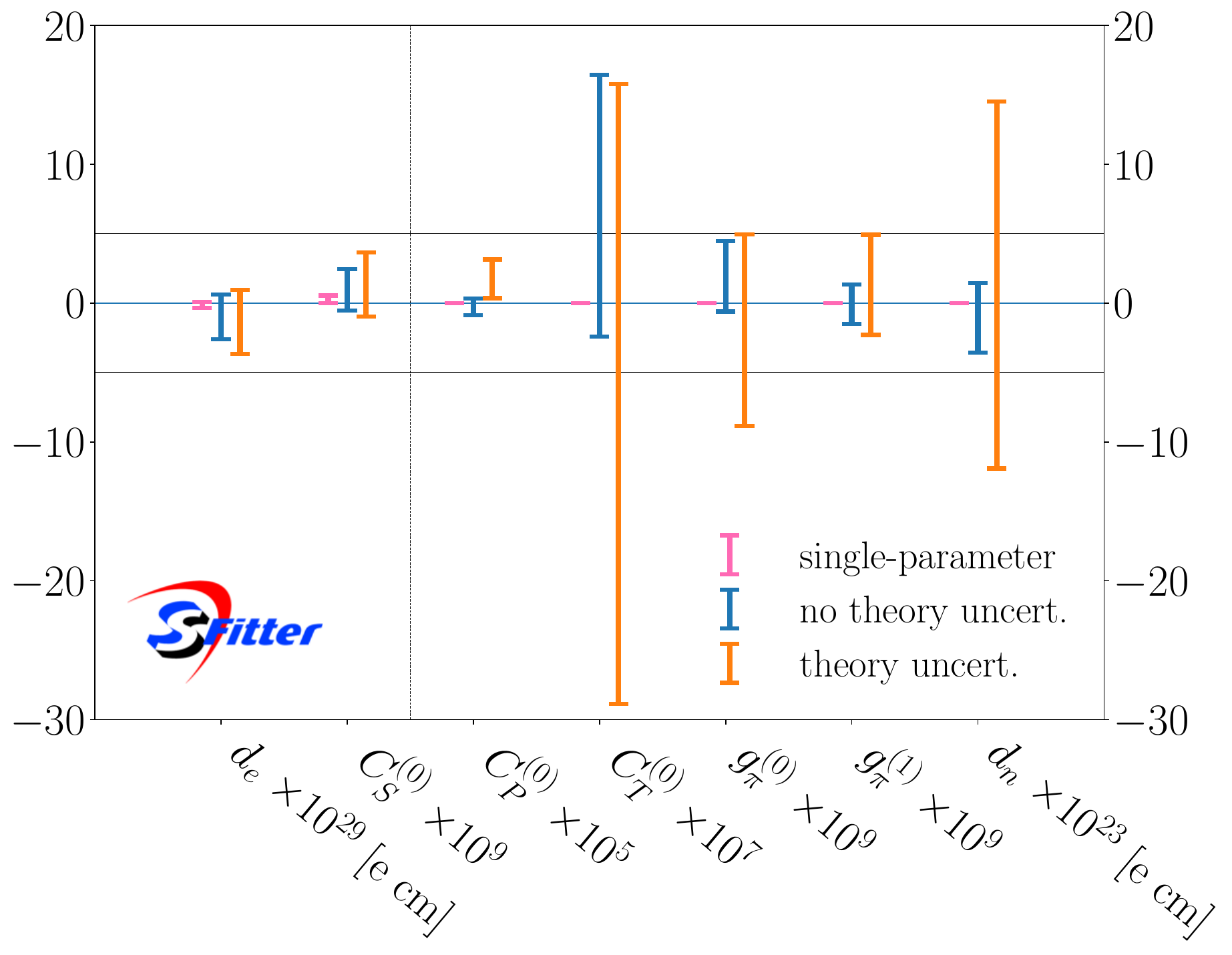}
\caption{$68\%$ CL constraints from the global EDM analysis on the 
  parameters of the hadronic-scale Lagrangian. We show (i) hugely
  over-constrained single-parameter ranges allowed by the best available measurement; (ii)
  over-optimistic allowed ranges for profiled single parameters,
  ignoring theory uncertainties; (iii) allowed ranges for profiled
  single parameters including experimental and theory uncertainties.}
\label{fig:bottomline}
\end{figure}

We can choose different Lagrangians to answer this question, starting
with UV-complete models versus EFTs. In the absence of a specific hint
for BSM physics we choose an EFT description. Next, we have a choice
of different energy scales with different degrees of freedom. For a
first \sfitter analysis we rely on the hadronic-scale Lagrangian, valid at
the GeV scale and describing the interactions of electrons and
nucleons. After relating the hadronic-scale Lagrangian to its weak-scale
SMEFT counterpart we constrain the seven Lagrangian parameters
given in Fig.~\ref{fig:bottomline} through 11 independent measurements
given in Tab.~\ref{tab:meas}.

As a toy analysis we look at single-parameter constraints from
individual EDM measurements. These limits are all driven by the same 
small set of highly constraining
measurements, like the open-shell molecular ion HfF$^+$, the neutron
EDM, or the closed-shell atom Hg. The extremely strong constraints indicated in Fig.~\ref{fig:bottomline} do not allow for a cancellation 
of contributions from two model parameters to a
given measurement, at the price of creating a numerical fine-tuning problem 
in the model parameters. 
While the single-parameter estimates 
indicate the strength of an experiment looking for a sign of CP violation,
they should not be confused with a measurement of a given parameter.

%-------------------------------------------
%-------------------------------------------

Our global analysis employs \sfitter, with a focus on the
statistical interpretation and a comprehensive uncertainty
treatment. First, we ignore all theory uncertainties and only consider experimental uncertainties
as uncorrelated and Gaussian (combining the statistical and systematic uncertainties reported in the
respective papers). In this case Bayesian marginalization and profile
likelihood give the same result.

In Sec.~\ref{sec:results_well} we find that a small set of powerful
measurements constrains the electron-hadron interactions as well as a
subset of the hadronic sector. The correlated limits on the electron
EDM parameter $d_e$ and the scalar coupling $C_S^{(0)}$ are especially
strong and factorize from the hadronic sector.  Next, we find in
Sec.~\ref{sec:results_dia} that the constraints on the hadronic sector
from the closed-shell systems show rich correlations, motivating our
global analysis described in Sec.~\ref{sec:results_poor}.  For the
hadronic parameters, the narrow correlations from the strong neutron
and Hg constraints do not appear in profiled 2-dimensional
correlations or single-parameter limits. As a result, the hadronic
model parameters are constrained much worse than the single-parameter
results suggest, as can be seen in Fig.~\ref{fig:bottomline}.

Finally, we show the same limits including theory uncertainties. Such
theory uncertainties always appear when we relate measurements to
fundamental Lagrangian parameters.  We assume a flat likelihood within
allowed ranges of the factors relating the Lagrangian parameters to
the EDM predictions. While the impact of the theory uncertainties on
the factorized $\{d_e, C_S^{(0)}\}$ sector is relatively mild, the correlated
analysis of the hadronic parameters leads to a significant weakening of
the constraints on all model parameters.

%%%%%%%%%%%%%%%%%%%%%%%%%%%%%%%%%%%%%%%%%%%%%%%%%%%%%%%%%%%%%%%%%%%%%%
\subsubsection*{Acknowledgments}

We gratefully acknowledge extremely helpful discussions and
correspondence with Robert Berger, Tim Chupp, Vincenzo Cirigliano, Jordy de Vries, Yohei Ema, Victor Flambaum, Timo Fleig, Konstantin Gaul, Maxim Pospelov, and Adam Ritz.  We also
thank Duarte Azevedo for contributions at an early phase of this
project.  We thank the ECT* in Trento and EURO-LABS (which received funding from the European Union’s Horizon Europe Research and Innovation programme under grant agreement No 101057511) for support and hospitality while submitting this work and presenting the results for the first time. 

NE is funded by the Heidelberg IMPRS \textsl{Precision Tests
  of Fundamental Symmetries}. This research is supported by the
Deutsche Forschungsgemeinschaft (DFG, German Research Foundation)
under grant 396021762 -- TRR~257: \textsl{Particle Physics
  Phenomenology after the Higgs Discovery} and through Germany's
Excellence Strategy EXC~2181/1 -- 390900948 (the \textsl{Heidelberg
  STRUCTURES Excellence Cluster}). Furthermore, we acknowledge support
by the state of Baden-W\"urttemberg through bwHPC and the German
Research Foundation (DFG) through grant no INST 39/963-1 FUGG
(bwForCluster NEMO).

\clearpage
\appendix
%%%%%%%%%%%%%%%%%%%%%%%%%%%%%%%%%%%%%%%%%%%%%%%%%%%%%%%%%%%%%%%%%%%%%%
\section{Alternative parametrization: short-range nucleon EDMs}
\label{sec:alternative}

In an alternative parametrization, we choose $d_{n,p}^\text{sr}$ as our model parameters and perform the global SFitter analysis for the set 
\begin{align}
  c_j \in \Big\{ \; 
  d_e, C_S^{(0)}, C_T^{(0)}, C_P^{(0)}, g_\pi^{(0)}, g_\pi^{(1)}, d_n^\text{sr}  \;  \; 
  \Big\} \; ,
\label{eq:low_paras1_app}
\end{align}
rather than Eq.\eqref{eq:low_paras1}. To remove the proton EDM parameter we now identify
\begin{align}
 d_p^\text{sr} \approx -d_n^\text{sr} 
 \label{eq:same_pn_app}
\end{align}
instead of Eq.\eqref{eq:same_pn}. We now employ the third line of Eq.\eqref{eq:Schiff}, dropping $g_\pi^{(2)}$ as before,
% The linear relations from Eq.\eqref{eq:Schiff} turn into 
% %
% \begin{align}
%    k_{i,S} S_i & = \sum_{c_j \in \{ d_{n,p}^\text{sr},g_\pi^{(0,1,2)} \}} \alpha_{i,c_j}c_j  \notag \\
%    &\approx k_{i,S}\left[ s_{i,n} d_n^\text{sr} + s_{i,p} d_p^\text{sr} + \frac{m_N g_A}{F_\pi}\left( a_{i,0}g_\pi^{(0)} + a_{i,1}g_\pi^{(1)}  \right)\right] \; ,
% \label{eq:Schiff_app}
% \end{align}
%
and noting that the coefficients $\tilde a_{i,0}$ and $\tilde a_{i,1}$ now include additional terms from the relation connecting $d_{n,p}$ and $d_{n,p}^\text{sr}$. 

In this appendix we collect the correspondingly adapted versions of all tables and figures from the main body of the paper, derived using this alternative model parameter choice.
This alternative parameterization comes with different central values and theory uncertainties for $g_\pi^{(0,1)}$ and $d_{n,p}^\text{sr}$. Thus, instead of, for example, $\alpha_{i,g_\pi^{(1)}}$ of the Yb measurement, $\alpha_{i,g_\pi^{(0)}}$ of Xe now includes the zero value and has no impact in the global analysis.

%-------------------------------------------
\begin{table}[b!]
\centering
\begin{small}
\begin{tabular}{c|rrrrrr}
\toprule
System $i$
& $k_{i,S} \left[ \text{cm}/\text{fm}^3 \right]$
& $s_{i,n} \left[ \text{fm}^2 \right]$
& $s_{i,p} \left[ \text{fm}^2 \right]$ 
%& $a_{i,0} \left[ \text{$e$ fm}^3 \right]$ 
%& $a_{i,1} \left[ \text{$e$ fm}^3 \right]$
%& $a_{i,2} \left[ \text{$e$ fm}^3 \right]$\\
\\ \midrule
Tl
%& $d_\text{Tl} = (-4.00\pm4.30) \cdot 10^{-25} \text{$e$cm}$
& $-4.2_{-1.8}^{+2.1}\cdot10^{-18}$ \cite{Gaul:2023hdd}
& $0.14^{\pm0.03}$
& $-0.38_{-0.45}^{+1.38}$ 
%& $0.113_{-0.008}^{+0.017}$
%& $-0.004_{-0.006}^{+0.0}$
%& $-0.226_{-0.03}^{+0.044}$ \cite{Flambaum:1985gv} % n/a
\\
Cs
%& $d_\text{Cs} = (-1.80\pm 6.70_\text{stat}\pm 1.80_\text{syst}) \cdot10^{-24}  \text{$e$cm}$
& $-9.99_{-4.1}^{+2.9}\cdot10^{-18}$ \cite{Gaul:2023hdd}
& 
& $0.1^{\pm0.1}$ 
%& $-0.006_{-0.074}^{+0.0}$
%& $-0.02^{\pm0.01}$ % - 
%& $-0.04_{-0.017}^{+0.0}$ \cite{Dmitriev:2004fk} % -
\\ \midrule
$^{199}$Hg
%& $d_\text{$^{199}$Hg} = (2.20\pm 2.75_\text{stat}\pm 1.48_\text{syst}) \cdot10^{-30}  \text{$e$cm}$
& $-2.26^{\pm0.23}\cdot10^{-17}$\cite{Hubert:2022pnl}
& $0.6_{-0.12}^{+1.33}$
& $0.06_{-0.01}^{+0.20}$ 
%& $0.01_{-0.005}^{+0.4}$
%& $0.02_{-0.05}^{+0.07}$
%& $0.02_{-0.01}^{+0.04}$ \cite{Engel:2013lsa}
%& $-1.56\cdot10^{-4}$ & $-1.56\cdot10^{-5}$
\\
$^{129}$Xe
%& $d_\text{$^{129}$Xe} = (1.4\pm 6.6_\text{stat}\pm 2.0_\text{syst}) \cdot10^{-28}  \text{$e$cm}$
& $3.62^{\pm0.25}\cdot10^{-18}$\cite{Hubert:2022pnl}
& $0.63_{-0.12}^{+0.16}$
& $0.14^{\pm0.03}$ 
%& $-0.008_{-0.042}^{+0.003}$
%& $0.006_{-0.003}^{+0.044}$
%& $-0.009_{-0.091}^{+0.004}$ \cite{Dmitriev:2004fk}
%& $1.70\cdot 10^{-5}$ & $3.51\cdot 10^{-6}$
\\
$^{171}$Yb
%& $d_\text{$^{171}$Yb} = (-6.8\pm 5.1_\text{stat}\pm 1.2_\text{syst}) \cdot10^{-27}  \text{$e$cm}$
& $-2.10^{+0.22}_{-0.0}\cdot10^{-17}$\cite{Dzuba:2009kn,Flambaum:2019kbn}
& $0.54_{-0.11}^{+0.13}$
& $0.054_{-0.014}^{+0.016}$ 
%& $0.01_{-0.0}^{+0.02}$
%& $0.02_{-0.027}^{+0.034}$
%& $0.02_{-0.01}^{+0.04}$ \cite{Dzuba:2007zz}
%& $-1.128\cdot10^{-4}$ & $-1.128\cdot10^{-5}$
\\
$^{225}$Ra
%& $d_\text{$^{225}$Ra} = (4\pm 6_\text{stat}\pm 0.2_\text{syst}) \cdot10^{-24}  \text{$e$cm}$
& $-8.5^{+0.25}_{-0.3}\cdot10^{-17}$\cite{Dzuba:2009kn,Flambaum:2019kbn,Chupp:2014gka}
& $0.63_{-0.12}^{+0.16}$
& $0.14_{-0.03}^{+0.04}$ 
%& $-1.5_{-4.5}^{+0.5}$
%& $6_{-2}^{+18}$
%& $-4_{-11}^{+3}$ \cite{Dobaczewski:2005hz}
%& $-5.36\cdot 10^{-4}$ & $-1.11\cdot 10^{-4}$
\\
TlF
%& $d_\text{TlF} = (-1.7\pm 2.9) \cdot10^{-23}  \text{$e$cm}$
& $-4.59^{\pm0.41}\cdot10^{-13}$\cite{Hubert:2022pnl}
& $0.14^{\pm0.03}$
& $-0.38_{-0.45}^{+1.38}$ 
%& $0.113_{-0.008}^{+0.017}$
%& $-0.004_{-0.006}^{+0.0}$
%& $-0.226_{-0.03}^{+0.044}$ \cite{Flambaum:1985gv}% n/a
%& $-9.47\cdot 10^{-2}$ & $-4.59\cdot 10^{-1}$
\\ \bottomrule
%& $k_{i,S} \left[ \text{cm}/\text{fm}^3 \right]$
%& $s_{i,n} \left[ \text{fm}^2 \right]$
%& $s_{i,p} \left[ \text{fm}^2 \right]$
& $a_{i,0} \left[ \text{$e$ fm}^3 \right]$ 
& $a_{i,1} \left[ \text{$e$ fm}^3 \right]$
& $a_{i,2} \left[ \text{$e$ fm}^3 \right]$\\
\midrule
Tl
%& $d_\text{Tl} = (-4.00\pm4.30) \cdot 10^{-25} \text{$e$cm}$
%& $-4.2_{-1.8}^{+2.1}\cdot10^{-18}$ \cite{Gaul:2023hdd}
%& $0.14^{\pm0.03}$
%& $-0.38_{-0.45}^{+1.38}$
& $0.113_{-0.008}^{+0.017}$
& $-0.004_{-0.006}^{+0.0}$
& $-0.226_{-0.03}^{+0.044}$ \cite{Flambaum:1985gv} % n/a
\\
Cs
%& $d_\text{Cs} = (-1.80\pm 6.70_\text{stat}\pm 1.80_\text{syst}) \cdot10^{-24}  \text{$e$cm}$
%& $-9.99_{-4.1}^{+2.9}\cdot10^{-18}$ \cite{Gaul:2023hdd}
%& 
%& $0.1^{\pm0.1}$
& $-0.006_{-0.074}^{+0.0}$
& $-0.02^{\pm0.01}$ % - 
& $-0.04_{-0.017}^{+0.0}$ \cite{Dmitriev:2004fk} % -
\\ \midrule
$^{199}$Hg
%& $d_\text{$^{199}$Hg} = (2.20\pm 2.75_\text{stat}\pm 1.48_\text{syst}) \cdot10^{-30}  \text{$e$cm}$
%& $-2.26^{\pm0.23}\cdot10^{-17}$\cite{Hubert:2022pnl}
%& $0.6_{-0.12}^{+1.33}$
%& $0.06_{-0.01}^{+0.20}$
& $0.01_{-0.005}^{+0.4}$
& $0.02_{-0.05}^{+0.07}$
& $0.02_{-0.01}^{+0.04}$ \cite{Engel:2013lsa}
%& $-1.56\cdot10^{-4}$ & $-1.56\cdot10^{-5}$
\\
$^{129}$Xe
%& $d_\text{$^{129}$Xe} = (1.4\pm 6.6_\text{stat}\pm 2.0_\text{syst}) \cdot10^{-28}  \text{$e$cm}$
%& $3.62^{\pm0.25}\cdot10^{-18}$\cite{Hubert:2022pnl}
%& $0.63_{-0.12}^{+0.16}$
%& $0.14^{\pm0.03}$
& $-0.008_{-0.042}^{+0.003}$
& $0.006_{-0.003}^{+0.044}$
& $-0.009_{-0.091}^{+0.004}$ \cite{Dmitriev:2004fk}
%& $1.70\cdot 10^{-5}$ & $3.51\cdot 10^{-6}$
\\
$^{171}$Yb
%& $d_\text{$^{171}$Yb} = (-6.8\pm 5.1_\text{stat}\pm 1.2_\text{syst}) \cdot10^{-27}  \text{$e$cm}$
%& $-2.10^{+0.22}_{-0.0}\cdot10^{-17}$\cite{Dzuba:2009kn,Flambaum:2019kbn}
%& $0.54_{-0.11}^{+0.13}$
%& $0.054_{-0.014}^{+0.016}$
& $0.01_{-0.0}^{+0.02}$
& $0.02_{-0.027}^{+0.034}$
& $0.02_{-0.01}^{+0.04}$ \cite{Dzuba:2007zz}
%& $-1.128\cdot10^{-4}$ & $-1.128\cdot10^{-5}$
\\
$^{225}$Ra
%& $d_\text{$^{225}$Ra} = (4\pm 6_\text{stat}\pm 0.2_\text{syst}) \cdot10^{-24}  \text{$e$cm}$
%& $-8.5^{+0.25}_{-0.3}\cdot10^{-17}$\cite{Dzuba:2009kn,Flambaum:2019kbn,Chupp:2014gka}
%& $0.63_{-0.12}^{+0.16}$
%& $0.14_{-0.03}^{+0.04}$
& $-1.5_{-4.5}^{+0.5}$
& $6_{-2}^{+18}$
& $-4_{-11}^{+3}$ \cite{Dobaczewski:2005hz}
%& $-5.36\cdot 10^{-4}$ & $-1.11\cdot 10^{-4}$
\\
TlF
%& $d_\text{TlF} = (-1.7\pm 2.9) \cdot10^{-23}  \text{$e$cm}$
%& $-4.59^{\pm0.41}\cdot10^{-13}$\cite{Hubert:2022pnl}
%& $0.14^{\pm0.03}$
%& $-0.38_{-0.45}^{+1.38}$
& $0.113_{-0.008}^{+0.017}$
& $-0.004_{-0.006}^{+0.0}$
& $-0.226_{-0.03}^{+0.044}$ \cite{Flambaum:1985gv}% n/a
%& $-9.47\cdot 10^{-2}$ & $-4.59\cdot 10^{-1}$
\\ \bottomrule
\end{tabular}
\end{small}
\caption{Inputs for computing the hadronic coefficients in Tabs.~\ref{tab:LEC2} and \ref{tab:LEC3_app}.}
%Ranges are inferred from the distribution of literature values, and should be taken as indicative; signs are adapted as necessary to match our conventions. In some cases where the valence nucleon is $n$ (respectively $p$) literature values are not available for the $s_{i,p}$ (respectively, $s_{i,n}$). For these cases we estimate based on the spin fractions of Tab.~\ref{tab:LEC}, as done previously for near-spherical nuclei within the shell model \cite{Dzuba:1984fw
%}. For some discussion of these estimations, and in particular difficulties associated with $s_{\text{Tl},p}$ and $s_{\text{Hg},n}$ see Ref.~\cite{Flambaum:2019kbn} and references therein. We do not report the coefficient $s_{\text{Cs},p}$ separately from $\alpha_{\text{Cs},p}$, since the latter includes also contributions from a nuclear magnetic quadrupole moment \cite{Khriplovich:1975dh,Khriplovich:1997ga}. This should be borne in mind if using Eq.\eqref{eq:Schiff} for $^{133}$Cs or nuclei with spin $I>1/2$; see also \cite{Flambaum:2014jta,GINGES200463} for relating a magnetic quadrupole moment to our coefficients $a_{i,j}$. References given for $a_{i,2}$ indicate the sources for central values for all three coefficients $a_{i,j}$; the ranges are inferred from the broader distribution of literature values, see e.g. \cite{Engel:2013lsa} for a related discussion.}
\label{tab:hadronic_details_app}
\end{table}
%-------------------------------------------

Table~\ref{tab:hadronic_details_app} provides the hadronic background to the 
input of our global analysis, given in Tabs.~\ref{tab:LEC2} and \ref{tab:LEC3_app}.
In some cases, where the valence nucleon is $n$ (respectively $p$) literature values are not available for the $s_{i,p}$ (respectively, $s_{i,n}$), we
given an estimate based on the spin fractions of Tab.~\ref{tab:LEC}, as done previously for near-spherical nuclei within the shell model \cite{Dzuba:1984fw
}. For some discussion of these estimations, and in particular difficulties associated with $s_{\text{Tl},p}$ and $s_{\text{Hg},n}$ see Ref.~\cite{Flambaum:2019kbn} and references therein. We do not report the coefficient $s_{\text{Cs},p}$ separately from $\alpha_{\text{Cs},p}$, since the latter includes also contributions from a nuclear magnetic quadrupole moment \cite{Khriplovich:1975dh,Khriplovich:1997ga}. This should be borne in mind if using Eq.\eqref{eq:Schiff} for $^{133}$Cs or nuclei with spin $I>1/2$; see also \cite{Flambaum:2014jta,GINGES200463} for relating a magnetic quadrupole moment to our coefficients $a_{i,j}$. References given for $a_{i,2}$ indicate the sources for central values for all three coefficients $a_{i,j}$; the ranges are inferred from the broader distribution of literature values, see e.g. \cite{Engel:2013lsa} for a related discussion.

%-------------------------------------------
\begin{sidewaystable}[ph!]
%\centering
\begin{footnotesize}
\resizebox{\textwidth}{!}{
\begin{tabular}{@{}c|rrrrrrrr}
\toprule
System $i$
& $\alpha_{i,d_e}$
& $\alpha_{i,C_S^{(0)}} \left[ \text{$e$ cm} \right]$ 
%& $\dfrac{\alpha_{i,C_S}}{\alpha_{i,d_e}} \left[ \text{$e$ cm} \right]$
& $\alpha_{i,C_P^{(0)}} \left[ \text{$e$ cm} \right]$
& $\alpha_{i,C_T^{(0)}} \left[ \text{$e$ cm} \right]$
& $\alpha_{i,g_\pi^{(0)}} \left[ \text{$e$ cm} \right]$
& $\alpha_{i,g_\pi^{(1)}} \left[ \text{$e$ cm} \right]$
& $\alpha_{i,d_n^\text{sr}}$
& $\alpha_{i,d_p^\text{sr}}$\\
\midrule
$n$
& $-$ & $-$ & $-$ & $-$
& $1.38^{\pm0.02}\cdot 10^{-14}$ 
& $2.73^{\pm0.02}\cdot 10^{-16}$ 
& $1$ & $(-1)$
\\ \midrule
$^{205}$Tl
%& $d_\text{Tl} = (-4.00\pm4.30) \cdot 10^{-25} \text{$e$cm}$
& $-558^{\pm28}$ \cite{sym12040498}
& $-6.77^{\pm0.34}\cdot10^{-18}$% $-6.77\cdot10^{-18}$
& $1.4^{+2.5}_{-0.8}\cdot10^{-19}$ %$1.46\cdot10^{-19}$ % $8\cdot10^{-24}$
& $8.8^{+4.0}_{-1.2}\cdot10^{-21}$ % $3\cdot10^{-21}$
& $-6.74^{+4.85}_{-5.12}\cdot10^{-18}$ 
& $2.20^{+6.53}_{-1.44}\cdot10^{-19}$ 
& $-5.75^{+3.59}_{-3.77}\cdot10^{-6}$ 
& $1.61^{+3.35}_{-7.56}\cdot10^{-5}$
\\
$^{133}$Cs
%& $d_\text{Cs} = (-1.80\pm 6.70_\text{stat}\pm 1.80_\text{syst}) \cdot10^{-24}  \text{$e$cm}$
& $123^{\pm4}$ \cite{Fleig:2018bsf,Chupp:2017rkp}
& $7.80^{+0.2}_{-0.8}\cdot10^{-19}$ % $7.8\cdot10^{-19}$ 
& $-1.4^{+0.8}_{-2.2}\cdot10^{-20}$ % $2.2\cdot10^{-23}$
& $1.7^{\pm0.4}\cdot10^{-20}$ % $9.2\cdot10^{-21}$
& $-$ & $-$ & $-$ & $-$ 
\\ \midrule
$^{199}$Hg
%& $d_\text{$^{199}$Hg} = (2.20\pm 2.75_\text{stat}\pm 1.48_\text{syst}) \cdot10^{-30}  \text{$e$cm}$
%& $-0.012^{+0.0094}_{-0.002}$ \cite{Fleig:2018bsf,Gaul:2023hdd}
& $11.6^{+10}_{-18}\cdot10^{-3}$ \cite{Martensson-Pendrill_1987,Gaul:2023hdd}
& $-1.26_{-1.6}^{+0.7}\cdot10^{-21}$ % $-2.8\cdot 10^{-22}$ 
& $6.6_{-1.6}^{+2.4}\cdot10^{-23}$ % $6\cdot 10^{-23}$
& $-6.4_{-4}^{+3}\cdot10^{-21}$ % $1.7\cdot 10^{-20}$
& $-4.70_{-19.4}^{+3.13}\cdot10^{-18}$
& $-6.13_{-25.7}^{+16.7}\cdot 10^{-18}$
& $-1.36^{+0.39}_{-3.45}\cdot10^{-4}$ % $-5.3\cdot10^{-4}$
& $-1.36^{+0.39}_{-5.11}\cdot10^{-5}$ % $-5.6\cdot10^{-4}$
\\
$^{129}$Xe
%& $d_\text{$^{129}$Xe} = (1.4\pm 6.6_\text{stat}\pm 2.0_\text{syst}) \cdot10^{-28}  \text{$e$cm}$
& $-8^{+0}_{-8}\cdot 10^{-4}$ \cite{Martensson-Pendrill_1987,Gaul:2023hdd} %$1.0\cdot 10^{-4}$ \cite{Dzuba:2009kn} 
& $-2.1_{-2.5}^{+1.2}\cdot10^{-22}$ % $0.71\cdot 10^{-23}$ \cite{PhysRevA.103.012807}
& $1.7_{-0.4}^{+0.5}\cdot10^{-23}$ % $1.6\cdot 10^{-23}$
& $1.24_{-0.61}^{+0.78}\cdot10^{-21}$ % $5.7\cdot 10^{-21}$ %$-6.1\cdot 10^{-21}$  \cite{PhysRevA.103.012807}
& $-1.53_{-24.5}^{+3.08}\cdot 10^{-19}$
& $2.98_{-1.71}^{+24.5}\cdot 10^{-19}$
& $2.29_{-0.58}^{+0.77}\cdot10^{-5}$ % $1.7\cdot 10^{-5}$
& $4.89^{+1.64}_{-1.25}\cdot10^{-6}$ % $3.5\cdot 10^{-6}$
\\
$^{171}$Yb
%& $d_\text{$^{171}$Yb} = (-6.8\pm 5.1_\text{stat}\pm 1.2_\text{syst}) \cdot10^{-27}  \text{$e$cm}$
& $1.44^{+165}_{-4.5}\cdot10^{-3}$ \cite{Gaul:2023hdd} %$5.5\cdot 10^{-4}$ \cite{Dzuba:2009kn}
& $-1.31_{-1.06}^{+0.55}\cdot10^{-21}$ % $-2.7\cdot 10^{-22}$
& $4.93_{-1.55}^{+3.54}\cdot10^{-23}$ % $4\cdot 10^{-23}$
& $-3.68_{-2.43}^{+1.86}\cdot10^{-21}$ % $1.2\cdot 10^{-20}$
& $-4.21^{+0.94}_{-6.52}\cdot 10^{-18}$
& $-5.70^{+7.77}_{-10.4}\cdot 10^{-18}$
& $-1.13^{+0.32}_{-0.28}\cdot10^{-4}$ % $-4.3\cdot 10^{-4}$
& $-1.13^{+0.32}_{-0.28}\cdot10^{-5}$ % $-4.3\cdot 10^{-5}$
\\
$^{225}$Ra
%& $d_\text{$^{225}$Ra} = (4\pm 6_\text{stat}\pm 0.2_\text{syst}) \cdot10^{-24}  \text{$e$cm}$
%& $-0.054^{\pm0.002}$ \cite{Gaul:2023hdd}%$-5.6 \cdot 10^{-3}$ \cite{Dzuba:2009kn}
& $-5.4_{-2.0}^{+134}\cdot10^{-2}$ \cite{Gaul:2023hdd}
& $1.13_{-0.51}^{+2.94}\cdot10^{-20}$ % $-8.6\cdot10^{-22}$ %TODO!
& $-7.63_{-3.88}^{+2.05}\cdot10^{-22}$ % $1.9\cdot10^{-22}$
& $-4.5_{-2.5}^{+2.0}\cdot10^{-20}$ % $5.3\cdot10^{-20}$
& $1.72_{-0.67}^{+5.73}\cdot10^{-15}$
& $-6.89_{-22.9}^{+2.66}\cdot10^{-15}$
& $-5.38^{+1.20}_{-1.58}\cdot10^{-4}$ % $-5.4\cdot 10^{-4}$
& $-1.19^{+0.27}_{-0.35}\cdot10^{-4}$ % $-1.1\cdot 10^{-4}$
\\
TlF
%& $d_\text{TlF} = (-1.7\pm 2.9) \cdot10^{-23}  \text{$e$cm}$
& $1.36_{-0.32}^{+0.36}\cdot 10^{3}$ \cite{Gaul:2023hdd} %\cite{Cho:1991ig,Barr:1992cm}
& $1.44_{-0.5}^{+0.8}\cdot10^{-16}$ % $2.9\cdot 10^{-18}$ %TODO
& $1.51_{-5.6}^{+2.2}\cdot10^{-18}$ % $3\cdot 10^{-19}$ %TODO
& $1.87_{-0.17}^{+0.19}\cdot10^{-15}$ % $2.7\cdot 10^{-16}$
& $-7.36_{-2.60}^{+2.51}\cdot 10^{-13}$
& $2.40^{+4.93}_{-0.53}\cdot 10^{-14}$
& $-6.28^{+2.01}_{-1.72}\cdot10^{-1}$ % $-0.095$
& $1.76^{+2.41}_{-6.76}$ % $-0.46$
\\ \midrule
HfF$^+$
& 1
& $9.17^{\pm0.06} \cdot 10^{-21}$
& $-$ & $-$ & $-$ & $-$ & $-$ & $-$
\\
ThO
& 1
& $1.51^{+0}_{-0.2} \cdot 10^{-20}$
& $-$ & $-$ & $-$ & $-$ & $-$ & $-$
\\
YbF
& 1
& $8.99^{\pm0.70} \cdot 10^{-21}$ 
& $-$ & $-$ & $-$ & $-$ & $-$ & $-$
\\ \midrule
& $\eta^{(m)}_{i,d_e} \left[  \dfrac{\text{mrad}}{\text{s $e$ cm}} \right]$
& $k^{(m)}_{i,C_S} \left[  \dfrac{\text{mrad}}{\text{s}} \right]$
& $\alpha_{i,C_P}$
& $\alpha_{i,C_T}$
& $\alpha_{i,g_\pi^{(0)}}$
& $\alpha_{i,g_\pi^{(1)}}$
& $\alpha_{i,d_n^\text{sr}}$ 
& $\alpha_{i,d_p^\text{sr}}$ 
\\ \midrule
%\multirow{3}{*}{molecules}
HfF$^+$
%& $2\pi (0.10 \pm 0.87 \pm 0.20) \dfrac{\text{mrad}}{\text{s}}$~\cite{Cairncross:2017fip}
& $3.49^{\pm0.14} \cdot 10^{28}$ \cite{Fleig:2018bsf,10.1063/1.4993622,Fleig:2017mls,Petrov:2007zz,PhysRevA.73.062108}
& $3.2^{+0.1}_{-0.2} \cdot 10^{8}$ \cite{Fleig:2018bsf,10.1063/1.4993622,Fleig:2017mls}
& $-$ & $-$ & $-$ & $-$ & $-$ & $-$
\\
ThO
%& $-510 \pm 373_\text{stat} \pm 310_\text{syst} \dfrac{\mu \text{rad}}{\text{s}}$~\cite{ACME:2018yjb} 
& $-1.21^{+0.05}_{-0.39} \cdot 10^{29}$ \cite{Fleig:2018bsf,10.1063/1.4968229,Meyer:2008gc,10.1063/1.4968597}$^\dagger$
& $-1.82^{+0.42}_{-0.27} \cdot 10^{9}$ \cite{10.1063/1.4968229,Fleig:2018bsf,PhysRevA.84.052108,PhysRevA.85.029901,10.1063/1.4968597}$^\dagger$
& $-$ & $-$ & $-$ & $-$ & $-$ & $-$
\\
YbF
%& $5.30 \pm 12.6 \pm 3.30 \dfrac{\text{mrad}}{\text{s}}$~\cite{Hudson:2011zz}
& $-1.96^{\pm0.15} \cdot 10^{28}$ \cite{Fleig:2018bsf,PhysRevA.84.052108,PhysRevA.85.029901,PhysRevA.93.042507,PhysRevA.90.022501}
& $-1.76^{\pm0.2} \cdot 10^{8}$ \cite{Fleig:2018bsf,PhysRevA.84.052108,PhysRevA.93.042507,PhysRevA.85.029901}
& $-$ & $-$ & $-$ & $-$ & $-$ & $-$
\\ \bottomrule
\end{tabular}
}
\end{footnotesize}
\caption{Central values and theory uncertainties for the $\alpha$-parameters defined in
  Eq.\eqref{eq:linear0}, now using Eq.\eqref{eq:d_np} to treat $d_{n,p}^{\text{sr}}$ as model parameters and assuming $d_{n}^{\text{sr}} = -d_{p}^{\text{sr}}$ in lieu of Eq.\eqref{eq:same_pn}. The coefficients $\alpha_{i,g_\pi^{(0,1)}}$ are accordingly modified via Eq.\eqref{eq:Schiff}, and differ from those in Tab.\ref{tab:LEC2}.  Here a ``$-$'' means that we neglect the dependence in
  our global analysis. $^\dagger$There appears to be an overall sign
  error in the coefficients reported for ThO in Table~4 of
  Ref.~\cite{Fleig:2018bsf}.   The values for Tl and Cs of
  $\alpha_{i,C_T^{(0)}}$ are estimated by simple analytical
  calculations~\cite{GINGES200463}, and the uncertainties quoted here
  are estimated as approximately twice those arising from the relevant
  hadronic matrix elements.}
\label{tab:LEC3_app}
\end{sidewaystable} 
%-------------------------------------------

%-------------------------------------------
\begin{table}[t]
\centering
\begin{small}
\begin{tabular}{c|rrrr}
\toprule
System $i$
& $d_e\left[ \text{$e$ cm} \right]$
& $C_S^{(0)}$ 
& $C_P^{(0)}$
& $C_T^{(0)}$ \\
\midrule
Tl
& $\left( 7.2 \pm 7.7 \right)\cdot 10^{-28}$
& $\left( 5.9 \pm 6.4 \right)\cdot 10^{-8}$
& $\left(-2.9 \pm 3.1 \right)\cdot 10^{-6}$
& $\left(-4.5 \pm 4.9 \right)\cdot 10^{-5}$
\\
Cs
& $\left(-1.5 \pm 5.6 \right)\cdot 10^{-26}$
& $\left(-2.3 \pm 8.9 \right)\cdot 10^{-6}$
& $\left( 1.3 \pm 5.0   \right)\cdot 10^{-4}$
& $\left(-1.1 \pm 4.1 \right)\cdot 10^{-4}$
\\ \midrule
$^{199}$Hg
& $\left(1.9 \pm 2.7 \right)\cdot 10^{-28}$
& $\left(-1.7 \pm 2.5 \right)\cdot 10^{-9}$
& $\left( 3.3 \pm 4.7 \right)\cdot 10^{-8}$
& $\left(-3.4 \pm 4.9 \right)\cdot 10^{-10}$
\\
$^{129}$Xe
& $\left( 2.2 \pm 2.3 \right)\cdot 10^{-25}$ 
& $\left( 8.4 \pm 8.7 \right)\cdot 10^{-7}$ 
& $\left(-1.0 \pm 1.1 \right)\cdot 10^{-5}$ 
& $\left(-1.4 \pm 1.5 \right)\cdot 10^{-7}$ 
\\
$^{171}$Yb
& $\left( -4.7 \pm 3.6 \right)\cdot 10^{-24}$
& $\left( 5.2 \pm 4.0 \right)\cdot 10^{-6}$
& $\left(-1.4 \pm 1.1 \right)\cdot 10^{-4}$
& $\left( 1.8 \pm 1.4 \right)\cdot 10^{-6}$
\\
$^{225}$Ra
& $\left(-0.7 \pm 1.1 \right)\cdot 10^{-22}$
& $\left( 3.5 \pm 5.3   \right) \cdot 10^{-4}$
& $\left(-5.2 \pm 7.9 \right)\cdot 10^{-3}$
& $\left(-0.9 \pm 1.3  \right)\cdot 10^{-4}$
\\
TlF
& $\left(-1.3 \pm 2.1 \right)\cdot 10^{-26}$
& $\left(-1.2 \pm 2.0 \right)\cdot 10^{-7}$
& $\left(-1.1 \pm 1.9  \right)\cdot 10^{-5}$
& $\left(-0.9 \pm 1.6 \right)\cdot 10^{-8}$
\\ \midrule
HfF$^+$
& $\left(-1.3 \pm 2.1 \right)\cdot 10^{-30}$
& $\left(-1.4 \pm 2.3 \right)\cdot 10^{-10}$
\\
ThO
& $\left( 4.3 \pm 4.0 \right)\cdot 10^{-30}$
& $\left( 2.8 \pm 2.7 \right)\cdot 10^{-10}$
\\
YbF
& $\left(-2.4 \pm 5.9 \right)\cdot 10^{-28}$
& $\left(-2.7 \pm 6.6 \right)\cdot 10^{-8}$
\\ \bottomrule
& $g_\pi^{(0)}$
& $g_\pi^{(1)}$
& $d_n^\text{sr}$ 
& $d_p^\text{sr}$\\
\midrule
$n$
& $\left(0 \pm 8.1 \right)\cdot 10^{-13}$ 
& $\left(0 \pm 4.1 \right)\cdot 10^{-11}$
& $\left(0 \pm 1.1 \right)\cdot 10^{-26}$ 
& $\left(0 \pm 1.1 \right)\cdot 10^{-26}$\\
\midrule
Tl
& $\left(5.9 \pm 6.4 \right)\cdot 10^{-8}$ 
& $\left(-1.8 \pm 2.0 \right)\cdot 10^{-6}$
& $\left(7.0 \pm 7.5 \right)\cdot 10^{-20}$ 
& $\left(-2.5 \pm 2.7  \right)\cdot 10^{-20}$
\\ 
\midrule
$^{199}$Hg
& $\left(-4.7 \pm 6.6 \right)\cdot 10^{-13}$
& $\left(-3.6 \pm 5.1 \right)\cdot 10^{-13}$
& $\left(-1.6 \pm 2.3 \right)\cdot 10^{-26}$
& $\left(-1.6 \pm 2.3 \right)\cdot 10^{-25}$
\\
$^{129}$Xe
& $\left( 1.2\pm 1.2 \right)\cdot 10^{-9}$ 
& $\left( 5.9   \pm 6.1 \right)\cdot 10^{-10}$
& $\left(-7.7 \pm 7.9 \right)\cdot 10^{-24}$
& $\left(-3.6 \pm 3.7 \right)\cdot 10^{-23}$
\\
$^{171}$Yb
& $\left( 1.6 \pm 1.2 \right)\cdot 10^{-9}$
& $\left(1.2 \pm 0.9 \right)\cdot 10^{-9}$
& $\left( 6.0 \pm 4.6 \right)\cdot 10^{-23}$
& $\left( 6.0 \pm 4.6 \right)\cdot 10^{-22}$
\\
$^{225}$Ra
& $\left( 2.3 \pm 3.5 \right)\cdot 10^{-9}$
& $\left(-5.8 \pm 8.7 \right)\cdot 10^{-10}$
& $\left(-0.7 \pm 1.1 \right)\cdot 10^{-20}$
& $\left(-3.4 \pm 5.0 \right)\cdot 10^{-20}$
\\
TlF
& $\left(2.3 \pm 3.9  \right)\cdot 10^{-11}$
& $\left(-0.7 \pm 1.2 \right)\cdot 10^{-9}$
& $\left( 2.7 \pm 4.6 \right)\cdot 10^{-23}$
& $\left( -1.0 \pm 1.6 \right)\cdot 10^{-23}$
\\ \bottomrule
\end{tabular}
\end{small}
\caption{Single-parameter ranges allowed by each of the EDM measurements given in Tab.~\ref{tab:meas}, using the coefficients from Tab.~\ref{tab:LEC3_app}.}
\label{tab:onepara_app}
\end{table} 
%------------------------------------------

%-------------------------------------------
\begin{figure}[t]
\centering
\includegraphics[width=0.90\textwidth]{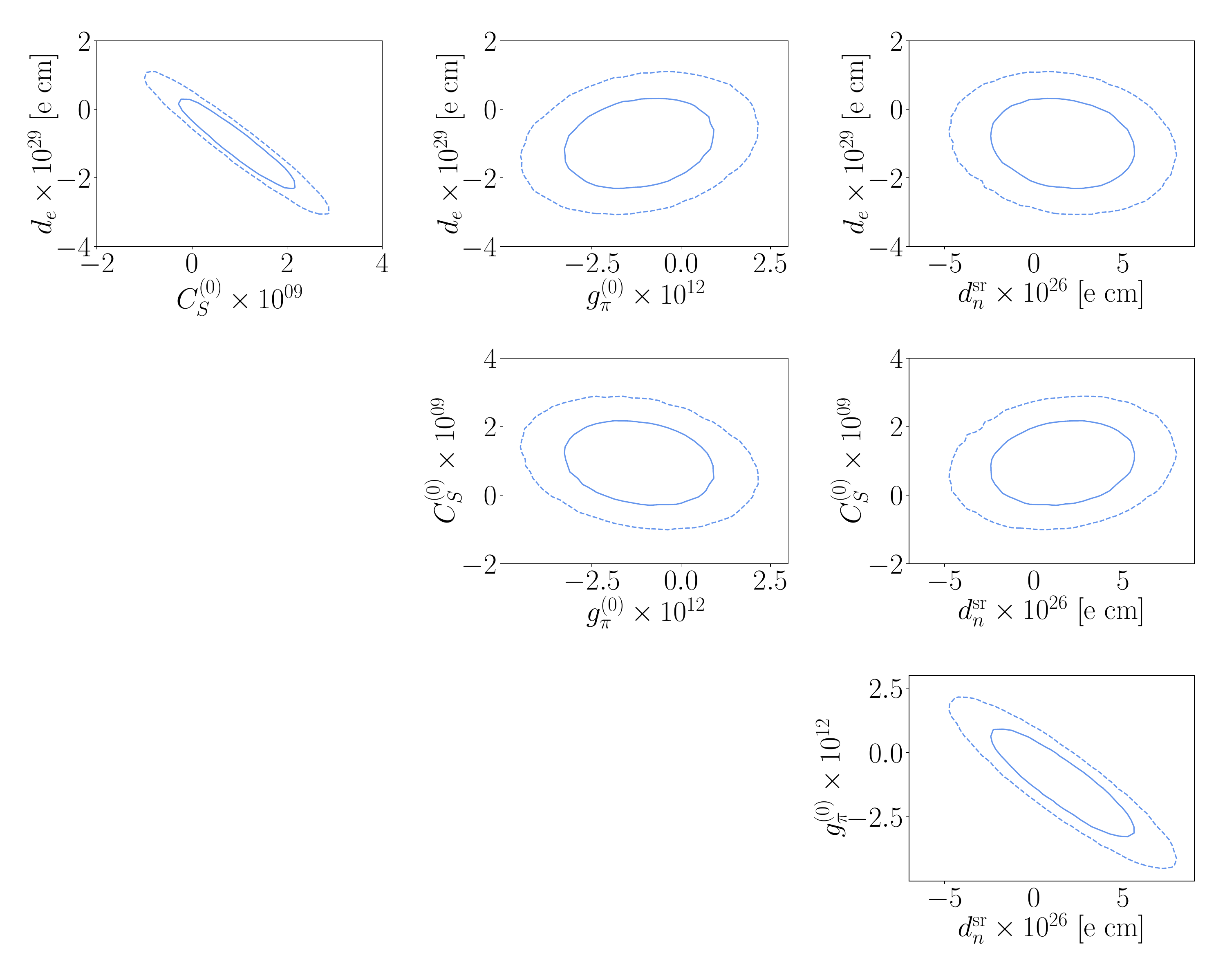}
\caption{Correlations from the 4-dimensional analysis of $\{ d_e, C_S^{(0)},
  g_\pi^{(0)}, d_n^\text{sr} \}$, based on all EDM measurements but
  neglecting theory uncertainties. The ellipses indicate $68\%$ and 
  $95\%$~CL. This figure corresponds to Fig.~\ref{fig:4d} for the $d_{n,p}$ parametrization.}
\label{fig:4d_app}
\end{figure}
%-------------------------------------------

%-------------------------------------------
\begin{figure}[t]
\includegraphics[width=0.99\textwidth]{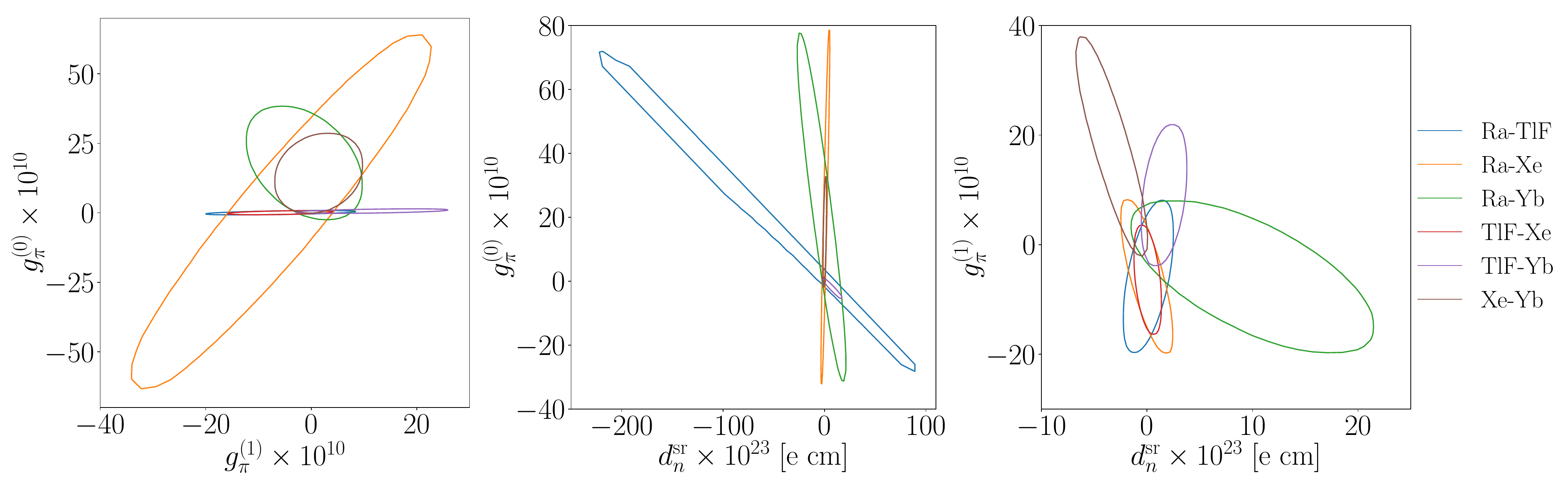}
\caption{Correlations from three 2-dimensional analyses in the $\{
  g_\pi^{(0)} ,g_\pi^{(1)}, d_n^\text{sr}\}$ parameter space, each
  based on a different pair of closed-shell EDM measurements, as indicated by the color. The
  ellipses indicate $68\%$~CL, neglecting theory uncertainties. This figure corresponds to Fig.~\ref{fig:diamag} for the $d_{n,p}$ parametrization.}
\label{fig:diamag_app}
\end{figure}
%-------------------------------------------

%-------------------------------------------
\begin{figure}[t]
\includegraphics[width=0.99\textwidth, page=1]{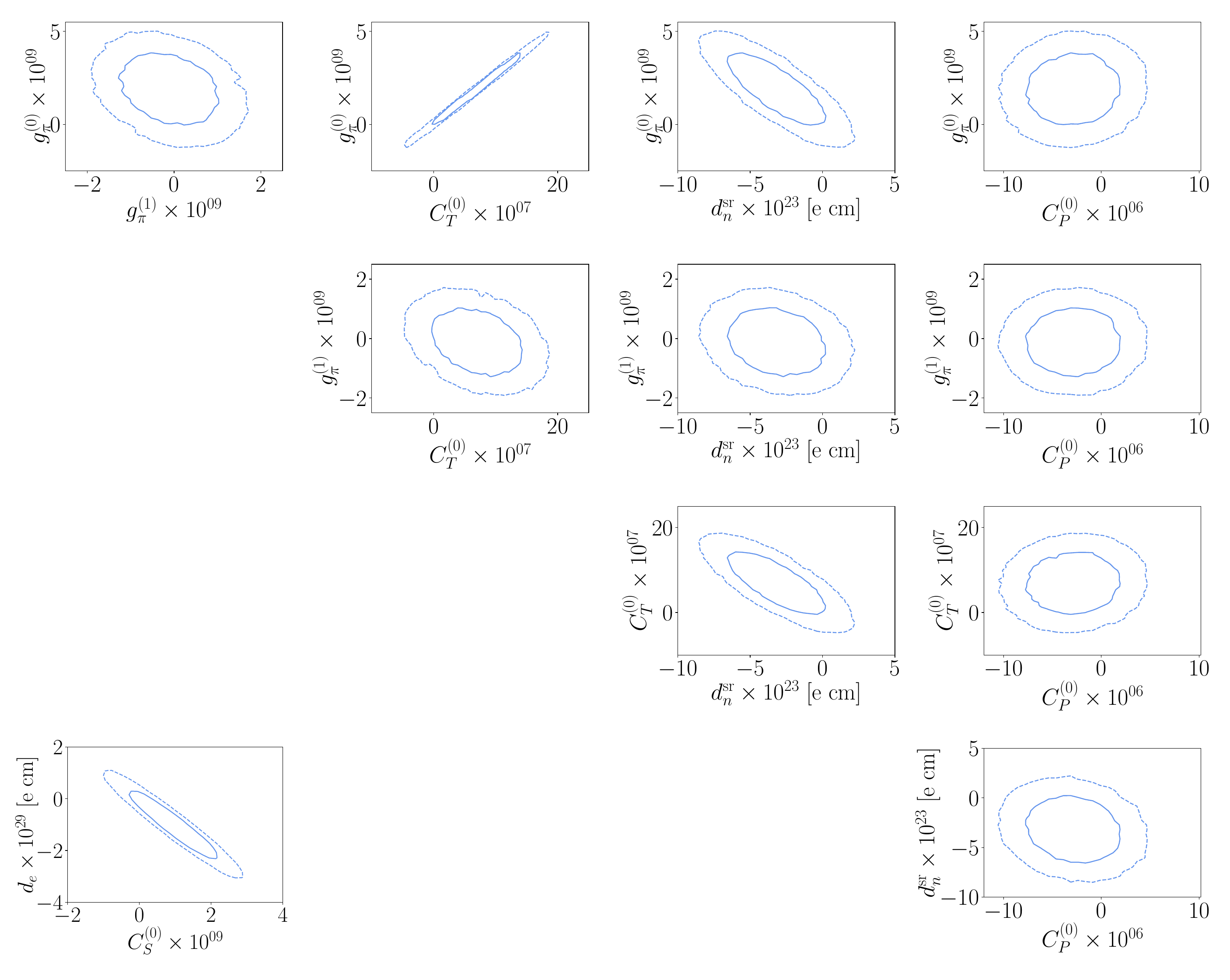}
\caption{Correlations from the 5-dimensional analysis of $\{ C_T^{(0)}, C_P^{(0)},
  g_\pi^{(0)}, g_\pi^{(1)}, d_n^\text{sr} \}$ and the factorized $d_e
  - C_S^{(0)}$ plane from Fig.~\ref{fig:4d}.  We ignore the neutron and Hg
  measurements, which induce narrow correlation patterns in the
  5-dimensional parameters space and do not affect the profiled
  2-dimensional correlations. The ellipses indicate $68\%$ and
  $95\%$~CL, neglecting theory uncertainties. This figure corresponds to Fig.~\ref{fig:badmeas} for the $d_{n,p}$ parametrization.}
\label{fig:badmeas_app}
\end{figure}
%-------------------------------------------

%-------------------------------------------
\begin{figure}[b!]
\centering
\includegraphics[width=0.90\textwidth]{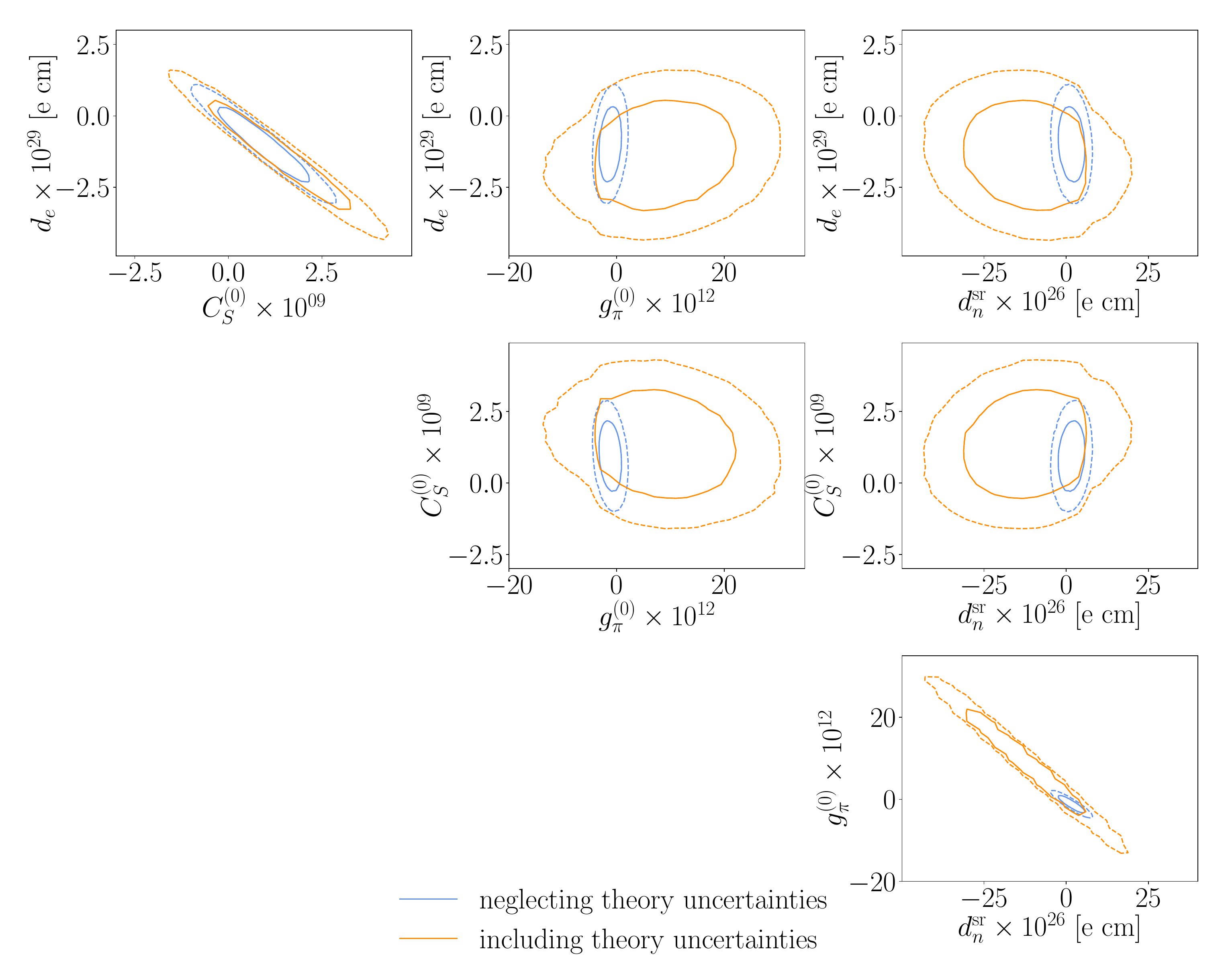}
\caption{Correlations from the 4-dimensional analysis of $\{ d_e, C_S^{(0)},
  g_\pi^{(0)}, d_n^\text{sr} \}$. The orange curves show the effect
  of theory uncertainties on the results of Fig.~\ref{fig:4d}. The
  ellipses indicate $68\%$ and $95\%$~CL. This figure corresponds to Fig.~\ref{fig:4d_th} for the $d_{n,p}$ parametrization.}
\label{fig:4d_th_app}
\end{figure}
%-------------------------------------------

%-------------------------------------------
\begin{figure}[t]
\includegraphics[width=0.99\textwidth, page=2]{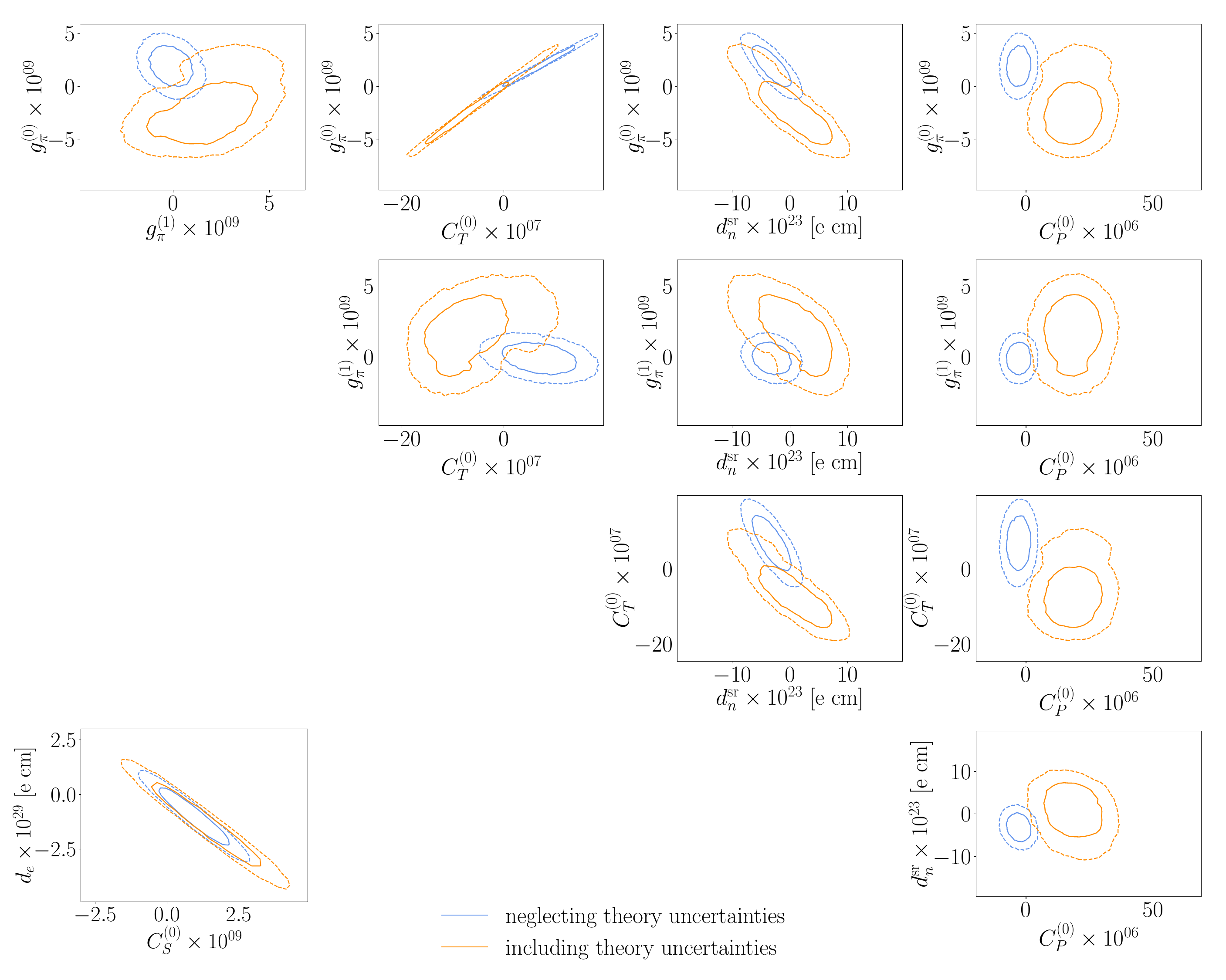}
\caption{Correlations from the 5-dimensional analysis of $\{ C_T^{(0)}, C_P^{(0)},
  g_\pi^{(0)}, g_\pi^{(1)}, d_n^\text{sr} \}$, and the factorized $d_e
  - C_S^{(0)}$ plane from Fig.~\ref{fig:4d_th}.  The orange curves show
  the effect of theory uncertainties on the results of
  Fig~\ref{fig:badmeas}. The ellipses indicate $68\%$ and $95\%$~CL. This figure corresponds to Fig.~\ref{fig:5d_th} for the $d_{n,p}$ parametrization.}
\label{fig:5d_th_app}
\end{figure}
%-------------------------------------------

%-------------------------------------------
\begin{figure}[t]
\centering
\includegraphics[width=0.60\textwidth]{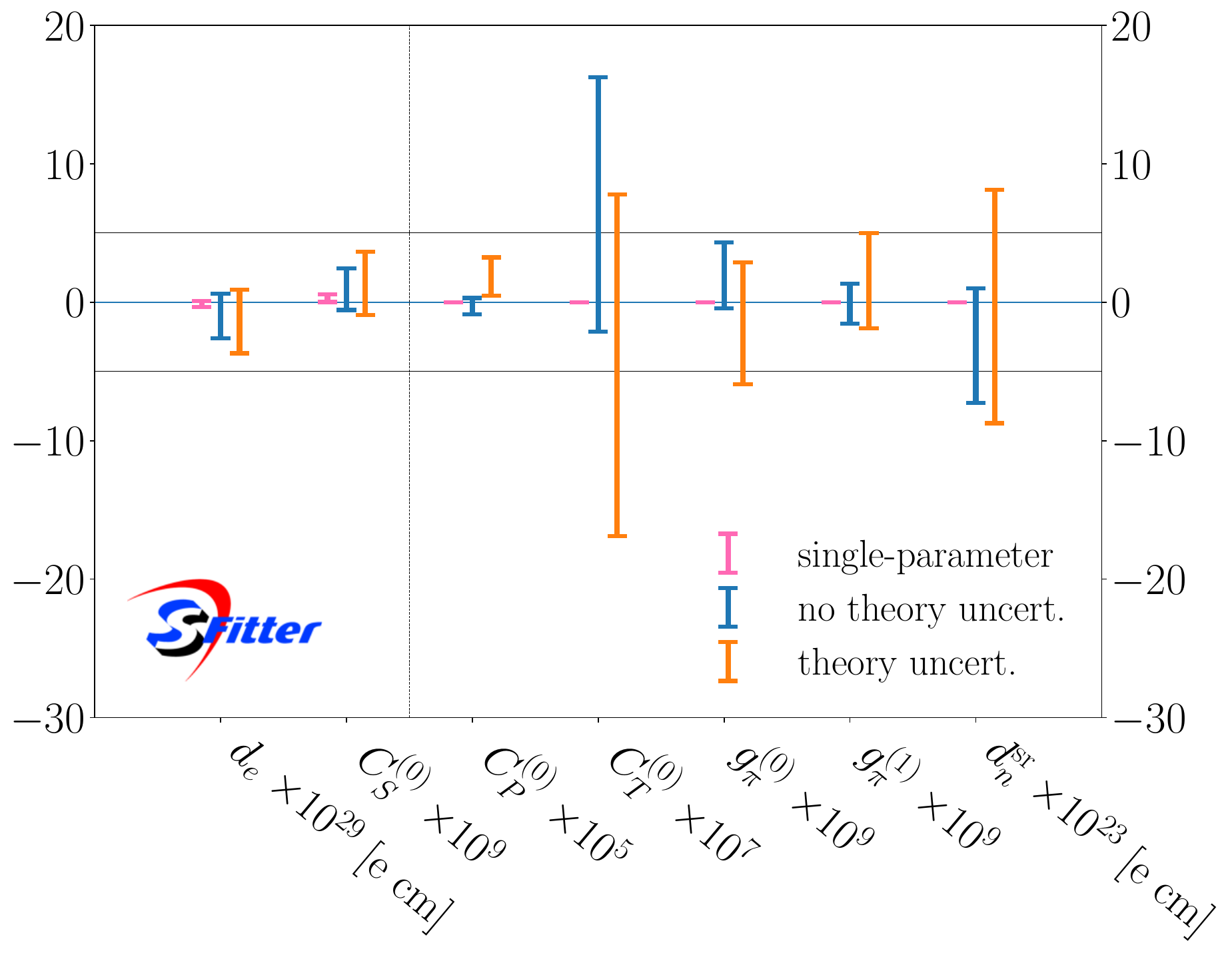}
\caption{$68\%$ CL constraints from the global EDM analysis on the 
  parameters of the hadronic-scale Lagrangian. We show (i) hugely
  over-constrained single-parameter ranges allowed by the best available measurement; (ii)
  over-optimistic allowed ranges for profiled single parameters,
  ignoring theory uncertainties; (iii) allowed ranges for profiled
  single parameters including experimental and theory uncertainties. This figure corresponds to Fig.~\ref{fig:bottomline} for the $d_{n,p}$ parametrization.}
\label{fig:bottomline_app}
\end{figure}
%-------------------------------------------

%
%\begin{align}
%\begin{pmatrix}
%C_T^{(0)} \\ C_P^{(0)} \\ g_\pi^{(0)} \\ g_\pi^{(1)} \\ d_n^\text{sr} 
%\end{pmatrix}
%=
%\begin{pmatrix}
%7.14\cdot 10^{18} & -4.15\cdot 10^{18} & 8.03\cdot 10^{19} & -1.01\cdot 10^{14} &% -2.42\cdot 10^{15}\\
%-1.25\cdot 10^{17} & -6.98\cdot 10^{19} & 1.35\cdot 10^{21} & -5.98\cdot 10^{14} %& -4.08\cdot 10^{16}\\
%-2.94\cdot 10^{14} & -1.80\cdot 10^{17} & 3.37\cdot 10^{18} & -2.85\cdot 10^{12} %& -1.03\cdot 10^{14}\\
%3.44\cdot 10^{14} & -7.65\cdot 10^{15} & -2.98\cdot 10^{18} & 1.95\cdot 10^{12} &% 6.84\cdot 10^{13}\\
%3.96 & 2.48\cdot 10^{3} & -4.57\cdot 10^{4} & 3.88\cdot 10^{-2} & 2.41\\
%\end{pmatrix} \; 
%\begin{pmatrix}
%d_\text{Tl} \\ d_\text{Hg}\\ d_\text{Xe}\\ d_\text{TlF}\\ d_\text{n}
%\end{pmatrix}
%\end{align}
%

\clearpage
%%%%%%%%%%%%%%%%%%%%%%%%%%%%%%%%%%%%%%%%%%%%%%%%%%%%%%%%%%%%%%%%%%%%%%
\bibliographystyle{tepml}
\bibliography{maggie,tilman,by_jordy,by_mrm,literature}
\end{document}